\documentclass[12pt]{iopart}

\usepackage{graphicx}
\usepackage{amssymb}
\usepackage{latexsym}
\usepackage{amsfonts,amssymb}

\def\refs{\marginpar {\scriptsize REFERENCES}}

\def\beq{\begin{equation}}
\def\eeq{\end{equation}} 
\def\bea{\begin{eqnarray}}
\def\eea{\end{eqnarray}}
\def\benu{\begin{enumerate}}
\def\eenu{\end{enumerate}}
\def\nn{\nonumber} 
\def\pa{{\partial}}
\def\f{\frac}
\def\l{\left}
\def\r{\right}
\def\d{{\rm d}}
\def\cR{{\cal R}}

\def\ee{\eta_{\rm e}}
\def\vx{{\bf x}}
\def\vk{{\bf k}}
\def\vka{{\bf k}_{1}}
\def\vkb{{\bf k}_{2}}
\def\vkc{{\bf k}_{3}}
\def\ka{k_{1}}
\def\kb{k_{2}}
\def\kc{k_{3}}
\def\cB{{\cal B}}
\def\cG{{\cal G}}
\def\fnl{f_{_{\rm NL}}}

\newcommand{\uPl}{\mathrm{Pl}}
\newcommand{\uin}{\mathrm{i}}

\newcommand{\usssPl}{\sss{\uPl}}

\newcommand{\Mp}{M_\usssPl}

\newcommand{\email}[1]{\ead{#1}}
\newcommand{\affiliation}[1]{\address{#1}}

\newcommand{\sss}[1]{{\scriptscriptstyle{#1}}}

\begin{document}

\title{The scalar bi-spectrum in the Starobinsky model:~The equilateral 
case}
\author{J\'er\^ome Martin}
\affiliation{Institut d'Astrophysique de Paris, UMR7095-CNRS, 
Universit\'e Pierre et Marie Curie, 98bis boulevard Arago,
75014 Paris, France.}
\email{jmartin@iap.fr}
\author{L.~Sriramkumar\footnote{Current address:~Department 
of Physics, Indian Institute of Technology Madras, Chennai~600036, India. 
E-mail:~sriram@physics.iitm.ac.in}}
\affiliation{Harish-Chandra Research Institute, Chhatnag Road,
Jhunsi, Allahabad~211019, India.}
\date{today} 
\begin{abstract}
  While a featureless, nearly scale invariant, primordial scalar power
  spectrum fits the most recent Cosmic Microwave Background (CMB) data
  rather well, certain features in the spectrum are known to lead to a
  better fit to the data (although, the statistical significance of such
  results remains an open issue). In the inflationary scenario,
  one or more periods of deviations from slow roll are necessary in
  order to generate features in the scalar perturbation spectrum.
  Over the last couple of years, it has been recognized that such
  deviations from slow roll inflation can also result in reasonably
  large non-Gaussianities. The Starobinsky model involves the
  canonical scalar field and consists of a linear inflaton potential
  with a sudden change in the slope. The change in the slope causes a
  brief period of departure from slow roll which, in turn, results in
  a sharp rise in power, along with a burst of oscillations in the
  scalar spectrum for modes that leave the Hubble radius just before
  and during the period of fast roll. The hallmark of the Starobinsky
  model is that it allows the scalar power spectrum to be evaluated
  analytically in terms of the three parameters that describe the
  model, viz. the two slopes that describe the potential on either
  side of the discontinuity and the Hubble scale at the time when the
  field crosses the discontinuity. In this work, we evaluate the
  bi-spectrum of the scalar perturbations in the Starobinsky model in
  the equilateral limit. Remarkably, we find that, just as the power 
  spectrum, all the different contributions to the the bi-spectrum
  too can be evaluated completely analytically and expressed in terms
  of the three paramaters that describe the model.  We show that the
  quantity~$\fnl$, which characterizes the extent of non-Gaussianity,
  can be expressed purely in terms of the ratio of the two slopes on
  either side of the discontinuity in the potential. Further, we find
  that, for certain values of the parameters, $\fnl$~in the
  Starobinsky model can be as large as the mean value that has been
  arrived at from the analysis of the recent CMB data. 
  We also demonstrate that the usual hierarchy of contributions to 
  the bi-spectrum can be altered for certain values of the parameters. 
  Altogether, we find that the Starobinsky model represents a unique
  scenario wherein, even when the slow roll conditions are violated,
  the background, the perturbations as well as the corresponding two 
  and three point correlation functions can be evaluated completely 
  analytically.
  As a consequence, the Starobinsky model can also be used to calibrate 
  numerical codes aimed at computing the non-Gaussianities.
\end{abstract}
\maketitle
\flushbottom


\section{Introduction}\label{sec:introduction}

A nearly scale invariant primordial scalar power spectrum, along with
the assumption of the concordant, background cosmological model
(i.e. the spatially flat, $\Lambda$CDM model), seems to be in
remarkable agreement with the data from the Wilkinson Microwave
Anisotropy Probe (WMAP) of the anisotropies in the Cosmic Microwave
Background (CMB) (for the most recent observations, see
Refs.~\cite{wmap-7}; for earlier results, see Refs.~\cite{wmap-1-5};
for constraints on different inflationary scenarios, see
Refs.~\cite{wmap3-2006,braneinflation-2008,kinflation-2008,reheat-2010}).
But, specific features in the primordial spectrum seem to improve the
fit to the data to a reasonable extent (for an inherently incomplete
list, see Refs.~\cite{rc,quadrupole,quadrupole-sic,pi,l2240,general,oip}), 
although the statistical relevance of such results remains difficult 
to assess.  Notably, while a sharp drop in power on scales
corresponding to the Hubble scale today has been found to fit the low
CMB quadrupole better~\cite{rc,quadrupole,quadrupole-sic,pi}, a burst
of oscillations of a suitable amplitude and over a certain range of
scales seem to provide a considerably improved fit to the outliers in
the CMB angular power spectrum near the multipoles of $l=22$ and
$40$~\cite{l2240}.  Moreover, interestingly, oscillating inflaton
potentials that generate small modulations over a wide range of scales
in the perturbation spectrum have also been found to perform better
against the data than the more conventional power law
spectrum~\cite{oip}.

\par

As is well known, a featureless, scale invariant perturbation spectrum
can be produced by a suitably long epoch of slow roll inflation (see
any of the following texts~\cite{texts} or reviews~\cite{reviews}).
However, generating features in the scalar power spectrum require one
or more periods of deviation from slow roll
inflation~\cite{starobinsky-1992,early-fips,recent-fips}.  For
instance, it is found that one needs a large deviation from slow roll
in order to produce a sharp drop in power, say, so as to fit the low
quadrupole~\cite{starobinsky-1992}.  The larger the deviation from
slow roll, the sharper the drop is found to be and, in fact, a brief
departure from inflation---which leads to a sharp rise in power 
{\it and}\/ a couple of oscillations before the spectrum turns nearly
scale invariant---has been found to provide a good fit to the data at
the lower multipoles~\cite{pi}.  In contrast, a small and short burst
of deviation from slow roll seems sufficient to generate the
oscillations that result in a much better fit to the outliers near the
multipoles of $l=22$ and $40$~\cite{l2240}.  While certain features
indeed lead to a better fit to the data than the standard power law
primordial spectrum, the improvement in the fit is often arrived at
with the introduction of extra parameters. As we pointed out above, 
the statistical significance of features achieved at the cost of
additional parameters, say, from the Bayesian point of
view~\cite{martin-2011a}, remains to be investigated satisfactorily.

\par

During the last few years, a variety of approaches have been developed
to determine the extent of non-Gaussianities in the WMAP data (for an
inexhaustive list, see Refs.~\cite{ng-da}; in this context, also see
the recent reviews~\cite{ng-da-reviews}).  Many of these analyses seem
to indicate that the CMB may possibly possess reasonably large amount
of non-Gaussianities.  For instance, the WMAP seven year data
constrains the parameter $\fnl$ that is often introduced to
characterize the extent of the non-Gaussianity to be $\fnl = 32 \pm
21$ in the local limit, at $68\%$ confidence level (see
Ref.~\cite{wmap-7}; also see the reviews~\cite{ng-da-reviews}).  While
a Gaussian primordial spectrum (which corresponds to a
vanishing~$\fnl$) lies within $2$-$\sigma$, evidently, the mean value
seems to indicate a rather large amount of non-Gaussianity.

\par

The ongoing Planck mission~\cite{planck} is expected to reduce the
above-mentioned uncertainty in the WMAP's determination of $\fnl$ by a
factor of about four or so.  On the theoretical side, it has been
realized that, if Planck indeed detects a reasonable amount of
non-Gaussianity, then it can act as a powerful tool to substantially
constrain the plethora of inflationary models that seem to be allowed
by the currently available data~\cite{ng-da-reviews}.  For example, it
has been established that the canonical scalar field models which lead
to a nearly scale invariant primordial spectrum contain only a
negligible amount of non-Gaussianity, with the magnitude of $\fnl$
turning out to be much smaller than
unity~\cite{earlyng,maldacena-2003}.  Hence, these models will cease
to be viable if non-Gaussianity turns out to be substantial.  However,
it is known that primordial spectra with features, generated due to
one or more deviations from slow roll inflation, can lead to
reasonably large non-Gaussianities~\cite{ng-f}.  (It should be
mentioned that, initial states other than the Bunch-Davies vacuum can
also lead to features and, possibly, relatively substantial
non-Gaussianities~\cite{ng-nvis}.  But, we shall not consider such
situations in this paper.)  Therefore, if non-Gaussianity indeed
proves to be large, then, either one has to reconcile with the fact
that the primordial spectrum contains features or one has to pay more
attention to non-canonical scalar field models such as, say, the
D-brane inflation models~\cite{ng-ncsf,ng-reviews}.

\par

Often, the efforts in evaluating the non-Gaussianities when features
arise in the primordial spectrum due to a departure from slow roll
have involved numerical computations~\cite{ng-f}, and it is
instructive to consider a model wherein it may be possible to evaluate
them analytically.  In this work, we shall consider the Starobinsky
model~\cite{starobinsky-1992}, which involves the canonical scalar
field and consists of a linear inflaton potential with an abrupt
change in the slope at a given point. The change in the slope causes
a brief period of fast roll, which leads to a sharp rise in power and
oscillations in the scalar spectrum for modes that leave the Hubble
radius just before and during the period of fast roll. The important
aspect of the Starobinsky model is that the scalar power spectrum can
be evaluated analytically and, under certain conditions on the
parameters, the analytic spectrum matches the exact spectrum, computed
numerically, extremely well.  We shall focus on the equilateral limit
and illustrate that, remarkably, under the same conditions, all
the contributions to the scalar bi-spectrum too can be calculated
completely analytically and expressed entirely in terms of the three
parameters that describe the Starobinsky model. We shall also discuss
the ranges of the parameters over which the non-Gaussianity in the
model can be large. 

\par 
At this point, we ought to mention that non-Gaussianities in 
the Starobinsky model have, in fact, been computed 
earlier~\cite{sasaki-2010}. But, since the formalism followed in 
the earlier work is rather different from the approach that we 
shall adopt, carrying out an exact comparison between the two 
efforts turns out to be difficult. While the non-Gaussianity 
parameter in the equilateral limit obtained in both the approaches 
lead to a sharp peak near the characteristic scale associated with 
the problem, we find the approach we have adopted here is able to 
capture the finer details as well. We believed that this can attributed
to the $\delta N$ formalism adopted in the earlier work, which is
essentially applicable on super Hubble scales. Despite the 
additional efforts that were made to take into account the 
contributions due to the decaying mode---which become important 
when deviations from slow roll occur (in this context, see, for example,
Refs.~\cite{e-dfsr})---we find that the approach still proves to
be insufficient to arrive at the exact form for the final result.

\par

This paper is organized as follows.  In the following section, we
shall outline the essential aspects of the Starobinsky model.  We
shall first describe the background evolution in the model, and then
go on to discuss the scalar power spectrum that arises in the model.
In particular, we shall highlight the assumptions and approximations
that are made in arriving at analytic expressions for the background
evolution and the scalar power spectrum.  In Sec.~\ref{sec:pebs}, we
shall rapidly sketch the by-now standard procedure for evaluating the
inflationary bi-spectrum of the scalar perturbations and the
non-Gaussianity parameter~$\fnl$.  In Sec.~\ref{sec:dc-bs}, using the
method outlined in Sec.~\ref{sec:pebs}, we shall evaluate the dominant
contribution to the scalar bi-spectrum in the Starobinsky model in the
equilateral limit.  In Sec.~\ref{sec:sdc-bs}, we shall evaluate all
the other sub-dominant contributions as well.  In
Sec.~\ref{sec:l-fnl}, we shall discuss the range of values of the
parameters over which the non-Gaussianity parameter~$\fnl$ can as
large as indicated by the currently observed mean values.  We shall
also touch upon an issue related to the hierarchy of the different
contributions to the bi-spectrum.  Finally, we shall close in
Sec.~\ref{sec:so} with a brief summary and outlook.  We shall relegate
some details concerning the evaluation of certain integrals to an
appendix.

\par

A few words on the conventions and notations that we shall adopt are
in order at this stage of our discussion.  We shall work in units such
that $\hbar=c=1$, and we shall set the Planck mass to be
$\Mp=(8\,\pi\,G)^{1/2}$.  Also, while Greek indices shall refer to the
spacetime coordinates, the Latin indices (barring the sub-script~$k$
which shall represent the wavenumber of the perturbations) shall
denote the three spatial coordinates.  Moreover, we shall work with
the metric signature of $(-,+,+,+)$.  We shall express the various
quantities in terms of either the cosmic time~$t$, the conformal
time~$\eta$ (also, at times, denoted as~$\tau$), or the number of
e-folds~$N$, as is convenient.  As usual, while an overdot shall
denote differentiation with respect the cosmic time, an overprime
shall denote differentiation with respect to the conformal time
coordinate.  Lastly, a plus or a minus sign in the sub-script or the
super-script of any quantity shall denote its value before and after
the field crosses the discontinuity in the potential, respectively.


\section{The Starobinsky model}\label{sec:sm}

In this section, we shall sketch the Starobinsky
model~\cite{starobinsky-1992} in some detail. 
We shall first describe the assumptions and the approximations that go 
into evaluating the background.  
We shall then go on to discuss the scalar modes and the resulting power 
spectrum that arise in the model.


\subsection{The background evolution in the Starobinsky model}
\label{sec:be-sm}

\begin{figure}
\begin{center}
\includegraphics[width=7.5cm]{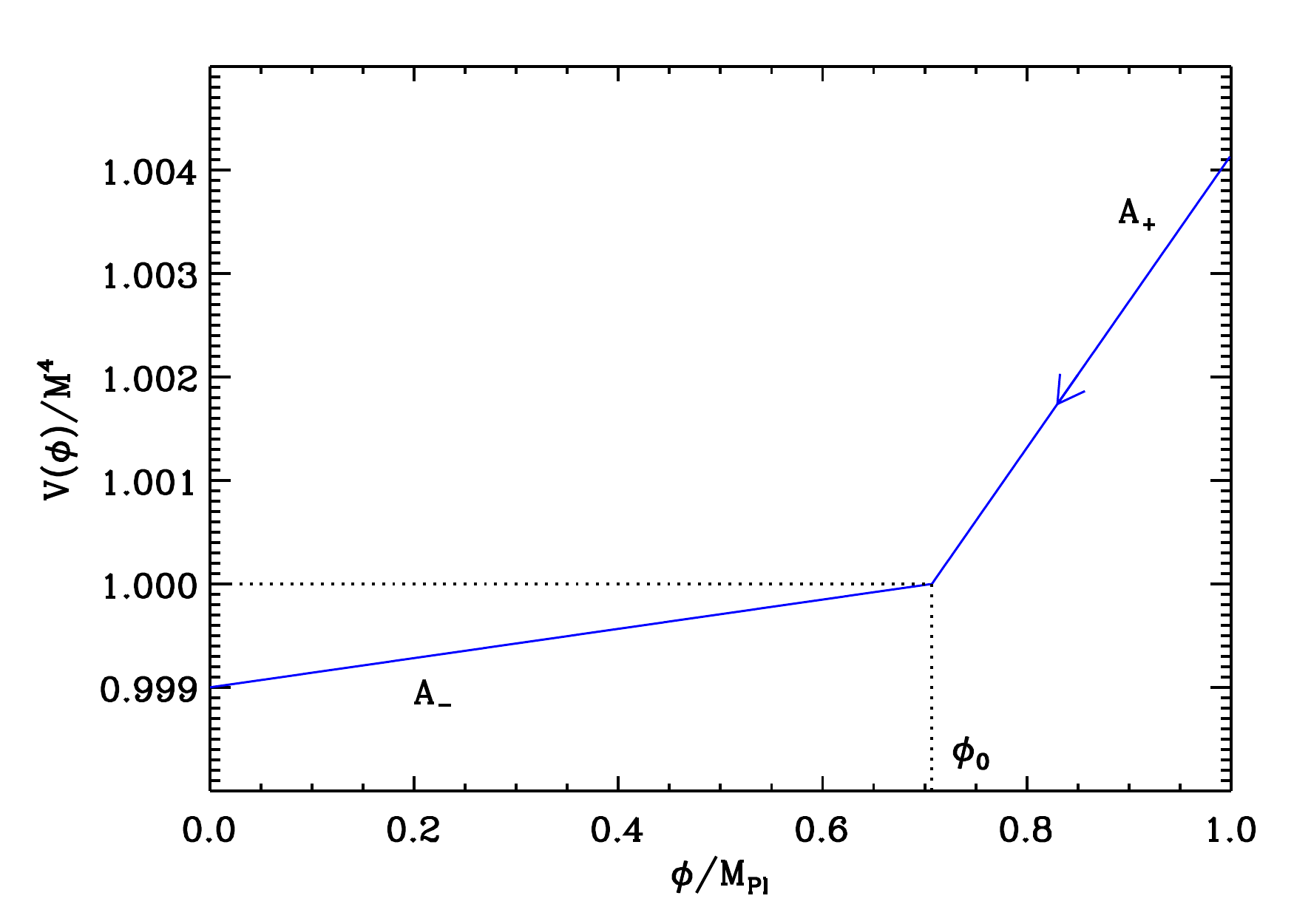}
\includegraphics[width=7.5cm]{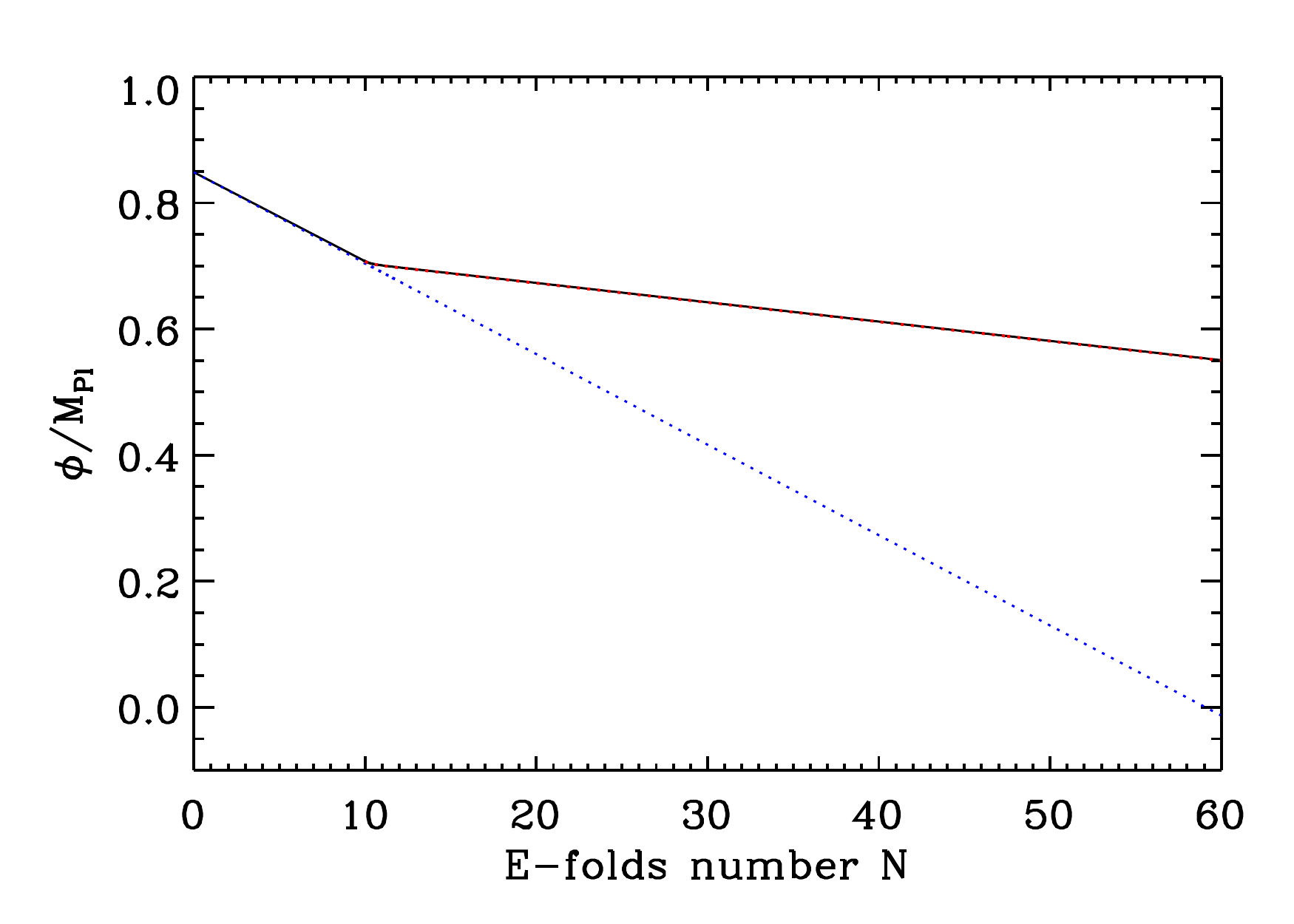}
\includegraphics[width=7.5cm]{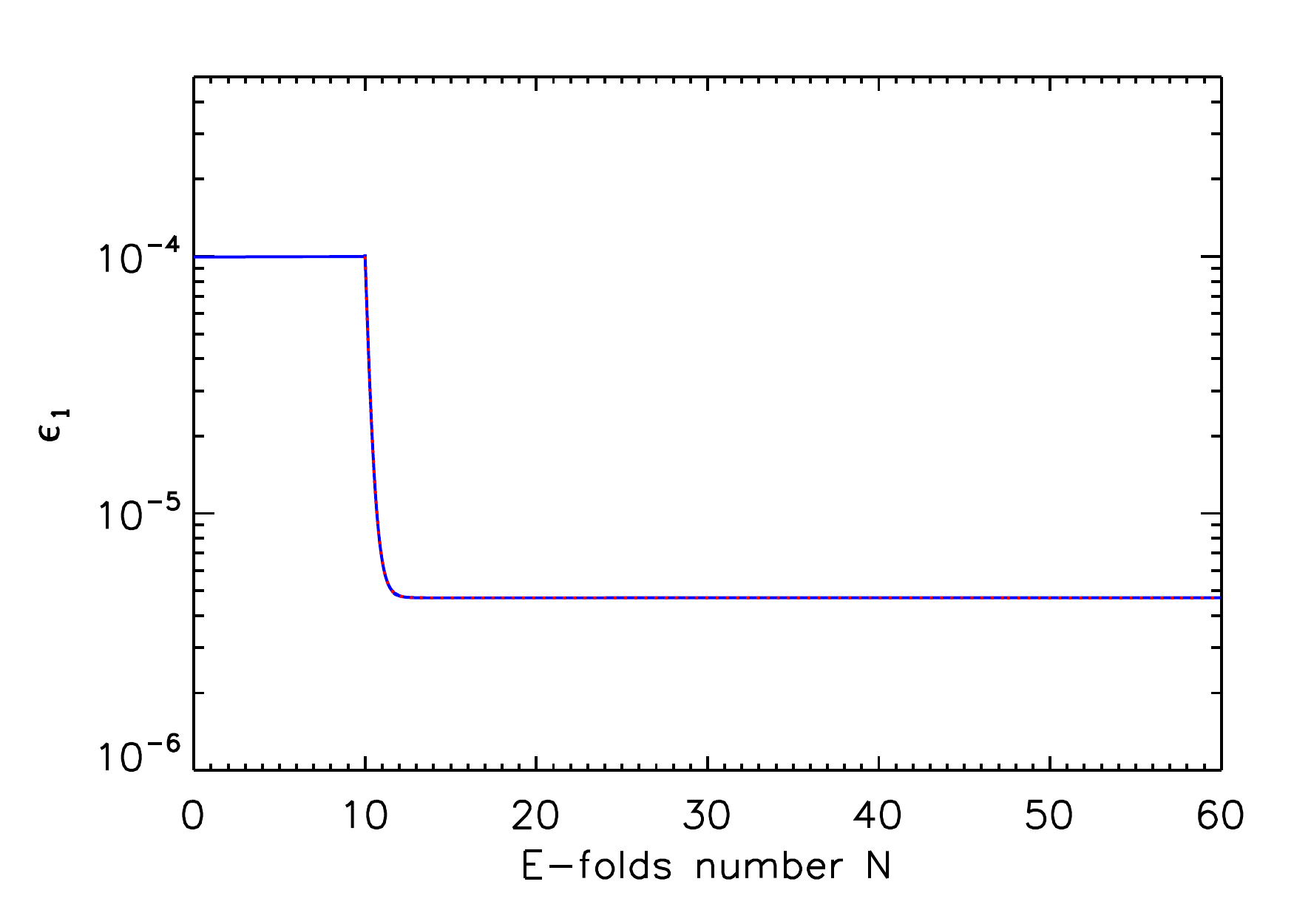}
\includegraphics[width=7.5cm]{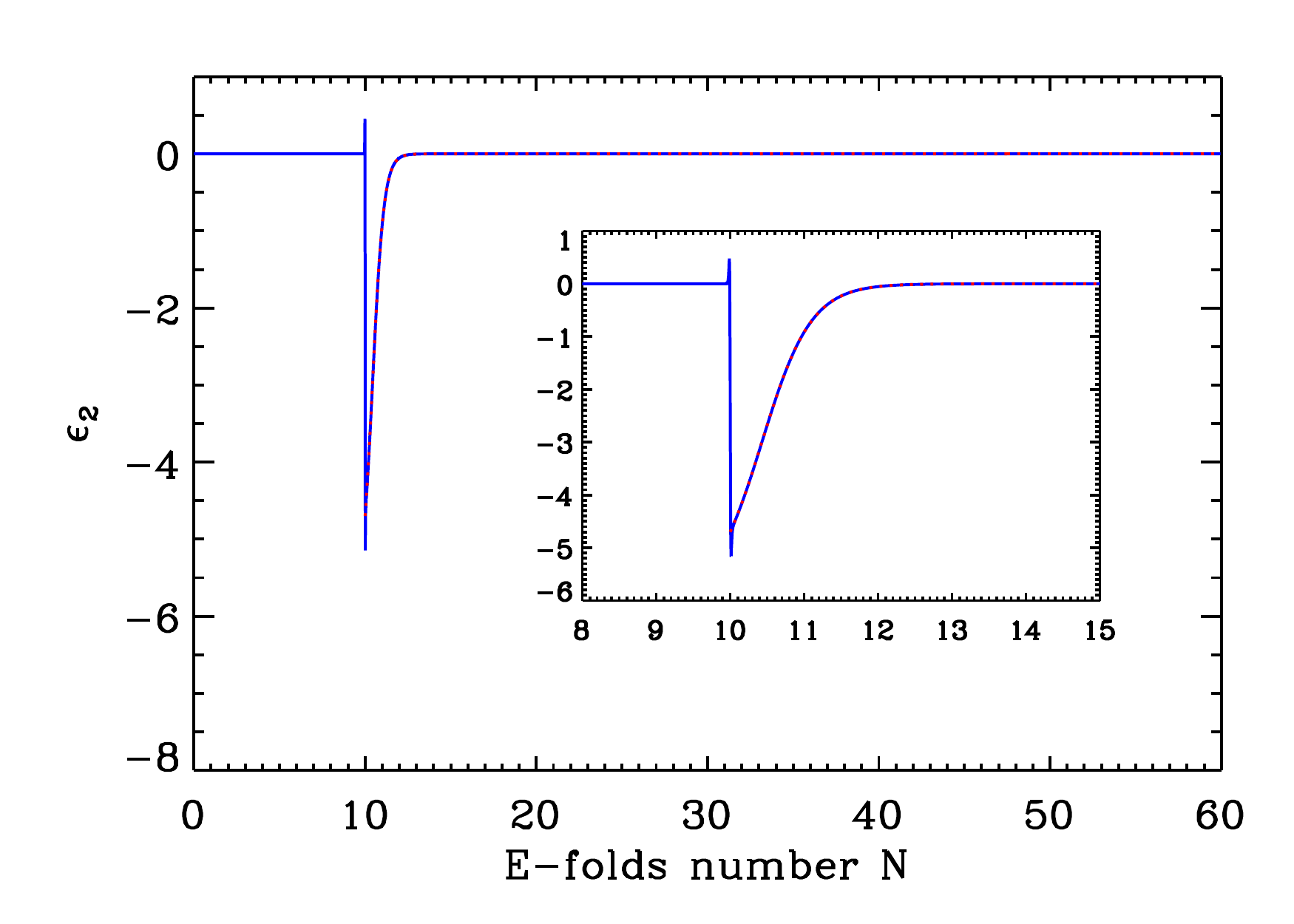}
\caption{The potential, the evolution of the scalar field and the
  first two slow roll parameters in the Starobinsky model.  {\sf Top
    left panel:}~The potential~(\ref{eq:p-sm}) with $\phi_0/\Mp\simeq
  0.707$, $V_0/\Mp^4=M^4/\Mp^4\simeq 2.37\times 10^{-12}$,
  $A_+/\Mp^3\simeq 3.35\times 10^{-14}$ and $A_-/\Mp^3\simeq
  7.26\times 10^{-15}$, which corresponds to $\Delta
  A/A_{-}=(A_--A_{+})/A_{-}\simeq -3.61$. The mass $M$ has been chosen
  such that the model is COBE normalized.  {\sf Top right panel:}~The
  exact, numerically computed, evolution of the field $\phi$ (for the
  above-mentioned values of the parameters) has been plotted as a
  function of the number of e-folds (the solid black line). The
  initial value of the field has been chosen to be
  $\phi_\uin/\Mp\simeq 0.849$, and it corresponds to $N_0=10$, where
  $N_0$ is the e-fold at which the field crosses the discontinuity
  [cf. Eq.~(\ref{eq:nzero})].  The dotted blue line corresponds to the
  slow roll expression~(\ref{eq:srfieldplus}) for the evolution of the
  field.  Clearly, the slow roll result is an excellent approximation
  to the actual result before the discontinuity. The dotted red line
  corresponds to Eq.~(\ref{eq:fieldminus}), and it too accurately
  matches the exact solution after the discontinuity.  {\sf Bottom
    left panel:}~The numerical result for the evolution of the first
  slow roll parameter $\epsilon_1$ (the solid blue curve).  The
  parameters that we work with lead to $\epsilon_{1+}\simeq 10^{-4}$
  [cf. Eq.~(\ref{eq:epsilon1plus})], as is confirmed by the plot.  The
  dotted red curve represents the approximate expression of
  $\epsilon_1$ after the discontinuity,
  viz. Eq.~(\ref{eq:epsilon1minus}). {\sf Bottom right panel:}~The
  solid blue curve represents the exact evolution of the second slow
  roll parameter $\epsilon_2$, obtained numerically. The dotted red
  curve represents the expression~(\ref{eq:epsilon2minus}), which is
  valid after the field has crossed the discontinuity in the
  potential.  The inset highlights the excellent agreement
  between the numerical and the analytical results.}
\label{fig:potstaro}
\end{center}
\end{figure}

Consider a spatially flat Friedmann universe that is described by the 
scale factor $a$, which is driven by the canonical scalar field~$\phi$.
If the scalar field is described by the potential~$V(\phi)$, then the 
evolution of such a system is governed by the following Friedman and 
Klein-Gordon equations:
\begin{equation}
H^2 = \frac{1}{3\,\Mp^2}\,\l[\frac{\dot \phi ^2}{2}+V(\phi)\r]
\quad{\rm and}\quad
\ddot{\phi}+3\,H\,\dot{\phi}+V_{\phi}= 0,\label{eq:be}
\end{equation}
where $H={\dot a}/a$ is the Hubble parameter, and $V_{\phi}\equiv
\d V/\d\phi$.
As we had mentioned, the Starobinsky model consists of a linear
potential with a sharp change in its slope at a given point.  
Let the value of the scalar field where the slope changes abruptly 
be $\phi_0$, and let the slope of the potential above and below 
$\phi_{0}$ be $A_{+}$ and $A_{-}$, respectively.  
Also, let the value of the potential at $\phi=\phi_0$ be, say, $V_{0}$.  
In other words, the potential $V(\phi)$ describing the canonical scalar 
field is given by
\begin{equation}
V(\phi) 
= \l\{\begin{array}{ll}
\displaystyle
V_0 + A_{+}\, \l(\phi-\phi_0\r)\ & {\rm for}\ \phi>\phi_0,\\
\displaystyle
V_0 + A_{-}\, \l(\phi-\phi_0\r)\ & {\rm for}\ \phi<\phi_0.
\end{array}\r.\label{eq:p-sm} 
\end{equation}
This potential is displayed in the top left panel of
Fig.~\ref{fig:potstaro}. 
It is useful to note here that the quantities $A_+$ and $A_-$ are of
dimension three.

Now, consider the case wherein the field is rolling down the above
potential from a value such that $\phi>\phi_{0}$.  
The exact trajectory of the field, computed numerically, is illustrated
in Fig.~\ref{fig:potstaro} (top right panel, solid black line). 
In the slow roll approximation, the number of e-folds, $N\equiv 
\ln \left(a/a_\uin\right)$, is given by
\begin{equation}
N\simeq 
-\frac{1}{\Mp^2}\,
\int_{\phi_\uin}^{\phi}{\rm d}\phi\;\frac{V}{V_{\phi}},
\end{equation}
where $\phi_\uin$ and $a_\uin$ are the initial values (i.e. at $N=0$) 
of the field and the scale factor, respectively.
If the field begins to roll down slowly, the trajectory of the 
field until it reaches the discontinuity at $\phi_{0}$ can be 
easily obtained by carrying out the above integral for~$N$.
One obtains that 
\begin{equation}
\label{eq:srfieldplus}
\phi_+=-\l(\frac{V_0}{A_+}-\phi_0\r)
+\l[\l(\phi_\uin-\phi_0+\frac{V_0}{A_+}\r)^2-2\,\Mp^2\,N\r]^{1/2},
\end{equation}
and, as can been seen in Fig.~\ref{fig:potstaro} (top right panel,
dotted blue line), this expression turns out to be an excellent
approximation to the actual numerical solution (as one would
expect, since the corresponding slow roll parameters are very small;
see discussion below).  This expression allows us to estimate the
e-fold, say, $N_{0}$, at which the field reaches the point where the
slope of the potential changes, and it is given by
\begin{equation}
\label{eq:nzero}
N_{0}=\l(\f{\phi_\uin-\phi_0}{2\Mp^2}\r)
\l(\phi_\uin-\phi_0+\frac{2\,V_0}{A_+}\right). 
\end{equation}
The `velocity' of the field before it reaches the discontinuity can 
be evaluated to be
\begin{equation}
\f{{\rm d}\phi_{+}}{{\rm d}N}
=-\Mp^2\,
\l[\l(\phi_\uin-\phi_0+\frac{V_0}{A_+}\r)^2-2\,\Mp^2\,N\r]^{-1/2}.
\end{equation}
Actually, the Starobinsky model assumes that the constant $V_{0}$ is
the dominant term in the potential for a range of $\phi$ near
$\phi_{0}$.  This regime is very similar to the so-called vacuum
dominated regime in hybrid inflation.  In such a case, we can set
$V\simeq V_{0}$ in the slow roll trajectory, and the expressions for
$\phi_+$ and $\d \phi_{+}/ \d N$ above simplify to
\begin{equation} 
\phi_{+}
\simeq \phi_{\uin}-\f{A_{+}\,\Mp^2}{V_{0}}\, N
\quad{\rm and}\quad
\f{{\rm d}\phi_{+}}{{\rm d}N}
\simeq -\f{A_{+}\Mp^2}{V_{0}}.\label{eq:dphidN-bt}
\end{equation}
The fact that, before the discontinuity, the slow roll trajectory can
be well approximated by a straight line with a negative slope is also
evident from Fig.~\ref{fig:potstaro} (top right panel). 
Further, in this limit, the above relation for $N_{0}$ reduces to
$N_{0}=[V_{0}\, (\phi_{\uin}-\phi_{0})/A_{+}\, \Mp^2]$.

\par

When the field crosses $\phi_{0}$, the slow roll approximation ceases 
to be valid, and we must return to the exact Klein-Gordon equation to 
understand the evolution of the field. 
Upon treating the number of e-folds as the independent variable, 
the Klein-Gordon equation in~(\ref{eq:be}) can be rewritten as
\begin{equation}
H^2\, \f{{\rm d}^2\phi}{{\rm d}N^2}
+\left(3-\epsilon_1\right)\, H^2\, 
\f{{\rm d}\phi}{{\rm d}N}+V_{\phi}=0,
\end{equation}
where $\epsilon_{1}\equiv -\dot{H}/H^2$ is the first slow roll
parameter. Since the constant term $V_{0}$ in the potential is
dominant, as a first step, it is a good approximation to assume that
the Hubble parameter~$H$ is constant.  In other words, one always has
$\epsilon_1\ll 1$, even when the field passes over the discontinuity
in the slope of the potential. This, in turn, implies that inflation
does not end even though the slow roll approximation breaks down
temporarily with the higher order slow roll parameters becoming large
(see discussion below).  In fact, this situation can be viewed as the
very definition of inflation with a transition.  The above
Klein-Gordon equation can be easily solved under these conditions and,
upon requiring the field and its derivative to be continuous at the
transition, we obtain the following solution for the field when
$\phi<\phi_{0}$:
\begin{equation}
\label{eq:fieldminus}
\phi_{-}
\simeq\phi_0+ \frac{\Delta A}{9\, H_0^2}
\biggl[1-{\rm e}^{-3\,\l(N-N_0\right)}\biggr]
-\frac{A_-}{3\,H_0^2}\, \l(N-N_0\r),
\end{equation}
where 
\begin{equation}
\Delta A \equiv A_{-}-A_{+}
\end{equation}
and we have set $H_0^2\simeq V_0/(3\,\Mp^2)$. We have plotted this
solution for the evolution of the field in Fig.~\ref{fig:potstaro}
(top right panel, dotted red line).  As can be noticed in the plot,
the agreement with the exact, numerical, solution is very good.

\par

Let us now evaluate the slow roll parameters on either side of the
transition due to the discontinuity in the slope of the 
potential.  We have already introduced the first slow roll
parameter~$\epsilon_{1}$.  As is well-known, the second and the higher
slow roll parameters are defined as~\cite{schwarz-2001,leach-2002}
\begin{equation}
\epsilon_{n+1}
=\f{{\rm d}\ln \left \vert \epsilon_n\right \vert}{{\rm d}N}
\quad{\rm for}\quad n\ge 1.\label{eq:d-e}
\end{equation}
When the slow roll approximation is valid, the first two slow roll 
parameters can be expressed in terms of the potential $V(\phi)$ and 
its derivatives as follows:
\begin{equation}
\epsilon_{1} \simeq \f{\Mp^2}{2}\, \l(\frac{V_{\phi}}{V}\r)^2
\quad{\rm and}\quad
\epsilon_{2} \simeq  2\,\Mp^2\, 
\l[\l(\f{V_{\phi}}{V}\r)^2-\frac{V_{\phi\phi}}{V}\r],
\end{equation}
where $V_{\phi\phi}=\d^{2} V/\d\phi^{2}$. Clearly, these relations can
be used until the field reaches $\phi_{0}$ and one obtains that
\begin{equation}
\label{eq:epsilon1plus}
\epsilon_{1+} \simeq  \frac{\Mp^2}{2}\,
\l[\frac{A_+}{V_0+A_+\left(\phi-\phi_0\right)}\r]^2
\simeq \f{A_+^2}{18\, \Mp^2H_0^4},
\end{equation}
where the last expression has been arrived at upon assuming that $V_0$
is dominant.  The fact that the first slow roll parameter is almost a
constant before the transition can be easily seen in
Fig.~\ref{fig:potstaro} (bottom left panel).  Further, since the
potential is linear in~$\phi$, we have $\epsilon_{2+} \simeq
4\,\epsilon_{1+}$, i.e. the second slow roll parameter is also a
constant until the time the field reaches the discontinuity
at~$\phi_{0}$.  Again, this behavior is confirmed by the corresponding
numerical result plotted in Fig.~\ref{fig:potstaro} (bottom right
panel).

\par

Once the field has crossed the break in the potential, one has to use 
the exact expressions to arrive at the behavior of the slow roll 
parameters. 
Upon using the solution~(\ref{eq:fieldminus}), we find that the first 
slow roll parameter can be obtained to be
\begin{equation}
\label{eq:epsilon1minus}
\epsilon_{1-}
=\f{1}{2\,\Mp^2}\, \left(\frac{{\rm d}\phi_{-}}{{\rm d}N}\right)^2
=\f{A_-^2}{18\,\Mp^2\,H_0^4}\,
\left[1-\f{\Delta A}{A_-}\, {\rm e}^{-3\,\l(N-N_{0}\r)}\right]^2.
\label{eq:e1-at}
\end{equation}
This expression proves to be an excellent approximation to the exact
result, as can be checked in Fig.~\ref{fig:potstaro} (it is 
represented by the dotted red curve in the bottom left panel). 
Let us now turn to the evolution of the second slow roll parameter
after the transition.
It can be written as
\begin{equation}
\epsilon_2=-6+2\,\epsilon_1-\frac{2\, V_{\phi}}{H^2}\, 
\l(\f{{\rm d}\phi}{{\rm d}N}\r)^{-1},
\end{equation}
which, we should emphasize, is an exact expression. 
In fact, we can arrive at the form of the second slow roll parameter
after the transition, upon using the expression~(\ref{eq:e1-at}) 
for $\epsilon_{1-}$ in Eq.~(\ref{eq:d-e}).
We find that, $\epsilon_{2-}$ is given by
\begin{equation}
\epsilon_{2-}
\simeq \f{6\,\Delta A}{A_-}\, 
\frac{{\rm e}^{-3\,\l(N-N_{0}\r)}}
{1-\l(\Delta A/A_{-}\r)\, {\rm e}^{-3\,\l(N-N_{0}\r)}}.
\end{equation}
However, at sufficiently late times after the transition, when the
slow roll evolution has been restored, we expect that $\epsilon_{2-}
\simeq 4\, \epsilon_{1-}$, as it was in the case before the transition
to fast roll (since the potential continues to be linear).  Careful
analysis points to the fact that, in order to achieve this relation at
late times, one needs to actually take into account the property
that, strictly, the Hubble parameter is not a constant. In
fact, this amounts to using the usual approximation that consists of
replacing the potential $V$ by $M^4$ while its derivative $V_{\phi}$
is still calculated exactly. Upon taking this fact into account,
one actually obtains that
\begin{equation}
\label{eq:epsilon2minus}
\epsilon_{2-}
\simeq \f{6\,\Delta A}{A_-}\, 
\frac{{\rm e}^{-3\,\l(N-N_{0}\r)}}
{1-\l(\Delta A/A_{-}\r)\, {\rm e}^{-3\,\l(N-N_{0}\r)}}
+4\, \epsilon_{1-},
\end{equation} 
with $\epsilon_{1-}$ in the second term being given by
Eq.~(\ref{eq:e1-at}).  Clearly, as the first term vanishes at late
times (i.e. at large $N$), the above expression satisfies the required
relation between the first two slow roll parameters when slow roll has
been restored much after the transition.  Yet again, we find that this
relation almost exactly matches the exact numerical result
for~$\epsilon_2$, as is evident from Fig.~\ref{fig:potstaro} (see the
bottom right panel).  In particular, the above expressions point to
the fact that the value of the second slow roll parameter soon after
the transition is controlled by only one quantity, viz. $\Delta
A/A_-$, and is given by $6\,(\Delta A/A_-)/(1-\Delta A/A_-)$.
Therefore, if $\vert \Delta A/A_-\vert $ is large, then $\epsilon_2$
quickly approaches $-6$ immediately after the transition, as can also
be checked in Fig.~\ref{fig:potstaro} (the figure actually corresponds
to $\vert \Delta A/A_-\vert \simeq 3.61$).

\par

\begin{figure}
\begin{center}
\includegraphics[width=15cm]{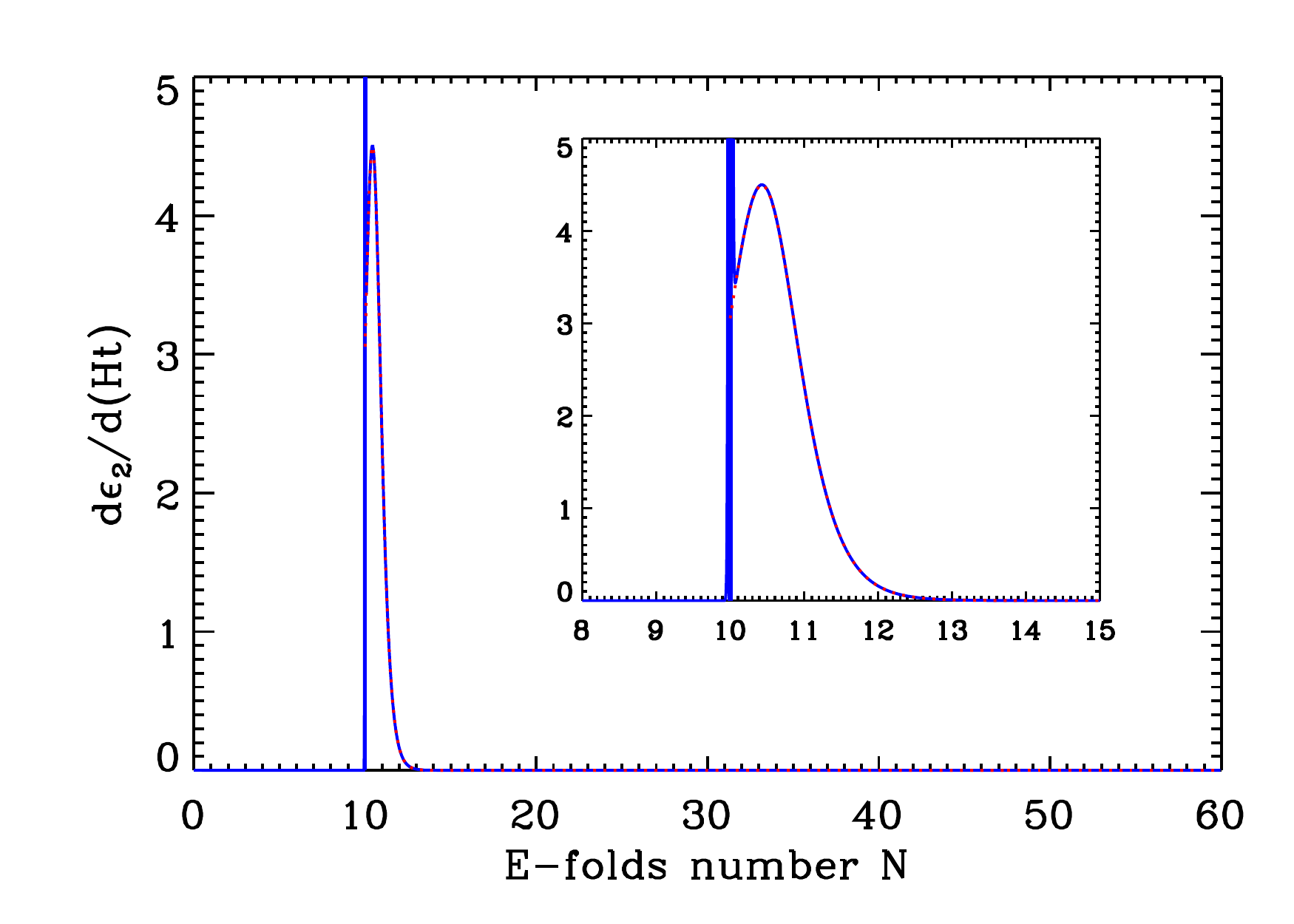}
\caption{The solid blue curve denotes the evolution of the quantity
  ${\rm d}\epsilon_2/{\rm d}(Ht)$, computed numerically, and plotted
  as a function of the number of e-folds.  Note that we have worked
  with the same set of parameters as in Fig.~\ref{fig:potstaro}.  The
  dotted red curve represents the analytical
  expression~(\ref{eq:e2d-at}).  As is clear from the inset, the
  analytical expression approximates the actual numerical result very
  well after the transition. The vertical blue line is just a
  numerical artifact of the parametrization used here and we have
  carefully checked that it plays no role in the following
  considerations.}
\label{fig:dotsr2}
\end{center}
\end{figure}

As we shall discuss in the next section, in order to evaluate the 
dominant contribution to the scalar bi-spectrum in the Starobinsky 
model, we shall also require the quantity~${\dot \epsilon_{2}}$.  
One can show that, it can be written as
\begin{equation}
\label{eq:de2dt}
\frac{{\rm d}\epsilon_{2}}{{\rm d}t}
=-\f{2\, V_{\phi\phi}}{H}
+12\,H\,\epsilon_{1}-3\,H\,\epsilon_{2}-4\,H\,\epsilon_{1}^2
+5\,H\,\epsilon_{1}\,\epsilon_{2}-\frac{H}{2}\, \epsilon_{2}^2,
\end{equation}
and, under the conditions of our interest, this relation reduces to
\begin{eqnarray}
\frac{{\rm d}\epsilon_{2-}}{{\rm d}t}
&\simeq& -3\,H_{0}\,\epsilon_{2-}-\frac{H_{0}}{2}\, 
\epsilon_{2-}^2\nn\\
&\simeq& -\f{18\, H_0\, \Delta A}{A_{-}}\,
\frac{{\rm e}^{-3\,\l(N-N_{0}\r)}}
{\l[1-\l(\Delta A/A_{-}\r)\, {\rm e}^{-3\l(N-N_{0}\r)}\r]^2}.
\label{eq:e2d-at}
\end{eqnarray}
We should mention here that, in arriving at the final equality, we
have ignored the second term involving $\epsilon_{1-}$ in
Eq.~(\ref{eq:epsilon2minus}). The fact that this is a valid
approximation is confirmed by the numerical analysis.  As in the case
of $\epsilon_{2-}$, we find that, during the period soon after the
transition, the value of ${\dot \epsilon_{2-}}$ is determined only by
ratio $\Delta A/A_-$.  In Fig.~\ref{fig:dotsr2}, we have plotted the
above expression for ${\dot \epsilon_{2}}$ as well as the exact
numerical result.  It is clear from the figure that the
expression~(\ref{eq:e2d-at}) is an excellent approximation to the
actual result.

\par

Let us now briefly pause to summarize the results we have obtained.
It is evident from the above expressions for the slow roll parameters
that the change in the slope in the potential leads to a departure
from slow roll for a short span of time, before slow roll is again
recovered when the friction on the scalar field due to the expansion
of the universe begins to dominate the force due to the potential.
While the first slow roll parameter~$\epsilon_{1}$ remains small
throughout the evolution (which essentially occurs due to $V_{0}$
being the dominant term in the potential around $\phi_{0}$), the
second slow roll parameter~$\epsilon_{2}$ and its time derivative
${\dot \epsilon_{2}}$ exhibit a sharp rise as the field crosses
$\phi_{0}$, before falling off.  We have been able to obtain simple
analytical expressions for the various background quantities even in
the regime where the conditions for slow roll are violated.  This
point turns out to be the key for our analysis.  As we shall see, the
simple expressions for the slow roll parameters permit us to evaluate
the bi-spectrum in the equilateral limit completely analytically.
However, before we turn to the evaluation of the scalar bi-spectrum,
let us discuss the power spectrum that arises in the Starobinsky
model.


\subsection{The scalar power spectrum in the Starobinsky model}
\label{subsec:ps}

Let ${\cal R}(\eta ,{\bf x})$ be the dimensionless curvature
perturbation induced by the scalar field, and let 
${\cal R}_{\vk}(\eta)$ denote the associated Fourier modes 
defined through the relation
\begin{equation}
{\cal R}(\eta ,{\bf x})=
\int \f{{\rm d}^{3}{\vk}}{(2\,\pi)^{3/2}}\, 
{\cal R}_{\vk}\, {\rm e}^{i\, \vk \cdot \vx}.
\end{equation}
On quantization, the operator corresponding to the curvature 
perturbation can be expressed as
\begin{equation}
{\hat \cR}(\eta ,\vx)
=\int \f{{\rm d}^{3}{\vk}}{(2\pi)^{3/2}}\, 
\l[{\hat a}_{\vk}\, f_{k}(\eta)\, 
{\rm e}^{i\,\vk\cdot \vx}
+{\hat a}_{\vk}^{\dagger}\, f_k^*(\eta)\, 
{\rm e}^{-i\,\vk \cdot \vx}\r],\label{eq:cR-d}
\end{equation}
where ${\hat a}_{\vk}$ and ${\hat a}_{\vk}^{\dagger}$ are the usual 
creation and annihilation operators which satisfy the following 
non-trivial commutation relation: $\l[{\hat a}_{\vk}, 
{\hat a}_{\bf p}^{\dagger}\r]=\delta^{(3)}\left(\vk -{\bf p}\right)$.
In the case of the canonical scalar field, the modes $f_k$ are governed 
by the differential equation~\cite{texts,reviews}
\begin{equation}
f_k''+2\, \f{z'}{z}\, f_k' + k^{2}\, f_{k}=0,\label{eq:defk}
\end{equation}
where $z \equiv a\, {\dot \phi}/H=a\,\Mp\,\sqrt{2\,\epsilon_1}$. 
The quantity ${\cal R}_{\vk}$ is of dimension $-3$, $f_k$ of 
dimension $-3/2$, while $z$ is of dimension one. It is useful to
note that, in terms of the Mukhanov-Sasaki variable $v_{k}=z\, f_k$
(which is of dimension $-1/2$), the above equation for $f_k$ 
reduces to
\begin{equation}
v_{k}''+\l(k^2-\f{z''}{z}\r)\, v_{k}=0.\label{eq:de-vk}
\end{equation}

\par

The dimensionless scalar power spectrum ${\cal P}_{_{\rm S}}(k)$ 
is defined in terms of the correlation function of the Fourier 
modes of the curvature perturbation as follows:
\begin{equation}
\label{eq:twopointfourier}
\langle 0\vert {\hat \cR}_{\vk}(\eta)\, 
{\hat \cR}_{\bf p}(\eta)\vert 0\rangle 
=\f{(2\, \pi)^2}{2\, k^3}\; {\cal P}_{_{\rm S}}(k)\;
\delta^{(3)}\l(\vk+{\bf p}\r),
\end{equation}
where the vacuum state $\vert 0\rangle$ is defined as ${\hat
  a}_{\vk}\vert 0 \rangle=0$, $\forall$ $\vk$.  Since we can write,
${\hat \cR}_{\vk}=({\hat a}_{\vk}\, f_k + {\hat a}_{-\vk}^{\dagger}\,
f_k^{*})$, one obtains that
\begin{equation}
{\cal P}_{_{\rm S}}(k)
=\frac{k^{3}}{2\, \pi^{2}}\, \vert f_{k}\vert^{2}
=\frac{k^{3}}{2\, \pi^{2}}\,\l(\f{\vert v_{k}\vert}{z}\r)^{2}
\label{eq:sps-d}
\end{equation}
with the right hand side (which depends on time) evaluated in the
super Hubble limit [i.e. when $k/(a\, H)\to 0$] or, more generically,
at the end of inflation.  The curvature perturbation is usually
assumed to be in the Bunch-Davies vacuum, which corresponds to
choosing $v_{k} \to {\rm e}^{i\,k\,\eta}/ \sqrt{2\,k}$ in the sub
Hubble limit, i.e. as $k/(a\, H)\to\infty$.  It is clear from
Eq.~(\ref{eq:de-vk}) that it is the `effective potential' $z''/z$
which determines the evolution of the scalar perturbations.  It
can be written in terms of the slow roll parameters as follows:
\begin{equation}
\label{eq:poteff}
\frac{z''}{z}
={\cal H}^2\, \l(2-\epsilon_1+\frac{3}{2}\, \epsilon_{2}
+\frac{1}{4}\, \epsilon_2^2
-\frac{1}{2}\, \epsilon_1\, \epsilon_{2}
+\frac{1}{2}\, \epsilon_2\, \epsilon_3\r),
\end{equation}
where ${\cal H}\equiv a'/a=a\, H$ is the conformal Hubble parameter, and
it should be emphasized that the above expression is an exact one.

\par

Our aim now is to derive the scalar power spectrum for modes that
leave the Hubble radius in the vicinity of the transition
at~$\phi_{0}$.  Before the transition to a brief period of fast roll
near~$\phi_{0}$, all the slow roll parameters remain small, and the
effective potential simplifies to $z''/z\simeq 2\, {\cal H}^2$.  We
should emphasize here that, for simplicity, we are neglecting terms
involving the first order slow roll parameters that would lead to a
non-vanishing spectral index for modes far from the characteristic
scale, viz. the mode which leaves the Hubble radius when the scalar
field crosses the break in the inflaton potential.  Recall that, since
the evolution is dominated by the constant term $V_0$ in the
potential, the expansion is of the de Sitter form corresponding to the
Hubble parameter~$H_{0}$.  Therefore, around the transition, the scale
factor can be expressed in terms of the conformal time as
$a(\eta)=-(H_{0}\, \eta)^{-1}$, so that we have $z''/z\simeq
2/\eta^{2}$.  In such a case, the solution to the Mukhanov-Sasaki
variable $v_{k}$ that satisfies the standard Bunch-Davies initial
condition is given by
\begin{equation}
v_{k}^{+}(\eta)
=\f{1}{\sqrt{2\, k}}\, 
\l(1-\frac{i}{k\,\eta}\r)\, {\rm e}^{-i\,k\,\eta}.\label{eq:vk-bt}
\end{equation}

\par

During the transitory fast roll regime, as the field
crosses~$\phi_{0}$, the slow roll parameters are no longer small and,
hence the solutions to the Mukhanov-Sasaki variable~$v_{k}$ can be
expected to look different.  In particular, we have seen that the
second slow roll parameter $\epsilon_2$ and its time derivative 
${\dot \epsilon}_2$ can be large immediately after the transition.  
From Eq.~(\ref{eq:poteff}), it is then clear that, a priori, one can 
no longer expect the effective potential to be just given by $z''/z
\simeq 2\, {\cal H}^2$.  However, remember that $\epsilon_1$ remains
small even during the transition.  Further, one has ${\dot
\epsilon_{2}}=H\, \epsilon_2\,\epsilon_3$.  Upon using these facts
and the result that ${\dot \epsilon_{2}}$ can be approximated 
using Eq.~(\ref{eq:de2dt}) (and, of course, the property that
$V_{\phi\phi}=0$), one finds that, after the transition
\begin{eqnarray}
\frac{z''}{z}
&\simeq &
{\cal H}^2\,\left(2+\f{3}{2}\,\epsilon_{2-}
+\f{1}{4}\,\epsilon_{2-}^2
+\frac{1}{2}\,\epsilon_{2-}\,\epsilon_{3-}\r)\nn\\ 
&\simeq &{\cal H}^2\,\left[2+\frac{3}{2}\,\epsilon_{2-}
+\f{1}{4}\,\epsilon_{2-}^2
+\f{1}{2}\,\l(-3\,\epsilon_{2-}-\frac{1}{2}\,\epsilon_{2-}^2\r)\r]
\simeq 2\, {\cal H}^2.
\end{eqnarray}
This implies that certain cancellations occur so that the effective
potential still retains the same form, i.e.  $z''/z \simeq 2\, {\cal
  H}^2$, even after the field has crossed~$\phi_{0}$. But, due to the
fast roll, post-transition, the modes~$v_{k}$ do not remain in the
Bunch-Davies vacuum and, as a result, the solution to~$v_{k}$ takes
the general form
\begin{equation}
v_{k}^{-}(\eta)
=\frac{\alpha _{k}}{\sqrt{2\, k}}\, 
\l(1-\frac{i}{k\,\eta}\r){\rm e}^{-i\,k\,\eta}
+\frac{\beta _{k}}{\sqrt{2\, k}}\, 
\l(1+\frac{i}{k\,\eta}\r){\rm e}^{i\,k\,\eta},
\label{eq:vk-at}
\end{equation}
where $\alpha_{k}$ and $\beta_{k}$ are the standard Bogoliubov 
coefficients.

\par

The Bogoliubov coefficients $\alpha_{k}$ and $\beta_{k}$ can now be
determined by carefully matching the mode~$v_{k}$ and its derivative
at the transition at $\phi_{0}$.  In order to do so, we first need to
understand the behavior of $z$ and the effective potential $z''/z$
across the transition.  In terms of the conformal time coordinate, the
transition occurs at $\eta_{0} =-(a_0\, H_{0})^{-1}$, where $a_0$ is
the scale factor at $\eta _0$.  Recalling that
$z=a\,\Mp\,\sqrt{2\,\epsilon_1}$, and upon using the expression of
$\epsilon_1$ before and after the transition, we arrive at
\begin{equation}
z(\eta) \simeq\l\{\begin{array}{ll}
\displaystyle
-\f{A_{+}}{3\, H_{0}^{3}\, \eta} & {\rm for}\ \eta<\eta_{0},\\
\displaystyle
-\f{A_{-}}{3\, H_{0}^{3}\, \eta}
-\frac{a_0^3\,\Delta A\, \eta ^2}{3} & {\rm for}\ \eta>\eta_{0}.
\label{eq:z-at}
\end{array}\r.
\end{equation}
As a consequence, one obtains that
\begin{equation}
z'(\eta) \simeq\l\{\begin{array}{ll}
\displaystyle
\f{A_{+}}{3\, H_{0}^{3}\, \eta^2}\ 
& {\rm for}\ \eta<\eta_{0},\\
\displaystyle
\f{A_{-}}{3\, H_{0}^{3}\, \eta^2} -\frac{2\,a_0^3\,\Delta A\, \eta}{3}\ 
& {\rm for}\ \eta>\eta_{0},
\end{array}\r.
\end{equation}
and it is clear that $z'$ jumps at the transition by the amount
\begin{equation}
[z'(\eta_0)]_\pm
\equiv z'_{-}(\eta_0)-z'_{+}(\eta_0)
=-\f{a_0^2\,\Delta A}{H_0}.
\end{equation}
So, at the transition, that is to say when $\eta \simeq \eta _0$ 
(and only at the transition!), we can write 
\begin{equation}
z'\simeq 
\frac{a_0^2\,A_+}{3\,H_0}
-\frac{a_0^2\,\Delta A}{H_0}\, \Theta\l(\eta -\eta_0\r),
\end{equation}
where $\Theta(x)$ is the step function. 
This implies that the effective potential $z''/z$ can be described 
as a Dirac delta function at the transition, and we have
\begin{equation}
\frac{z''}{z}
\simeq \frac{3\,a_0\,H_{0}\,\Delta A}{A_{+}}\, 
\delta^{(1)}\left(\eta -\eta_0\right).
\end{equation}

\par

\begin{figure}
\begin{center}
\includegraphics[width=15cm]{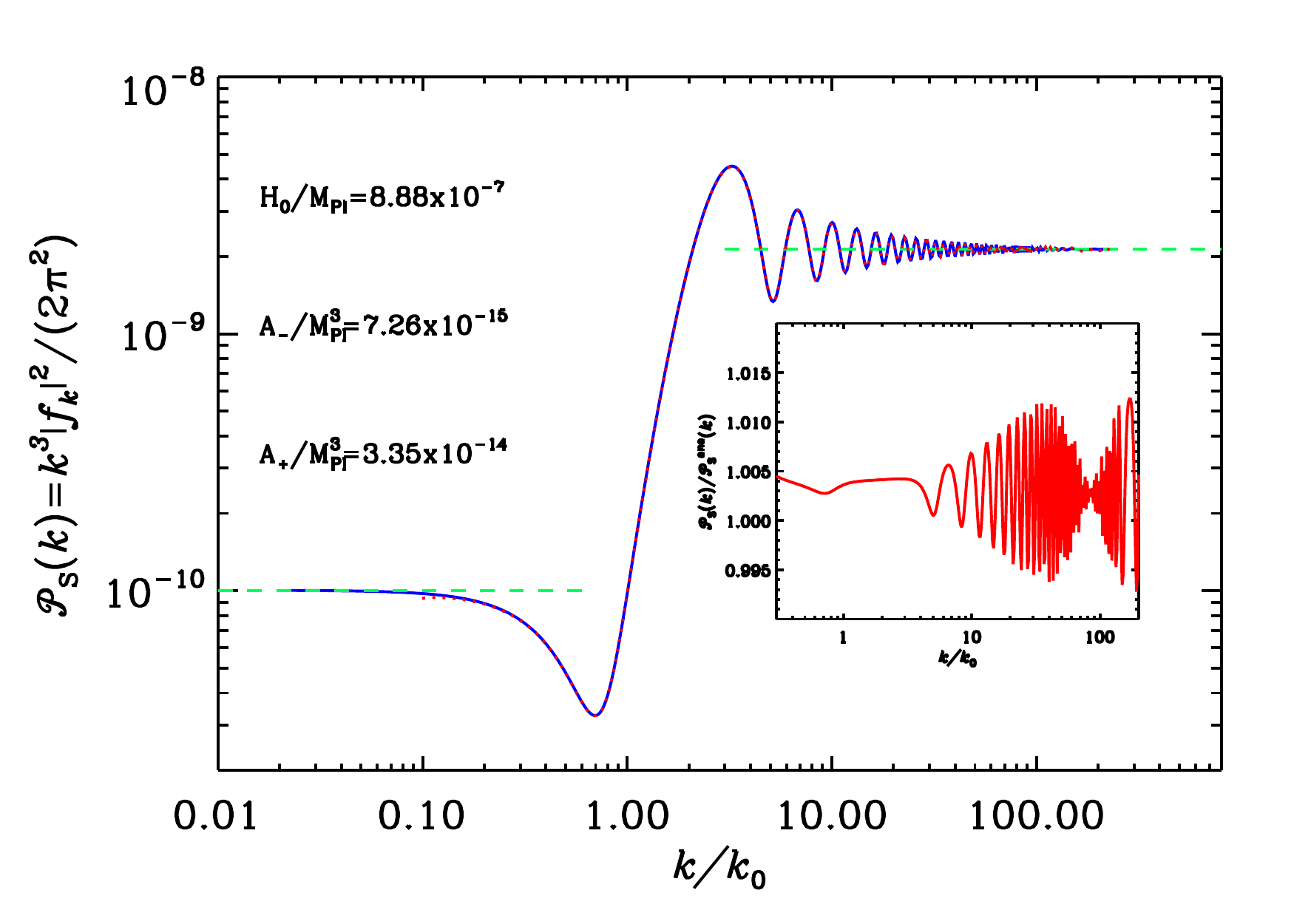}
\caption{The scalar power spectrum in the Starobinsky model.  While
  the blue solid curve denotes the analytic spectrum~(\ref{eq:sps}),
  the red dots represent the corresponding numerical scalar power
  spectrum that has been obtained through an exact numerical
  integration of the background as well as the perturbations. The
  green dashed lines indicate the asymptotic scale invariant values
  [cf. Eqs.~(\ref{eq:limps}) and~(\ref{eq:ps-ss})]. And, the
  inset highlights the small difference between the exact numerical 
  spectrum and the analytic one over a certain window in
  the wavenumber.  In plotting these spectra, we have worked with the
  same values of the parameters that we had mentioned in the first
  figure. The first scale (i.e. the largest one) for which the power
  spectrum has been numerically computed corresponds to the physical
  wavenumber given by $k/a_\uin=500\, H_\uin$ at the beginning of
  inflation, which is an arbitrary but convenient choice. In terms of
  this scale, $k_0$ is given by $k_0=y\, k_\uin$, where $500\,
  H_\uin\, y\, {\rm e}^{-N_0}=H_0$.  Given that $N_0=10$ and
  $H_\uin\simeq H_0$ (as can be easily checked in the slow roll
  approximation and as can be verified with great accuracy
  numerically), one obtains $k_0\simeq 44\, k_{\uin}$.  Evidently, the
  analytically evaluated spectrum is in remarkable agreement with the
  numerical result.  We should mention that an inaccurate
  determination of $k_0$ would have resulted in a `phase shift'
  between the numerical and analytical spectra.}
\label{fig:spec}
\end{center}
\end{figure}

The above expression for $z''/z$ allows us to establish the matching
conditions on the modes at the transition, which read as
\begin{equation}
[v_{k}(\eta _0)]_\pm
=v_{k}^{-}\l(\eta _0\r) - v_{k}^{+}\l(\eta _0\r)=0
\end{equation}
and
\begin{equation}
[v_{k}'(\eta _0)]_\pm
=v_{k}^{-}{}'\l(\eta _0\r) 
- v_{k}^{+}{}'\l(\eta _0\r)
=\f{3\,a_0\,H_0\,\Delta A}{A_{+}}\; v_{k}\left(\eta _0\right).
\end{equation}
The modes~(\ref{eq:vk-bt}) and~(\ref{eq:vk-at}), along with these two
conditions, then lead to the following expressions for the Bogoliubov
coefficients~$\alpha_{k}$ and $\beta_{k}$:
\begin{eqnarray}
\alpha_{k} 
&=& 1+\frac{3\,i\,\Delta A}{2\,A_{+}}\;\frac{k_0}{k}\,
\left(1+\frac{k_0^2}{k^2}\right),
\label{eq:alphak-sm}\\
\beta_{k} 
&=& -\frac{3\,i\,\Delta A}{2\,A_+}\;\frac{k_0}{k}\,
\l(1+\frac{i\, k_0}{k}\r)^2\, {\rm e}^{2\,i\,k/k_{0}},
\label{eq:betak-sm}
\end{eqnarray}
where $k_0=-1/\eta_{0}=a_0H_0$ corresponds to the mode that leaves the
Hubble radius at the transition.  The ratio $k/k_0=k/(a_0\, H_0)$ can
also be viewed as the ratio of the physical wavenumber at the
transition, viz. $k/a_0$, to the characteristic physical wave number
$k_0^{\rm phys}\equiv H_0$.  Therefore, upon using the
modes~(\ref{eq:vk-at}) post-transition, the resulting power spectrum,
evaluated as $k\,\eta\to 0$, is found to be~\cite{starobinsky-1992}
\begin{eqnarray}
\label{eq:sps}
{\cal P}_{_{\rm S}}(k)
&=&\l(\f{H_{0}}{2\,\pi}\r)^{2}\, \l(\f{3\, H_{0}^{2}}{A_{-}}\r)^{2}\,
\vert\alpha_{k}-\beta_{k}\vert^{2} \nonumber \\
&=&\l(\f{H_{0}}{2\,\pi}\r)^{2}\, 
\l(\f{3\, H_{0}^{2}}{A_{-}}\r)^{2}
\Biggl\{1-\f{3\, \Delta A}{A_+}\f{k_0}{k}
\Biggl[\l(1-\f{k_0^2}{k^2}\r)\, \sin \l(\f{2\, k}{k_0}\r)
\nonumber \\
& &+\f{2\,k_0}{k}\cos\l(\f{2\,k}{k_0}\r)\Biggr]
+\f{9\,\Delta A^2}{2\,A_+^2}\, \f{k_0^2}{k^2}\,
\left(1+\f{k_0^2}{k^2}\right)
\Biggl[\l(1+\f{k_0^2}{k^2}\r) 
\nonumber \\
& & -\f{2\,k_0}{k}\, \sin \l(\f{2\,k}{k_0}\r)
+\l(1-\frac{k_0^2}{k^2}\r)\, \cos\l(\frac{2\,k}{k_0}\r)\Biggr]\Biggr\},
\label{eq:ps-sm}
\end{eqnarray}
and it should be emphasized that the spectrum depends on the
wavenumber only through the ratio $k/k_0$.  Also, as one would expect,
this power spectrum turns scale invariant far away from $k_0$ on
either side. While, for $k /k_0\to 0$, its scale invariant value is
given by
\begin{equation}
\label{eq:limps}
\lim_{k/k_0\,\to\, 0}\;
{\cal P}_{_{\rm S}}(k)
= \l(\f{H_{0}}{2\,\pi}\r)^{2}\, \l(\f{3\, H_{0}^{2}}{A_{+}}\r)^{2},
\end{equation}
for $k /k_0\to \infty$, it simplifies to
\begin{equation}
\lim_{k/k_0\,\to\, \infty}\;
{\cal P}_{_{\rm S}}(k)
= \l(\f{H_{0}}{2\,\pi}\r)^{2}\, \l(\f{3\, H_{0}^{2}}{A_{-}}\r)^{2}.
\label{eq:ps-ss}
\end{equation}
Given a $H_0$, or equivalently, $V_0$, COBE normalization (i.e. the
amplitude of the power spectrum at suitably small scales) determines
the value of $A_{-}$.  In Fig.~\ref{fig:spec}, we have plotted the
above analytical power spectrum and the corresponding numerical result
for certain values of the parameters for which COBE normalization is
achieved and the assumptions of the Starobinsky model are valid.
The two results are clearly in good agreement, which indicates the remarkable 
extent of validity of the assumptions and approximations that are used to 
arrive at analytic forms for the background evolution, the modes as well as 
the power spectrum in the Starobinsky model.


\section{A rapid outline of the procedure for evaluating the 
scalar bi-spectrum}\label{sec:pebs}

In this section, we shall quickly sketch the by-now commonly used
procedure for evaluating the scalar bi-spectrum generated during
inflation (for the original discussion, see
Ref.~\cite{maldacena-2003}; for further discussions, see
Refs.~\cite{ng-f,ng-ncsf} and also the recent
review~\cite{ng-reviews}).  Due to the constrained nature of the
Einstein equations, the scalar perturbations during inflation can be
basically described by a single function, say, the curvature
perturbation~$\cR$.  Just as the scalar power spectrum can be
expressed in terms of the two point function of the curvature
perturbation [using Eqs.~(\ref{eq:cR-d})
and~(\ref{eq:twopointfourier})], the scalar bi-spectrum is related to
the three point function through a suitable Fourier transform
[cf.~Eqs.~(\ref{eq:tpf-cR}) and~(\ref{eq:bi-s})].  It is well known
that, at the linear order in the perturbations, the curvature
perturbation~$\cR$ is governed by a quadratic action~\cite{reviews},
which leads to the equation of motion~(\ref{eq:defk}) that we had
considered in order arrive at the the power spectrum in the previous
section.  Evidently, in a linear and Gaussian theory, the
three point function will be identically zero\footnote{It may be 
useful to point out here that, even a linear theory can lead to 
non-Gaussian distributions, when one works with, say, non-vacuum
initial states.}. Therefore, in order to
evaluate the bi-spectrum during inflation, the first step that needs
to be taken is to arrive at the action describing the curvature
perturbation at the next higher, i.e.  the cubic, order.  Then, based
on the cubic order terms, one evaluates the corresponding three point
function of the curvature perturbation (and, thence the bi-spectrum)
using the standard techniques of perturbative quantum field theory.
In what follows, we shall first describe as to how the cubic order
action is arrived at, and then discuss the contributions to the scalar
bi-spectrum due to the various terms in the cubic order action.


\subsection{The cubic order action describing the curvature 
perturbation}\label{subsec:action3}

The action that describes the curvature perturbation at 
the cubic order is usually arrived at using the 
Arnowitt-Deser-Misner (ADM) formalism~\cite{arnowitt-1960}.  
Recall that, in the ADM formalism, the metric is expressed in 
terms of the lapse and the shift functions~$N$ and~$N^{i}$ as 
follows:
\begin{equation}
\d s^2 = -N^2\, \d(x^0)^2 
+ h_{ij}\, \l(N^{i}\, \d x^0
+ \d x^{i}\r)\, \l(N^{j}\, \d x^0 +\d x^{j}\r),
\label{eq:m-admf} 
\end{equation}
where $x^0$ and $x^i$ denote the time and the spatial coordinates.  As
is common knowledge, the lapse and the shift functions~$N$ and~$N^{i}$
turn out to be Lagrangian multipliers in the action for the complete
system, and the variation of the action with respect to them leads to
the Hamiltonian and the momentum constraints. The remaining equations
govern the dynamics of the spatial metric~$h_{ij}$.  These six
equations (corresponding to the independent components of the spatial
metric) and the four constraint equations essentially constitute the
ten Einstein equations corresponding to the complete spacetime
metric.

\par

The system of our interest is gravity described by the
Einstein-Hilbert action and a scalar field that is governed by the
canonical action.  For such a case, in terms of the
metric~(\ref{eq:m-admf}), the action describing the complete system
can be written as~\cite{maldacena-2003,ng-reviews} 
\begin{eqnarray}
{\cal S}[N,N_i,h_{ij},\phi]
&=&\int \d x^0\int\d^{3}\vx\; N\,\sqrt{h}\;
\Biggl\{\f{\Mp^2}{2} 
\l[\frac{1}{N^2} \l(E_{ij}\, E^{ij} - E^2\r)+\, {}^{(3)}R\r]\nn\\
& & +\,\Biggl[\frac{1}{2\,N^2}\, \l(\f{\pa \phi}{\pa x^0}\r)^2
-\frac{N^i}{N^2}\,\frac{\partial \phi}{\partial x^0}\, \partial _i\phi
-\frac{1}{2}\,h^{ij}\,\partial _i\phi\, \partial_j\phi
\nonumber \\
& &+\frac{N^i\,N^j}{2\,N^2}\,\partial _i\phi\, \partial_j\phi
-V(\phi)\Biggr]\Biggr\},\label{eq:a-cmg}
\end{eqnarray}
where $h\equiv \det\,(h_{ij})$, $^{(3)}R$ is the spatial curvature
associated with the metric $h_{ij}$, and the scalar field is assumed
to be, in general, dependent on time as well as space.  The quantity
$E_{ij}$ is the rescaled second fundamental form and is given by
\begin{equation} 
E_{ij} =N\,K_{ij}=
\f{1}{2}\,\l[\frac{\partial h_{ij}}{\partial x^0} 
- \l({}^{(3)}\nabla_i\, N_j + {}^{(3)}\nabla_j\, N_i\r)\r], 
\end{equation}
while $E = h_{ij}\,E^{ij}$. Evidently, in the above action, the
quantities within the first and the second square brackets correspond
to the gravitational and the scalar field parts, respectively.

\par

In the Hamiltonian formulation that we are working in, the conjugate
momenta corresponding to the spatial metric $h_{ij}$ and the scalar
field $\phi$, are given by
\begin{equation}
\pi^{ij}
=\f{\Mp^2\, \sqrt{h}}{2}\,\l(K^{ij}-h^{ij}\,K\r)\quad{\rm and} 
\quad 
\pi_{\phi}
=\frac{\sqrt{h}}{N}\,\l(\f{\partial \phi}{\partial x^0}
-N^i\,\partial _i\phi\right).
\end{equation}
The Hamiltonian density, say,~${\sf H}$, of the entire system can be 
expressed in terms of the generalized coordinates $h_{ij}$ and $\phi$,
and the corresponding conjugate momenta $\pi_{ij}$ and $\pi_\phi$ as 
follows:
\begin{eqnarray}
{\sf H}
&=&\f{2\, N}{\Mp^2\,\sqrt{h}}\;
\l(\pi^{ij}\pi_{ij}-\f{\pi^2}{2}\right)
-\f{\Mp^2\, N\,\sqrt{h}}{2}\; {}^{(3)}R
-2\,N_j\;{}^{(3)}\nabla _i\pi^{ij}\nn\\ 
& & +\,N\,\l(\frac{\pi_{\phi}^2}{2\,\sqrt{h}}
+\frac{\sqrt{h}}{2}\,h^{ij}\,\pa_i\phi\,\pa_j\phi
+\sqrt{h}\,V(\phi)\r)+\pi_{\phi}\,N^i\,\pa_i\phi,
\end{eqnarray}
where, clearly, the first line corresponds to gravity, while the
second line corresponds to that of the scalar field.  Moreover, it is
evident from the above Hamiltonian that the lapse and the shift
functions indeed act as Lagrange multipliers, as we mentioned above.
Varying with respect to $N$ and $N_i$ respectively leads to the
following Hamiltonian and momentum constraints: 
\begin{eqnarray}
& &\f{\Mp^2}{2}\,
\l[\l(K^{ij}-h^{ij}\,K\r)\,\l(K_{ij}-h_{ij}\,K\r)
-2\,K^2-{}^{(3)}R\right]
\nonumber \\
& & +\,\f{1}{2\,N^2}\,\l(\f{\pa \phi}{\pa x^0}
-N^i\,\partial_i\phi\r)^2
+\f{1}{2}\, h^{ij}\,\partial _i\phi\;\partial _j\phi
+V(\phi)=0,\label{eq:hc}
\end{eqnarray}
and 
\begin{eqnarray}
-2\,{}^{(3)}\nabla _i\pi^{ij}
+\pi_{\phi}\, h^{ij}\,\partial _i\phi=0.\label{eq:mc}
\end{eqnarray}
Let us emphasize that, so far, the formalism has been general.
Nothing has yet been said about the symmetries and the form of the
metric tensor, nor has any approximation been made.

\par

In the case of the Friedmann metric, based on the time coordinate that
one works with, the lapse function $N$ can either be set to unity if
one chooses $x^0$ to be the cosmic time $t$ or we can set $N=a$ when
working with the conformal time $x^0=\eta$.  At this stage, it is
instructive to briefly compare the ADM approach with the more 
common method of arriving at the equations governing the perturbations 
from the Einstein equations.  In the latter approach, when the scalar
perturbations are taken into account, the spatially flat, Friedmann
metric in an arbitrary gauge is usually written as
\begin{eqnarray}
{\rm d}s^2
&=&-\l(1+2\,\varphi\r)\,{\rmd}t ^2
+2\,a(t)\, \partial_iB\, {\rm d}t\,{\rm d}x^i
+\, a^2(t)\, \biggl[\l(1-2\,\psi\r)\,\delta _{ij}\,
{\rm d}x^i\,{\rm d}x^j\,
\nonumber \\
& & +2\,\partial_i\partial_j{\cal E}\;{\rm d}x^i\,{\rm d}x^j\biggr],
\end{eqnarray} 
where $\varphi$, $B$, $\psi$ and ${\cal E}$ are the scalar 
functions that describe the perturbations. Upon comparing this 
line-element with Eq.~(\ref{eq:m-admf}), it is clear that, we have
\begin{eqnarray}
N_i\,N^i-N^2=-(1+2\,\varphi), \quad N_i=a\partial _iB,\nn\\ 
h_{ij}=a^2\,(1-2\,\psi)\,\delta _{ij}
+2\,a^2\,\partial_i\partial_j{\cal E}.
\end{eqnarray}
At the first order in perturbations, these equations reduce to
\begin{equation}
N\simeq 1+\varphi, \quad N_i\simeq a\,\partial _iB, \quad 
h_{ij}\simeq a^2\,\left(\delta _{ij}-2\,\psi\,\delta _{ij}
+2\,\partial_i\partial_j{\cal E}\right),
\end{equation}
and the lapse function $N$ is just unity at the zeroth order because
we have chosen to work in terms of the cosmic time coordinate.

\par

From the above expressions for the different components of the metric
tensor, one can evaluate the second fundamental force and the spatial
curvature, and one obtains that
\begin{eqnarray}
K_{ij} 
= \left(H-2H\psi-\dot \psi -H\varphi\right)a^2\delta _{ij}
-a\partial _i\partial _jB
+a^2\partial_i\partial_j\l(\dot{\cal E}+2\,H\, {\cal E}\r),\\
{}^{(3)}R = \frac{4}{a^2}\,\partial _i\partial^i\psi.
\end{eqnarray}
For a consistent perturbation theory, we must, of course, perturb the
scalar field as well.  Let us write the scalar field as a homogeneous
term plus a perturbative part, i.e. as $\phi +\delta \phi$.  If we
then use the above expressions for $K_{ij}$ and ${}^{(3)}R$ in the
constraint equations~(\ref{eq:hc}) and~(\ref{eq:mc}), we arrive at
\begin{eqnarray}
& & -3\,H\,\l(H\,\varphi+\dot\psi\r)
+\frac{1}{a^2}\,\partial_i\partial^i\psi
-\frac{H}{a}\,\partial_i\partial^iB
+H\,\partial_i\partial^i\dot{\cal E}
\nonumber \\
& & =\frac{1}{2\,\Mp^2}\,\biggl(-\dot \phi^2\,\varphi
+\dot \phi\, \dot{\delta \phi}+V_{\phi}\,\delta\phi\biggr),\\
& & \partial ^i\dot \psi+H\,\partial ^i\varphi
 =\frac{\dot \varphi }{2\,\Mp^2}\,\partial ^i\delta\phi.
\end{eqnarray} 
These are nothing but the time-time and time-space perturbed Einstein
equations.

\par

Our main goal now is to evaluate the action~(\ref{eq:a-cmg}) at the
cubic order in the perturbed quantities.  Ideally, one would have
liked to work in an arbitrary gauge, as is done to arrive at the
standard action that describes the curvature perturbations at the
quadratic order~\cite{reviews}.  However, as one can imagine, the
calculation in an arbitrary gauge turns out to be rather complicated
and, to the best of our knowledge, so far, the action at the cubic
order seems to have been arrived at by working in a specific gauge.
In fact, the gauge that proves to be the most convenient is the one
wherein ${\cal E}=0$ and $\delta\phi=0$, i.e. the gauge wherein 
perturbations in the inflaton are assumed to
vanish~\cite{maldacena-2003,ng-ncsf,ng-reviews}.  Then, using the
perturbed equations derived above, it is easy to show that
\begin{eqnarray}
\varphi=-\frac{\dot \psi}{H}, \quad 
a\,B=\frac{\psi}{H}-\frac{a ^2\,\dot \phi^2}{2\,\Mp^2\,H^2}\,
\partial ^{-2}\dot \psi,
\end{eqnarray}
where the operator $\partial ^{-2}$ is defined by the relation
$\partial^{-2}\partial_i\partial ^i\psi\equiv \psi$. The first
of these two equations can be obtained from the momentum constraint, 
while the second arises due to the Hamiltonian constraint. 
Let us now introduce the curvature perturbation~${\cal R}$, a 
quantity that is conserved at super Hubble scales, and is 
given by~\cite{lyth-1985,schwarz-1998}
\begin{equation}
{\cal R}=-\psi -\frac{H}{\dot \phi}\,\delta \phi,
\end{equation}
In the gauge that we are working in, we have ${\cal R}=-\psi$, and 
the above equations simplify to
\begin{eqnarray}
\varphi=\frac{\dot {\cal R}}{H}, \quad 
a\,B=-\frac{{\cal R}}{H}
+\frac{a^2\,\dot \phi^2}{2\,\Mp^2\,H^2}\,
\partial ^{-2}\dot {\cal R}.
\end{eqnarray}
These expressions match Eqs.~(28) and~(29) in the first 
article of Refs.~\cite{ng-ncsf}, if one notices that, in the 
case of a canonical scalar field, the quantity $\Sigma$ introduced 
in the article is given by $\Sigma={\dot \phi}^2/2$.

\par

We shall assume that the spatial metric~$h_{ij}$ is given by
\begin{equation}
h_{ij} = a^2(t)\; e^{2\,\cR(t,{\bf x})}\; \delta_{ij}
\end{equation}
and, it is worth noting that, in the gauge that we are working in, the
quantity~$\cR$ that appears in the exponential above is essentially
the curvature perturbation.  One finds that, it suffices to solve the
two constraint equations for the lapse and the shift functions at the
linear order in perturbations.  This is due to the fact that, at the
cubic order in~$\cR$, two type of terms arise in the action which
involve the lapse and shift functions at the second and the third
orders.  However, it is found that, the coefficient of the second
order term contains a first order constraint, while the third order
term proves to be proportional to the zeroth order constraint, both of
which, evidently, vanish~\cite{maldacena-2003,ng-ncsf,ng-reviews}.

\par

For the canonical scalar field of our interest, the action that
governs the curvature perturbation~$\cR$ at the quadratic and the
cubic orders can be arrived at from the original
action~(\ref{eq:a-cmg}).  After a considerable amount of
manipulations, most of which involve using the solutions to the
constraint equations, repeatedly integrating the action by 
parts and systematically throwing away the surface
terms, one can show that, at the quadratic order in $\cR$, the
action~(\ref{eq:a-cmg}) simplifies
to~\cite{maldacena-2003,ng-ncsf,ng-reviews}
\begin{equation}
{\cal S}_{2}[\cR]
=\f{1}{2}\, \int \d \eta\; \int \d^{3}\vx\;\, z^{2}\,
\l[{\cR'}^2-\l(\pa\cR\r)^{2}\r].
\label{eq:a-cmg-so}
\end{equation}
It is straightforward to check that the variation of this action with
respect to $\cR$ leads to the differential equation~(\ref{eq:defk})
that we had considered earlier.

\par

The action at the cubic order in the curvature perturbation can be
obtained in a similar fashion, and is found to
be~\cite{maldacena-2003,ng-ncsf,ng-reviews}
\begin{eqnarray}
{\cal S}_{3}[\cR] 
&=& \Mp^2\int \d \eta\; \int\d^3{\bf x}\; 
\Biggl[a^2\, \epsilon_{1}^2\, {\cal R}\, {\cR'}^2
+a^2\,\epsilon_{1}^2\, {\cal R}\, (\pa {\cal R})^2
\nonumber \\
& &-2\,a\, \epsilon_{1}\, {\cal R}'\, (\pa^{i}{\cal R})\, 
(\pa_{i}\chi)
+\,\frac{a^2}{2}\,\epsilon_{1}\, {\epsilon_{2}'}\, 
{\cal R}^2\, {\cal R}'
+\frac{\epsilon_{1}}{2}\, (\partial^{i}\cR)\, 
(\pa_{i}\chi)\, (\pa^2 \chi)
\nonumber \\
& &+\f{\epsilon_{1}}{4}\, (\pa^2{\cal R})\, (\pa \chi)^2
+a\, {\cal F}\l(\f{\delta {\cal L}_{2}}{\delta \cR}\r)\Biggr],
\label{eq:a-cmg-to}
\end{eqnarray}
where the quantity $\chi$ is defined through the relation
\begin{equation}
\chi \equiv \partial ^{-2}\Lambda,
\end{equation}
with $\Lambda$ being given by
\begin{equation}
\Lambda \equiv \frac{a^2\,\dot\phi^2}{2\,\Mp^2\,H^2}\,\dot \cR
=a\,\epsilon_1\,\cR'. 
\end{equation}
Note that, while $\Lambda $ is of dimension one, $\chi$ is of
dimension $-1$. The quantity $\delta {\cal L}_{2}/\delta\cR$
denotes the variation of the Lagrangian density corresponding to the
quadratic action~(\ref{eq:a-cmg-so}), and can be written as
\begin{equation}
\f{\delta {\cal L}_{2}}{\delta\cR}
={\dot \Lambda}+H\, \Lambda 
-\epsilon_{1}\, (\partial^2\cR).
\end{equation}
The term ${\cal F}(\delta {\cal L}_{2}/\delta \cR)$ that has been
introduced in the above cubic order action refers to the following
expression:
\begin{eqnarray}
\label{eq:calf}
{\cal F}\l(\f{\delta {\cal L}_{2}}{\delta \cR}\r)
&=& \f{1}{2aH}\,\Biggl\{\l[a^{2}H\epsilon_2\cR^2
+4\,a\,\cR\,\cR'+(\partial^{i}\cR)(\pa_{i}\chi)
-\f{1}{H}(\pa\cR)^2\r]
\f{\delta {\cal L}_{2}}{\delta\cR} 
\nonumber \\
& & +\, \l[\Lambda\, (\pa_{i}\cR)+(\pa^{2}\cR)\, (\pa_{i}\chi)\r]\,
\delta^{ij}\, \pa_{j}\l[\pa^{-2}\,
\l(\f{\delta {\cal L}_{2}}{\delta\cR}\r)\r]
\nonumber \\
& & +\,\frac{1}{H}\,\delta^{im}\delta^{jn}\,(\pa_{i}\cR)\, (\pa_{j}\cR)\;
\pa_{m}\,\pa_{n}\,\l[\pa^{-2}\,
\l(\f{\delta {\cal L}_{2}}{\delta\cR}\r)\r]\Biggr\}.
\end{eqnarray}

\par 

Before we proceed, four remarks are required to be made.
Firstly, we should point out that, in the above expressions 
that constitute the action ${\cal S}_3$, the spatial indices 
are to be raised or lowered with the Kr\"onecker symbol. 
Secondly, notice that, all the terms in the 
expression~(\ref{eq:calf})
for ${\cal F}(\delta {\cal L}_{2}/\delta \cR)$, barring the 
first one, contain a derivative of the curvature perturbation 
(either a time or a spatial derivatives or both).
Since the quantities have to be evaluated at the end of 
inflation, the terms which involve the derivatives will not 
contribute to the final results.
Thirdly, we have checked that our expression for ${\cal S}_3$ 
is consistent with the results obtained earlier (such as in the 
first three articles of Refs.~\cite{ng-ncsf}). 
Lastly, it is useful to note that, because the speed of sound 
is unity for the canonical scalar field, no term involving 
$\dot{{\cal R}}^3$ arises, as it happens in cases involving 
non-canonical scalar fields.


\subsection{The definition of the bi-spectrum and the different 
contributions}\label{subsec:dbs-dc}

The goal now is to treat the third order action~(\ref{eq:a-cmg-to}) as
the interaction term and evaluate the three point function of the
curvature perturbation using the standard techniques of perturbative
quantum field theory. 
To begin with, it can be shown that the last term in the 
action~(\ref{eq:a-cmg-to}) which involves $\delta {\cal L}_{2}/\delta \cR$ 
can be removed by a field redefinition of~${\cal R}$ of the following form:
\begin{equation}
\cR \rightarrow \cR_n+F(\cR_n),\label{eq:frd}
\end{equation}
where $F=\epsilon_2\,{\cal R}^2/4$ (for further details, see
Refs.~\cite{maldacena-2003,ng-ncsf,ng-reviews}).  With such a
redefinition, the interaction Hamiltonian corresponding the action
${\cal S}_{3}[\cR]$ above can be written in terms of the conformal
time coordinate as
\begin{eqnarray}
H_{\rm int}(\eta) 
& = &-\Mp^2\, \int \d^{3}{\bf x}\;
\Biggl[a^2\epsilon_{1}^2\cR\cR'^2  
+ a^2\epsilon_{1}^2\cR(\pa \cR)^2 
- 2a\epsilon_{1}\cR'(\pa^{i}\cR)(\pa_{i} \chi)
\nonumber \\ 
& & +\f{a^2}{2}\epsilon_{1}\epsilon_{2}'\cR^2\cR' 
+ \f{\epsilon_{1}}{2}(\pa^{i} \cR)(\pa_{i}\chi) 
\l(\pa^2 \chi\r)
+ \f{\epsilon_{1}}{4}\l(\pa^2 \cR\r)(\pa \chi)^2\Biggr].
\label{eq:Hint}
\end{eqnarray}

\par

The three point correlation function of the curvature perturbation 
can be expressed in terms of the Fourier modes as follows:
\begin{eqnarray}
\langle {\hat \cR}(\eta,\vx)\, {\hat \cR}(\eta,\vx)\, 
{\hat \cR}(\eta,\vx)\rangle
&= &\int \f{\d^3 \vka}{(2\,\pi)^{3/2}}\; 
\int \f{\d^3 \vkb}{(2\,\pi)^{3/2}}\;
\int \f{\d^3 \vkc}{(2\,\pi)^{3/2}}\nn\\ 
&\times &
\langle {\hat \cR}_{\vka}(\eta)\, {\hat \cR}_{\vkb}(\eta)\, 
{\hat \cR}_{\vkc}(\eta)\rangle\; 
{\rm e}^{i\,\l(\vka+\vkb+\vkc\r)\cdot \vx}.\label{eq:tpf-cR}
\end{eqnarray}
At the leading order in the perturbations, one then finds that 
the three point correlation in Fourier space is described by the 
integral~\cite{maldacena-2003,ng-ncsf,ng-reviews}
\begin{eqnarray}
& &\langle {\hat \cR}_{\vka}(\eta_{\rm e})
{\hat \cR}_{\vkb}(\eta_{\rm e})
{\hat \cR}_{\vkc}(\eta_{\rm e})\rangle
\nonumber \\
&=& -i\int_{\eta_{\rm i}}^{\eta_{\rm e}}\d\tau
\l\langle\l[{\hat \cR}_{\vka}(\eta_{\rm e}) 
{\hat \cR}_{\vkb}(\eta_{\rm e})
{\hat \cR}_{\vkc}(\eta_{\rm e}), 
{\hat H}_{\rm int}(\tau)\r]\r\rangle,
\label{eq:tpc-fs}
\end{eqnarray}
where ${\hat H}_{\rm int}$ is the operator corresponding to the 
interaction Hamiltonian~(\ref{eq:Hint}), while $\eta_{\rm i}$ is 
the time at which the initial conditions are imposed on the modes 
when they are well inside the Hubble radius, and $\eta_{\rm e}$ 
denotes a very late time, say, close to when inflation ends.
Moreover, while the square brackets imply the commutation of the 
operators, the angular brackets denote the fact that the correlations 
are evaluated in the initial vacuum state (viz. the Bunch-Davies 
vacuum in the situation of our interest).

\par

The scalar bi-spectrum $\cB_{_{\rm S}}(\vka,\vkb,\vkc)$ is related 
to the three point correlation function of the Fourier modes of the 
curvature perturbation, evaluated at the end of inflation, say, 
$\eta_{\rm e}$, as follows~\cite{wmap-7}: 
\begin{equation}
\langle {\hat \cR}_{\vka}(\eta _{\rm e})\, 
{\hat \cR}_{\vkb}(\eta _{\rm e})\, {\hat \cR}_{\vkc}(\eta _{\rm e})\rangle 
=\l(2\,\pi\r)^3 \cB_{_{\rm S}}(\vka,\vkb,\vkc)\,
\delta^{(3)}\l(\vka+\vkb+\vkc\r).\label{eq:bi-s}
\end{equation}
For convenience, we shall set
\begin{equation}
\cB_{_{\rm S}}(\vka,\vkb,\vkc)=
\l(2\,\pi\r)^{-9/2}\, G(\vka,\vkb,\vkc)
\end{equation}
and, it is useful to note that, the two quantities $\cB_{_{\rm S}}$ 
and $G$ are of dimension $-6$.
We can arrive at the quantity $G(\vka,\vkb,\vkc)$ from the
expressions~(\ref{eq:tpc-fs}) and~(\ref{eq:Hint}) for the three point
function in Fourier space and the interaction Hamiltonian 
$H_{\rm int}$, respectively. Upon using the decomposition~(\ref{eq:cR-d}),
as well as the Wick's theorem, which applies to the correlation
functions of free quantum fields, we find that $G(\vka,\vkb,\vkc)$ can
be written as
\begin{eqnarray}
G(\vka,\vkb,\vkc)
&\equiv & \sum_{C=1}^{7}\; G_{_{C}}(\vka,\vkb,\vkc)\nn\\
&\equiv & \Mp^2\; \sum_{C=1}^{6}\; 
\Biggl\{\l[f_{\ka}(\ee)\, f_{\kb}(\ee)\,f_{\kc}(\ee)\r]\; 
\cG_{_{C}}(\vka,\vkb,\vkc)\nn\\ 
& &+\l[f_{\ka}^{\ast}(\ee)\, f_{\kb}^{\ast}(\ee)\,f_{\kc}^{\ast}(\ee)\r]\;
\cG_{_{C}}^{\ast}(\vka,\vkb,\vkc)\Biggr\}
\nonumber \\ & &
+ G_{7}(\vka,\vkb,\vkc).\label{eq:G}
\end{eqnarray}
The quantities $\cG_{_{C}}(\vka,\vkb,\vkc)$ with $C =(1,6)$, 
which are of dimension $-7/2$, correspond to the six terms in the
interaction Hamiltonian~(\ref{eq:Hint}), and are described by the
integrals~\cite{maldacena-2003,ng-f,ng-ncsf}
\begin{eqnarray}
\cG_{1}(\vka,\vkb,\vkc)
&=&2\,i\,\int_{\eta_\uin}^{\eta_{\rm e}} \d\tau\, a^2\, 
\epsilon_{1}^2\, \l(f_{\ka}^{\ast}\,f_{\kb}'^{\ast}\,
f_{\kc}'^{\ast}+{\rm two~permutations}\r),\label{eq:cG1}\\
\cG_{2}(\vka,\vkb,\vkc)
&=&-2\,i\;\l(\vka\cdot \vkb + {\rm two~permutations}\r)\,
\nonumber \\ & & \times
\int_{\eta_\uin}^{\eta_{\rm e}} \d\tau\, a^2\, 
\epsilon_{1}^2\, f_{\ka}^{\ast}\,f_{\kb}^{\ast}\,
f_{\kc}^{\ast},\label{eq:cG2}\\
\cG_{3}(\vka,\vkb,\vkc)
&=&-2\,i\,\int_{\eta_\uin}^{\eta_{\rm e}} \d\tau \, a^2\,
\epsilon_{1}^2\, \Biggl[\l(\f{\vka\cdot\vkb}{\kb^{2}}\r)\,
f_{\ka}^{\ast}\,f_{\kb}'^{\ast}\, f_{\kc}'^{\ast} \nonumber \\ & &
+ {\rm five~permutations}\Biggr],\label{eq:cG3}\\
\cG_{4}(\vka,\vkb,\vkc)
&=&i\,\int_{\eta_\uin}^{\eta_{\rm e}} \d\tau\, a^2\,\epsilon_{1}\,
\epsilon_{2}'\, \l(f_{\ka}^{\ast}\,f_{\kb}^{\ast}\,
f_{\kc}'^{\ast}+{\rm two~permutations}\r),\label{eq:cG4}\\
\cG_{5}(\vka,\vkb,\vkc)
&=&\frac{i}{2}\,\int_{\eta_\uin}^{\eta_{\rm e}} \d\tau\, 
a^2\, \epsilon_{1}^{3}\, \Biggl[\l(\f{\vka\cdot\vkb}{\kb^{2}}\r)\,
f_{\ka}^{\ast}\,f_{\kb}'^{\ast}\, f_{\kc}'^{\ast}
\nonumber \\ & & 
+ {\rm five~permutations}\Biggr],\label{eq:cG5}\\
\cG_{6}(\vka,\vkb,\vkc) 
&=&\frac{i}{2}\,\int_{\eta_\uin}^{\eta_{\rm e}} \d\tau\, a^2\, 
\epsilon_{1}^{3}\,
\Biggl\{\l[\f{\ka^{2}\,\l(\vkb\cdot\vkc\r)}{\kb^{2}\,\kc^{2}}\r]\, 
f_{\ka}^{\ast}\, f_{\kb}'^{\ast}\, f_{\kc}'^{\ast}
\nonumber \\ & &
+ {\rm two~permutations}\Biggr\}.\label{eq:cG6}
\end{eqnarray}
The additional, seventh term $G_{7}(\vka,\vkb,\vkc)$ arises due 
to the field redefinition~(\ref{eq:frd}), and its contribution 
to $G(\vka,\vkb,\vkc)$ is found to be given 
by~\cite{maldacena-2003,ng-ncsf,ng-reviews}
\begin{equation}
G_{7}(\vka,\vkb,\vkc)
=\frac{\epsilon_{2}(\eta_{\rm e})}{2}\,
\l(\vert f_{\kb}(\eta_{\rm e})\vert^{2}\, 
\vert f_{\kc}(\eta_{\rm e})\vert^{2} 
+ {\rm two~permutations}\r).\label{eq:G7}
\end{equation}

\par

Note that, whereas the first three integrals $\cG_{1}$, $\cG_{2}$ and
$\cG_{3}$ involve $\epsilon_{1}^{2}$, the next three, viz. $\cG_{4}$,
$\cG_{5}$ and $\cG_{6}$, depend on either
$\epsilon_{1}\,\epsilon_{2}'$ or $\epsilon_{1}^{3}$.  And, evidently,
the term $G_{7}$ is proportional to $\epsilon_{2}$.  During slow roll,
the parameter $\epsilon_{1}$ is almost constant and is typically of
the order of $10^{-2}$ or so, while the second and higher slow roll
corrections are even smaller than~$\epsilon_{1}$.  Therefore,
in a slow roll inflationary scenario driven by the canonical scalar
field, it is found that the first three terms $G_{1}$, $G_{2}$ and
$G_{3}$ (that involve the integrals $\cG_{1}$, $\cG_{2}$ and
$\cG_{3}$) and the term $G_7$, which arises due to the field
redefinition, that contribute the most to the
bi-spectrum~\cite{maldacena-2003,ng-ncsf,ng-reviews}.  However, when
there are deviations from slow roll, it has been noticed that the
contribution due to the term $G_{4}$ proves to be the largest to the
bi-spectrum as the corresponding integral $\cG_{4}$ contains the
quantity $\epsilon_{1}\,\epsilon_{2}'$~\cite{ng-f}.


\subsection{The non-Gaussianity parameter $\fnl$}

The observationally relevant dimensionless non-Gaussianity
parameter~$\fnl$ that we had discussed about in the introductory
section is related to the three point correlation function of the
curvature perturbation as follows.  It is introduced through the
equation~\cite{maldacena-2003,ng-f}
\begin{equation}
\cR(\eta, \vx)=\cR^{\rm G}(\eta, \vx)
-\frac{3\,\fnl}{5}\, \l[\cR^{\rm G}(\eta, \vx)\r]^2,
\end{equation}
where $\cR^{\rm G}$ denotes the Gaussian quantity, and the factor of
$3/5$ arises due to the relation between the Bardeen potential and the
curvature perturbation during the matter dominated epoch.  In Fourier
space, the above equation can be written as
\begin{equation}
\cR_\vk
=\cR^{\rm G}_\vk-\frac{3\,\fnl}{5}\, 
\int \f{{\rm d}^{3}{\bf p}}{(2\,\pi)^{3/2}}\, \cR^{\rm G}_{\bf p}\;
\cR^{\rm G}_{\bf k-p}.
\end{equation} 
Upon using this relation, the Wick's theorem, which applies to 
the correlation functions of free quantum fields or, equivalently, 
to Gaussian random fluctuations, and the 
definition~(\ref{eq:twopointfourier}) of the power spectrum, one  
can arrive at the three point correlation of the curvature 
perturbation in Fourier space in terms of the parameter $\fnl$.
It is found to be
\begin{eqnarray}
\langle \hat \cR_{\vka} \hat \cR_{\vkb} \hat \cR_{\vkc} \rangle
&=& -\frac{3\,\fnl}{10}\, (2\,\pi)^{4}\; (2\,\pi)^{-3/2}\;
\f{1}{k_{1}^3\, k_{2}^3\,k_{3}^3\,}\,
\delta^{(3)}(\vka+\vkb+\vkc)\nn\\
& &
\times\l[k_1^{3}\; {\cal P}_{_{\rm S}}(k_2)\; {\cal P}_{_{\rm S}}(k_3)
+{\rm two~permutations}\r].
\end{eqnarray}
Using this expression for the three point function and the 
definition~(\ref{eq:bi-s}) of the bi-spectrum, we can then 
arrive at the following relation between the non-Gaussianity 
parameter $\fnl$ and the bi-spectrum 
$\cB_{_{\rm S}}(\vka,\vkb,\vkc)$:
\begin{eqnarray}
\fnl
&=&-\frac{10}{3}\, (2\,\pi)^{-4}\; (2\,\pi)^{9/2}\;
k_{1}^3\, k_{2}^3\,k_{3}^3\,
\cB_{_{\rm S}}(\vka,\vkb,\vkc)\nn\\
& &
\times\l[k_1^{3}\; {\cal P}_{_{\rm S}}(k_2)\; {\cal P}_{_{\rm S}}(k_3)
+{\rm two~permutations}\r]^{-1}\nn\\
&=&-\frac{10}{3}\, (2\,\pi)^{-4}\; 
k_{1}^3\, k_{2}^3\,k_{3}^3\, G(\vka,\vkb,\vkc)\nn\\
& &
\times\l[k_1^{3}\; {\cal P}_{_{\rm S}}(k_2)\; {\cal P}_{_{\rm S}}(k_3)
+{\rm two~permutations}\r]^{-1}.
\end{eqnarray}

\par 

In this paper, we shall restrict our attention to evaluating the
bi-spectrum in the Starobinsky model in the equilateral limit (i.e.
when $\vka=\vkb=\vkc$).  In such a case, the above expression for
$\fnl$ simplifies to
\begin{equation}
\fnl^{\rm eq}
=-\frac{10}{9}\, (2\,\pi)^{-4}\; 
\f{k^6\; G(k)}{{{\cal P}_{_{\rm S}}^{2}(k)}},\label{eq:fnl-el}
\end{equation}
with $G(k)$ given by [cf. Eq~(\ref{eq:G})]
\begin{eqnarray}
G(k)
&\equiv& \sum _{C=1}^7 G_{_{_C}}(k)\nn\\
&=&\Mp^2\, \sum_{C=1}^6\left[f_k^3\l(\eta_{\rm e}\right)\,
\cG_{_{C}}(k) 
+f_k^{\ast}{}^3\l(\eta_{\rm e}\r)\, \cG_{_{C}}^{\ast}(k)\r]
+G_7(k).\label{eq:G-el}
\end{eqnarray}
Using this expression and the dimensions of the quantities
involved, which we had pointed out before, it is straightforward 
to check that $\fnl$ is indeed a dimensionless quantity.
Further, in the Starobinsky model, due to
the difference in the dynamics before and after the transition, the
integrals in Eqs.~(\ref{eq:cG1})--(\ref{eq:cG6}) need to carried out
separately on either side of the transition.  Therefore, we can write
\begin{eqnarray} 
\label{eq:defG}
G(k) 
&=& \Mp^2\,\sum_{C=1}^6\biggl\{f_k^3\left(\eta_{\rm e}\r)\,
\l[\cG_{_{C}}^+(k)+\cG_{_{C}}^-(k)\r]
+f_k^{\ast}{}^3\l(\eta_{\rm e}\r)\,
\l[\cG_{_{C}}^+{}^{\ast}(k) +\cG_{_{C}}^-{}^{\ast}(k)\r]\biggr\}
\nonumber \\ & &
+G_7(k)\nn\\
&=& \Mp^2\, \sum_{C=1}^6 \l[f_k^3\left(\eta_{\rm e}\r)\,
\cG_{_{C}}^+(k)+f_k^{\ast}{}^3\left(\eta_{\rm e}\r)
\cG_{_{C}}^+{}^{\ast}(k)\r]\nn\\
& &+\,\Mp^2\, \sum_{C=1}^6\l[f_k^3\left(\eta_{\rm e}\r)\,\cG_{_{C}}^-(k)
+f_k^{\ast}{}^3\left(\eta_{\rm e}\r)\,
\cG_{_{C}}^-{}^{\ast}(k)\r]+G_7(k)\nn\\
&=&  \sum_{C=1}^6 \l[G_{C}^{+}(k)+G_{C}^{-}(k)\r]+G_7(k),
\end{eqnarray}
where, as we had mentioned earlier, the plus and the minus signs in
the super-scripts refer to the quantities before and after the
transition, respectively. In the following section, we shall first
evaluate the contribution to the bi-spectrum due to the supposedly
dominant term~$G_{4}$ in the Starobinsky model.  And, in the
subsequent section, we shall evaluate all the remaining contributions
as well. As we have pointed out before, remarkably, we are able
to evaluate the bi-spectrum in the equilateral limit completely
analytically without any further assumptions or approximations.  We
shall then discuss the range of values for the parameters of the
Starobinsky model for which the non-Gaussianity parameter $\fnl$ can
be large.


\section{The dominant contribution to the bi-spectrum in the 
Starobinsky model}\label{sec:dc-bs}

In this section, we shall evaluate the contribution due to the
term~$G_{4}$ to the bi-spectrum in the Starobinsky model in the
equilateral limit.

\par

To begin with, recall that, in the Starobinsky model, the second roll
parameter $\epsilon_{2}$ is constant before the field reaches the
discontinuity at $\phi_{0}$ when $\eta=\eta_{0}=-\l(a_0\,
H_{0}\r)^{-1} =k_0^{-1}$.  As a result, the integral $\cG_{4}$ which
involves $\epsilon_{2}'$ [cf.~Eq.~(\ref{eq:cG4})] vanishes before
$\eta_{0}$ and, hence, we are left with only the following
contribution after the transition:
\begin{equation}
\cG_{4}(\vka,\vkb,\vkc)
=i\int_{-k_0^{-1}}^{0} \d\tau\, a^2\epsilon_{1-}
\epsilon_{2-}' \l(f_{k_1}^{\ast} 
f_{k_2}^{\ast}\, f_{k_3}'^{\ast} 
+{\rm two~permutations}\r).
\end{equation}
Also, post-transition, the mode $v_k$ and the quantity~$z$ are given 
by Eqs.~(\ref{eq:vk-at}) and~(\ref{eq:z-at}), respectively.
Since $f_k =v_k/z$, one obtains that 
\begin{eqnarray}
f_k^{-}(\eta)
=\frac{iH_0\alpha_k}{2\Mp\sqrt{{k^3}\epsilon_{1-}}}
\left(1+ik\eta\right) {\rm e}^{-i\,k\,\eta}
-\frac{iH_0\beta_k}{2\Mp\sqrt{{k^3}\,\epsilon_{1-}}}
\l(1-ik\eta\right) {\rm e}^{i\,k\,\eta},\label{eq:fk-at}
\end{eqnarray}
while the corresponding derivative with respect to the conformal time
coordinate is given by 
\begin{eqnarray}
f_k^{-}{}'(\eta)
&=&\frac{iH_0\alpha_k}{2\Mp\sqrt{{k^3}\epsilon_{1-}}}
\l[-{\cal H}\epsilon_{1-}\l(1+ik\eta\right)
-\f{{\cal H}\,\epsilon_{2-}}{2}\l(1+ik\eta\right)
+k^2\eta\r]
{\rm e}^{-i\,k\,\eta}\nn\\
&-&
\frac{iH_0\beta_k}{2\Mp\sqrt{{k^3}\epsilon_{1-}}}
\l[-{\cal H}\epsilon_{1-}\l(1-ik\eta\right)
-\f{{\cal H}\,\epsilon_{2-}}{2}\l(1-ik\eta\right)
+k^2\eta\r]{\rm e}^{i\,k\,\eta}.\label{eq:fkp-at}
\nn \\
\end{eqnarray}
In this expression, one can ignore the first term in the square
bracket since it is proportional to the first slow roll parameter
which, as demonstrated before, always remains small even when the
field is crossing the discontinuity in the slope of the potential.
However, we should stress here that, in doing so, we are in fact
ignoring contributions to $G_{4}$ that are possibly of the same order
as the sub-dominant terms, such as $G_{1}$, $G_{2}$, $G_{3}$ and
$G_7$.  We shall comment more on this point in the next section.
 
\par

\begin{figure}
\begin{center}
\includegraphics[width=15cm]{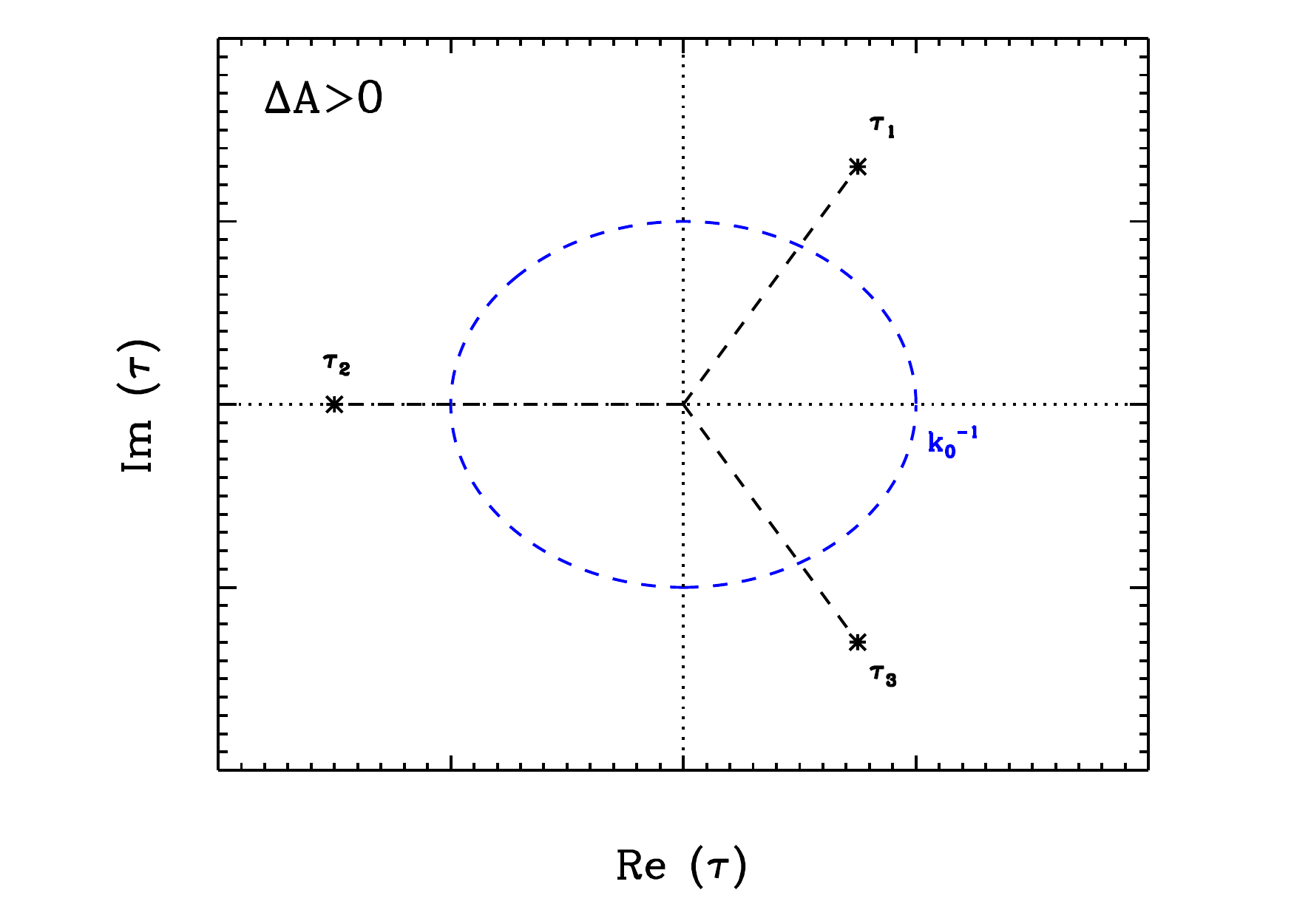}
\caption{The three poles corresponding to the factor
  $\l(1-\rho^3\,\tau^3\r)^{-1}$ that appears in the integrals $I_4$
  and $J_4$ [cf.~Eqs.~(\ref{eq:I4}) and~(\ref{eq:J4})] are marked with
  black asterisks in the complex $\tau$-plane for the case wherein
  $\Delta A > 0$.  Evidently, they will all fall on a circle of radius
  $\vert\rho\vert^{-1} =(A_-/\vert\Delta A\vert)\,k_0^{-1}$ which is
  centered at the origin [cf.~Eq.~(\ref{eq:rho3})].  Note that, in the
  integrals $I_4$ and $J_4$, the path of integration runs from
  $-k_0^{-1}$ to zero along the negative $\tau$-axis.  Amongst the
  three poles, viz. $\tau_1$, $\tau_2$ and $\tau_3$ [in this context,
  see Eq.~(\ref{eq:taup}) below], only $\tau_2$ lies on the axis.  The
  dashed blue curve represents a circle of radius $k_0^{-1}$ about the
  origin.  However, since $\vert\Delta A\vert/A_- < 1$ when $\Delta
  A > 0$, all the poles, including $\tau_2$, lie outside the blue curve.
  Therefore, no pole falls on the path of integration.}
\label{fig:poleplus}
\end{center}
\end{figure}

We find that, upon using the expressions~(\ref{eq:e1-at})
and~(\ref{eq:e2d-at}) for $\epsilon_{1}$ and ${\dot \epsilon_{2}}$
after the transition, in the equilateral limit, we can write
$\cG_{4}$ as follows:
\begin{eqnarray}
\cG_{4}(k)
&=&\f{81\, \Delta A\,\,k_0^3\, H_{0}^{3}}{2\, 
\sqrt{2\, k^9}\,\Mp^2\, A_{-}^{2}}\,
\biggl[{\alpha_{k}^{\ast}}^{3}\, I_{4}(k) 
-{\beta_{k}^{\ast}}^{3}\, I_{4}^{\ast}(k)
-{\alpha_{k}^{\ast}}^{2}\, \beta_{k}^{\ast}\,J_{4}(k)
\nonumber \\ & &
+\alpha_{k}^{\ast}\, {\beta_{k}^{\ast}}^{2}\, J_{4}^{\ast}(k)\biggr].
\label{eq:cG4-pv}
\end{eqnarray}
The quantities $I_{4}$ and $J_{4}$ in the above expression are 
described by the integrals
\begin{eqnarray}
I_{4}(k)
&=&\int_{-k_{0}^{-1}}^{0} 
\f{\d\tau \; \tau}{\l(1-\rho^{3}\, \tau^{3}\r)^{4}}\, 
\l(1-i\, k\, \tau\r)^{2}\,
\biggl[k^{2}+3\, \rho^{3}\, \l(1-i\,k\,\tau\r)\,\tau
\nonumber \\ & &
-k^{2}\,\rho^{3}\,\tau^{3}\biggr]\, {\rm e}^{3\, i\,k\,\tau},
\label{eq:I4}\\
J_{4}(k)
&=&\int_{-k_0^{-1}}^{0} 
\f{\d\tau\; \tau}{\l(1-\rho^{3}\, \tau^{3}\r)^{4}} 
\l(1-i\, k\, \tau\r)
\biggl[3\,k^{2}+9\, \rho^{3}\, \tau
+i\, k^{3}\,\tau+6\, \rho^{3}\,k^{2}\tau^{3}
\nonumber \\ & &
-i\, k^{3}\rho^{3}\tau^{4}\biggr]\,{\rm e}^{i\,k\,\tau},
\label{eq:J4}
\end{eqnarray}
where the quantity $\rho$ is defined by the following expression
\begin{equation}
\rho^{3}\equiv -\f{\Delta A}{A_{-}}\, k_0^{3}.\label{eq:rho3}
\end{equation}
A concern could arise that, in the above integrals, one may encounter
poles corresponding to the factor $\l(1-\rho^3\, \tau^3\r)^{-1}$ along
the path of integration, viz. from $-k_0^{-1}$ to zero on the negative
$\tau$-axis.  It is easy to establish that this does not occur.  Since
$\tau$ is negative over the region of interest, when $\Delta A < 0$,
$\rho^3>0$, so that $\rho^3\,\tau^3<0$.  So,
$\l(1-\rho^3\,\tau^3\r)>0$ for all $\tau<0$ and, hence, no pole occurs
on the negative $\tau$ axis in such a case. [The poles when $\Delta A
< 0$ are given by Eq.~(\ref{eq:taum}), and their actual locations in
the complex $\tau$-plane can be found represented in
Fig.~\ref{fig:poleminus}.]  Whereas, when $\Delta A > 0$, $\rho^3 <
0$, and $\rho^3\,\tau^3>0$ for $\tau<0$.  It is then clear that, under
such conditions, a pole will indeed arise on the negative $\tau$-axis.
However, as we have illustrated in Fig.~\ref{fig:poleplus}, the pole
on the axis of integration only occurs at $\tau <-\,k_0^{-1}$, which
lies beyond the lower limit of the integrals.
Therefore, we encounter no poles on the path of integration.

\par

As no poles arise, we find that the integrals $I_{4}$ and $J_{4}$ can
be evaluated easily, and we obtain that
\begin{eqnarray}
I_{4}(k)
&=&\f{1}{3}-\frac13\l[\f{1+i\, k/k_{0}}{1+(\rho/k_{0})^{3}}\r]^{3}\,
\,{\rm e}^{-3\, i\,k/k_{0}},\\
J_{4}(k)
&=&1-\f{(1-i\, k/k_{0})\, 
\l(1+i\, k/k_{0}\r)^{2}}{\l[1+(\rho/k_{0})^{3}\r]^3}\,
\,{\rm e}^{-i\,k/k_{0}}.
\end{eqnarray}
Upon substituting the above results for the integrals $I_4$ and $J_4$
in Eq.~(\ref{eq:cG4-pv}), we obtain that 
\begin{eqnarray}
\cG_{4}(k)
&=&\f{81\, \Delta A\,\,k_0^3\, H_{0}^{3}}{2\, 
\sqrt{2\, k^9}\,\Mp^2\, A_{-}^{2}}
\Biggl\{\f{\alpha_k^{\ast}{}^3}{3}
-\alpha_k^{\ast}{}^2\,\tilde{\beta}_k^{\ast}\,{\rm e}^{-2\,i\,k/k_0}
-\f{\l[1+(i\,k/k_0)\r]^3}{3\,\l[1+(\rho/k_0)^3\r]^3}
\,{\rm e}^{-3ik/k_0}
\nonumber \\ & &
+\,\alpha ^{\ast}\,\tilde{\beta}_k^{\ast}{}^2\,{\rm e}^{-4\,i\,k/k_0}
-\f{\tilde{\beta}_k^{\ast}{}^3}{3}\,{\rm e}^{-6ik/k_0}\Biggr\},
\label{eq:cG4-sm}
\end{eqnarray}
where, for convenience, we have introduced the quantity
\begin{equation}
{\tilde \beta}_k\equiv \beta_k\, {\rm e}^{-2\,i\,k/k_0}.
\end{equation}
As we shall see, working in terms of ${\tilde \beta}_k$ allows us to
isolate exponential factors easily.

\par

\begin{figure}
\begin{center}
\includegraphics[width=15.0cm]{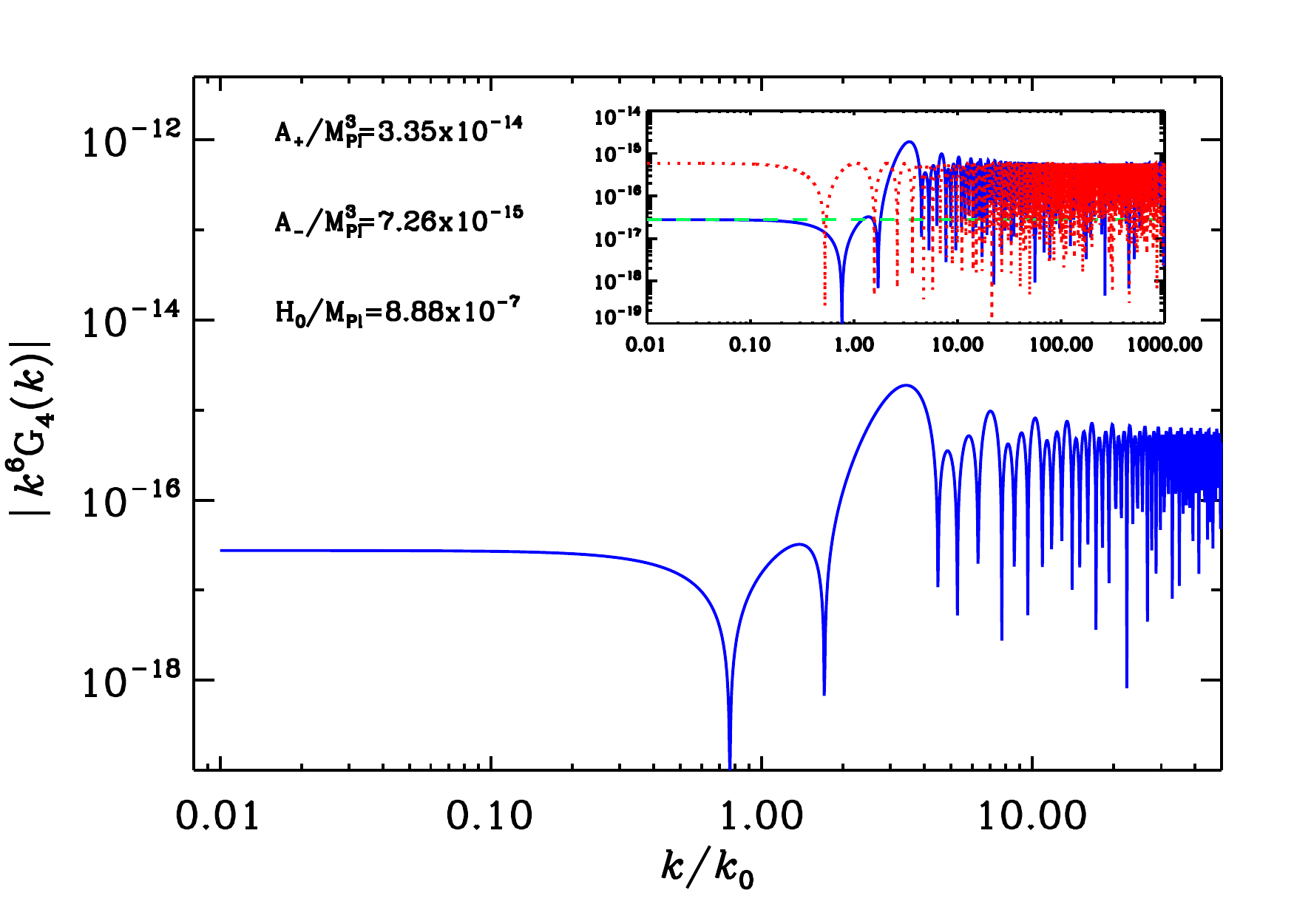}
\end{center}
\caption{The absolute value of the quantity $k^6\, G_4$, as given by
  Eq.~(\ref{eq:G4}), has been plotted as a function of $k/k_0$ (the
  blue curve).  We have worked with the same of values of $A_{+}$,
  $A_{-}$ and $H_0$ as in the earlier figures.  We should stress the
  fact that, as in the case of the power spectrum, the quantity $k^6\,
  G_4$ depends on the wavenumber only through the ratio $k/k_0$.  The
  green and the red curves in the inset represent the asymptotic
  behavior for $k\ll k_0$ and $k\gg k_0$,
  respectively.}\label{fig:abs-G4}
\end{figure}

We next need to evaluate the quantity $G_4(k)$. 
According to our prior discussions [see in particular 
Eqs.~(\ref{eq:defG})], this quantity is given by
\begin{equation}
G_4(k)= \Mp^2\, \l[f_k^3(\eta_{\rm e})\, \cG_4(k)
+f_k^{\ast}{}^3(\eta_{\rm e})\, \cG_4^{\ast}(k)\r].
\end{equation}
Towards the end of inflation, i.e. as $\eta\to 0$, the mode 
$f_{k}$ reduces to
\begin{equation}
f_k(\eta_{\rm e})
=\frac{i\, H_0}{2\, \Mp\, \sqrt{{k^3}\,\epsilon_{1-}(\eta_{\rm e})}}\,
\l(\alpha_k-\beta _k\r),\label{eq:fk-lt}
\end{equation}
where $\epsilon_{1-}(\eta _{\rm e})$ denotes the value of the first
slow roll parameter at late times.  On using this expression and the
relation~(\ref{eq:cG4-sm}) for $\cG_4$, we find that we can write
$G_4$ as
\begin{eqnarray}
k^6\, G_4(k)
&=& \frac{81}{8\sqrt{2\epsilon_{1-}^{3}(\eta_{\rm e})}} 
\l(\f{k_0}{k}\r)^3\;
\f{\Delta A\, H_0^6}{A_-^2\,\Mp^3}\,
\Biggl[{\cal A}_1(k)\,\sin\l(\f{k}{k_0}\right)
\nonumber \\ 
&+&{\cal A}_2(k)\cos\l(\f{k}{k_0}\r)
+{\cal A}_3(k)\sin\l(\f{3 k}{k_0}\r)
+{\cal A}_4(k)\cos\l(\f{3k}{k_0}\r)\Biggr],\label{eq:G4}
\end{eqnarray} 
where the four coefficients ${\cal A}_1$, ${\cal A}_2$, 
${\cal A}_3$ and ${\cal A}_4$ are found to be
\begin{eqnarray}
{\cal A}_1(k) 
&=& \frac{3\,A_-^3\,\Delta A}{4\,A_+^5}
\l(1+\frac{k^2}{k_0^2}\right)^2\,
\Biggl[9\,A_-\l(1+\frac{k^2}{k_0^2}\r)
\nonumber \\ & &
+\,A_+\l(-9-\f{9\,k^2}{k_0^2}+\f{2\,k^4}{k_0^4}\r)\Biggr]\,
\l(\frac{k}{k_0}\r)^{-6},\label{eq:cA1}\\
{\cal A}_2(k) 
&=& -\frac{3\,A_-^3\, \Delta A}{4\,A_+^5}\,
\l(1+\f{k^2}{k_0^2}\r)^2\,
\Biggl[9\,A_-\l(1+\frac{k^2}{k_0^2}\r)
\nonumber \\ & &
-A_+\,\l(9+\frac{11\, k^2}{k_0^2}\right)
\Biggr]\, \l(\frac{k}{k_0}\right)^{-5},\\
{\cal A}_3(k) 
&=& -\frac{A_-^3}{12\,A_+^5}\,
\Biggl[27\, \Delta A^2\l(1-\f{k^2}{k_0^2}\r)
-27\, \Delta A\,\l(5\,A_--7\,A_+\r)\,\frac{k^4}{k_0^4}
\nonumber \\ & &
-\l(9A_--11A_+\r)^2\frac{k^6}{k_0^6}
+6 A_+\l(-3A_-+5A_+\r)\frac{k^8}{k_0^8}\Biggr]
\left(\frac{k}{k_0}\right)^{-6},\\
{\cal A}_4(k) 
&=& \frac{A_-^3}{12\,A_+^5}
\Biggl[-27\,A_-^2\,\l(-3+\frac{k^2}{k_0^2}\r)\,
\l(1+\frac{k^2}{k_0^2}\right)^2
+\,18\,A_-\,A_+\l(1+\frac{k^2}{k_0^2}\right)\,
\nonumber \\ & & \times
\l(-9-\frac{7\,k^2}{k_0^2}+\frac{6\,k^4}{k_0^4}\right)
+\,A_+^2\Biggl(81+\frac{153\,k^2}{k_0^2}-\frac{9\,k^4}{k_0^4}
-\frac{93\,k^6}{k_0^6}
\nonumber \\ & &
+\frac{4\,k^8}{k_0^8}\Biggr)\Biggr]\,
\left(\frac{k}{k_0}\right)^{-5}.\label{eq:cA4}
\end{eqnarray}
Let us now understand the behavior of $G_4$ at large and small scales.
Experience with the slow roll results suggest that, far from~$k_0$, on
either side, one can expect $k^6\, G_4$ to turn scale invariant. As
$k/k_0\to 0$, we find that
\begin{equation}
\lim_{k/k_0\,\to\, 0}\;
k^6\,G_4(k)
=\frac{27}{8\, \sqrt{2\,\epsilon_{1-}^{3}(\eta_{\rm e})}}\,
\frac{\Delta A\, A_-^3\, H_0^6}{A_+^5\,\Mp^3}
\end{equation}
and, in the limit $k/k_0\to\infty$, one arrives at the result
\begin{equation}
\lim_{k/k_0\,\to\, \infty}\;
k^6\,G_4(k)
=\frac{27}{8\,\sqrt{2\,\epsilon_{1-}^{3}(\eta_{\rm e})}}\,
\frac{H_0^6\,\Delta A\,A_-}{A_+^3\,\Mp^3}\,
\cos\l(\frac{3\,k}{k_0}\right).
\end{equation}
In Fig~\ref{fig:abs-G4}, we have plotted the absolute value of the
quantity $k^6\, G_{4}$ given by the expression~(\ref{eq:G4}). Clearly,
while the quantity is strictly scale invariant for $k\ll k_0$, the
scale invariant amplitude is modulated by superimposed oscillations
for $k\gg k_0$.

\par

It is interesting to note that, while the power spectrum had depended
on trigonometric functions involving $2\, k/k_0$
[cf. Eq.~(\ref{eq:ps-sm})], the bi-spectrum depends on trigonometric
functions involving $3\, k/k_0$.
Definitely, these behavior can be attributed to the fact that 
the power spectrum and the bi-spectrum depend on the second 
and the third powers of the curvature perturbation, respectively. 
In fact, that such a behavior can broadly be expected to 
occur has been pointed out previously in the literature 
while investigating a model that, in some aspects, is 
similar to the Starobinsky model~\cite{ng-reviews}. 
The model considered earlier contains a discontinuity in 
the potential itself rather than in the derivative of the 
potential. 
Despite this difference, the slow roll parameters seem to 
behave just as in the Starobinsky model, with the first slow 
roll parameter remaining small through the transition as the 
field crosses the discontinuity, whereas $\epsilon_2$ and 
$\dot{\epsilon_2}$ grow large for a short period around the 
time of the transition. 
Since the background behavior is rather similar, it is possible
for the two cases to be meaningfully compared. 
We find that, in order to evaluate the $\fnl$ in the model, the 
earlier work~\cite{ng-reviews} simply represents $\epsilon_2'$ 
as a Dirac delta function at the transition which, then, permits 
an easy integration of the dominant contribution. 
However, such an approach cannot seem to reproduce the detailed 
pattern that we have obtained in our calculation here, and one 
arrives at only a rough oscillatory behavior on the small scales 
(in this context, see Eq.~(6.32) of Ref.~\cite{ng-reviews}). 
Moreover, the method does not seem to allow the calculation of 
the other, sub-dominant, contributions, as we are able to carry
out in the case of the Starobinsky model (see our discussion 
in the following section). 
With regards to the overall amplitude of the effect, it is not  
straightforward to compare the results, since the models are not 
exactly similar. 
However, the order of magnitude of $\fnl$ is found to be in 
agreement with what we obtain (see Sec.~\ref{sec:l-fnl}). 
This leads us to conclude that our results are broadly consistent 
with those of Ref.~\cite{ng-reviews}.


\section{The sub-dominant contributions to the bi-spectrum}
\label{sec:sdc-bs}

In this section, we shall arrive at analytic expressions for the
other, sub-dominant contributions to the bi-spectrum in the
equilateral limit.  Unlike the dominant term $G_4$ wherein no
contribution arose before the transition (due to the vanishing 
${\dot \epsilon_{2}}$), the other five terms involving
integrals---viz. $G_1$, $G_2$, $G_3$, $G_5$ and $G_6$---contribute
before as well as after the transition.  As one would expect, the
calculation of the contribution until the transition to fast roll
largely follows the computations in the slow roll case.  In contrast,
the computation of the contribution post-transition proves to be more
involved, but, as we shall see, tractable.  The last term $G_7$ is,
obviously, straightforward to evaluate since it does not involve any
integrals, and only requires the amplitude of the curvature
perturbation at late times.


\subsection{The contribution due to the second term}

We shall first evaluate the contribution due to the second term in the
interaction Hamiltonian~(\ref{eq:Hint}), viz. the quantity $G_2$ arising 
due to the integral $\cG_2$ [cf.~Eq.~(\ref{eq:cG2})], as it involves 
simpler integrals.


\subsubsection{Before the transition}

Since $\epsilon_{1}$ remains a constant before the transition,
the integral~(\ref{eq:cG2}) can be written as
\begin{eqnarray}
\cG_{2}^{+}(\vka,\vkb,\vkc)
&=&-2\,i\;\l(\vka\cdot \vkb + {\rm two~permutations}\r)\,
\epsilon_{1+}^2\,
\nonumber \\ & &\times
\int_{-(1-i\,\gamma)\,\infty}^{-k_0^{-1}}
 \d\tau\, a^2\, 
f_{\ka}^{\ast}\,f_{\kb}^{\ast}\, f_{\kc}^{\ast},\label{eq:cG2p}
\end{eqnarray}
where the $\gamma$ which appears in the lower limit is an
infinitesimal positive quantity that has been introduced by hand, as
is done in the standard slow roll case~\cite{maldacena-2003,ng-ncsf}.
The procedure can also be viewed as though the integration contour has
been rotated by a small angle $\gamma$ in the complex $\tau$ plane.
The inclusion of $\gamma$ ensures the selection of the correct choice
for the perturbative vacuum. Also, algebraically, its introduction
allows us to avoid the increasingly rapid oscillations at very early
times (i.e. as $\eta\to -\infty$), as it acts as an exponential
cut-off that leads to the rapid convergence of the integral.

\par

Note that, before the transition, the mode $v_k$ and the quantity~$z$
are given by Eqs.~(\ref{eq:vk-bt}) and~(\ref{eq:z-at}), so that we
have
\begin{equation}
f_k^{+}(\eta)
=\frac{i\, H_0}{2\, \Mp\, \sqrt{{k^3}\,\epsilon_{1+}}}\,
\l(1+i\,k\,\eta\right)\,{\rm e}^{-i\,k\,\eta},\label{eq:fk-bt}
\end{equation}
Upon substituting this mode in the above expression for $\cG_{2}^{+}$,
one finds that the resulting integral can the easily evaluated even in
the most generic case of $\vka\ne\vkb\ne\vkc$.  We obtain the
contribution until the transition to be
\begin{eqnarray}
\cG_2^{+}(\vka,\vkb,\vkc)
&=&\frac{H_0\,}{4\,\Mp^3}
\sqrt{\frac{\epsilon_{1+}}{(k_1\,k_2\,k_3)^{3}}}\,
\l(\vka\cdot\vkb + \vka\cdot\vkc + \vkb\cdot\vkc\r)\,
{\rm e}^{-i\,k_{_{\rm T}}/k_0}
\nonumber \\ & \times &
\l[k_0-\frac{k_1k_2k_3}{k_{_{\rm T}}\, k_0}
+\frac{i\,k_1\,k_2\,k_3}{k_{_{\rm T}}^2}
+\frac{i}{k_{_{\rm T}}}\left(k_1\,k_2+k_2\,k_3+k_1\,k_3\right)\r],
\nonumber  \\
\end{eqnarray}
where $k_{_{\rm T}}\equiv \ka+\kb+\kc$. It can be easily checked that
this expression reduces to the standard slow roll result as $k_0\to
\infty$ \cite{maldacena-2003,ng-ncsf}, which is basically equivalent
to assuming that the transition to fast roll does not take place at
all.  On restricting to the equilateral case, the above expression
simplifies to
\begin{eqnarray}
\cG_2^+(k)=\frac{3\,H_0\,k^2\,k_0}{4\,\Mp^3}\,
\sqrt{\f{\epsilon_{1+}}{k^{9}}}\,
\left(1+\frac{10\,i\,k}{9\, k_0}-\frac{k^2}{3\,k_0^2}\right)
{\rm e}^{-3\,i\,k/k_0}.
\end{eqnarray}

\par

With the help of the above $\cG_2^+$, we are now in a position to 
calculate $G_{2}^{+}$, which is given by
\begin{eqnarray}
G_{2}^{+}(k)
&=&\Mp^2\,\left[f_k^3\l(\eta_{\rm e}\r)\,\cG_2^+(k)
+f_k^{\ast}{}^3\l(\eta_{\rm e}\r)\,\cG_2^{\ast}{}^+(k)\r]\nn\\
&=& \frac{-i H_0^3}{8 \Mp \sqrt{k^9\,
\epsilon_{1-}^{3}(\eta _{\rm e})}}
\left[\left(\alpha_k-\beta_k\right)^3 \cG_2^+(k)
-\left(\alpha_k^{\ast}-\beta_k^{\ast}\right)^3
\cG_2^+{}^{\ast}(k)\right].
\qquad
\end{eqnarray}
Upon using the expressions~(\ref{eq:alphak-sm}) and (\ref{eq:betak-sm})
for $\alpha_k$ and $\beta_k$, straightforward manipulations allow us to 
write $G_{2}^{+}$ as follows:
\begin{eqnarray}
k^6\,G_{2}^{+}(k)
&=& \frac{1}{16\, \sqrt{2\,\epsilon_{1-}^{3}(\eta_{\rm e})}}\,
\frac{k_0}{k}\, \frac{A_+\, H_0^2}{\Mp^5}\nn\\
& \times &
\Biggl\{3\,\Re\Biggl[
\left(\alpha_k^2\,\tilde{\beta}_k
+\alpha_k^{\ast}\,\tilde{\beta}_k^{\ast}{}^2\right)
\left(1-\frac{k^2}{3\,k_0^2}\right)
\nonumber \\ &+&
\frac{10\,i\,k}{9\,k_0}\;
\left(\alpha_k^2\,\tilde{\beta}_k
-\alpha_k^{\ast}\,\tilde{\beta}_k^{\ast}{}^2\right)\Biggr]
\sin \left(\frac{k}{k_0}\right)\nn\\
& -&
3\,\Im\Biggl[\left(\alpha_k^2\,\tilde{\beta}_k
+\alpha_k^{\ast}\,\tilde{\beta}_k^{\ast}{}^2\right)
\left(1-\frac{k^2}{3\,k_0^2}\right)
\nonumber \\ & +&
\frac{10\,i\,k}{9\,k_0}\,
\left(\alpha _k^2\,\tilde{\beta}_k
-\alpha_k^{\ast}\,\tilde{\beta}_k^{\ast}{}^2\right)\Biggr]
\cos \left(\frac{k}{k_0}\right)\nn\\
&- &
\Re~\Biggl[
\left(\alpha_k^3+\tilde{\beta}_k^{\ast}{}^3\right)
\left(1-\frac{k^2}{3\,k_0^2}\right)+\frac{10\,i\,k}{9\,k_0}
\left(\alpha _k^3-\tilde{\beta}_k^{\ast}{}^3\right)\Biggr]
\sin \left(\frac{3\,k}{k_0}\right)\nn\\
&+ &
\Im\Biggl[
\left(\alpha_k^3+\tilde{\beta}_k^{\ast}{}^3\right)
\left(1-\frac{k^2}{3 k_0^2}\right)+\frac{10 i k}{9 k_0}\,
\left(\alpha _k^3-\tilde{\beta}_k^{\ast}{}^3\right)\Biggr]
\cos \left(\frac{3\,k}{k_0}\right)\Biggr\},
\nonumber \\
\end{eqnarray}
where $\Re$ and $\Im$ denote the real and the imaginary parts of the
arguments. 
Notice that the above expression contains trigonometric functions 
of arguments $k/k_0$ and $3\,k/k_0$, as the contribution due to 
the fourth term had.
Therefore, this expression can be expected to lead to a similar (but, 
in some ways, different!) oscillatory pattern as the dominant 
contribution.

\par

Let us now understand the asymptotic forms of $k^6\, G_2^{+}$. As
$k/k_0\rightarrow 0$, we find that it goes to a constant given by
\begin{equation}
\lim _{k/k_0\to 0} k^6\, G_{2}^{+}(k)
=-\frac{17}{144\,\sqrt{2\,\epsilon_{1-}^{3}(\eta_{\rm e})}}\;
\frac{A_-^3\, H_0^2}{\Mp^5\, A_+^2}.
\end{equation}
While, in the limit $k/k_0\to \infty$, one obtains the following
behavior
\begin{eqnarray}
\lim _{k/k_0\to \infty} k^6\,G_{2}^{+}(k)
&=& \frac{1}{16\,\sqrt{2\,\epsilon_{1-}^{3}(\eta_{\rm e})}}\,  
\f{A_+\, H_0^2}{\Mp^5}
\Biggl[\frac{3}{2}\,\l(1-\frac{A_-}{A_+}\r)\,
\cos\l(\frac{k}{k_0}\r)
\nonumber \\ & +&
\frac{1}{3}\,\frac{k}{k_0}\,\sin\l(\frac{3\,k}{k_0}\r)
+\l(\frac{47}{18}-\frac{3\,A_-}{2\,A_+}\r)\,
\cos\left(\frac{3\,k}{k_0}\right)\Biggr].\label{eq:G2p-lk}
\end{eqnarray}
It should be pointed out that the second term grows with $k$ as
$(k/k_0)\, \sin\,(3\,k/k_0)$. In other words, $k^6\, G_2^{+}$ diverges
linearly at large $k$. Interestingly, as we shall illustrate, this
diverging term will be canceled by a similar term but with the
opposite sign that arises due to the contribution post-transition,
ensuring that the complete $k^6\,G_2$ remains finite at large~$k$.


\subsubsection{After the transition}

Let us now turn to the evaluation of the contribution $\cG_{2}^{-}$
after the transition. The calculation proceeds in a manner very
similar to the computation of $\cG_4^{-}$. But, in contrast to 
$\cG_4^{-}$, for evaluating $\cG_2^{-}$, we only require the behavior 
of the first slow roll parameter after the transition, which is given 
by Eq.~(\ref{eq:e1-at}). 
Upon substituting the modes~(\ref{eq:fk-at}) after the transition in 
Eq.~(\ref{eq:cG2}) and, after some manipulations, one obtains that
\begin{eqnarray}
\cG_2^{-}(k) 
&=& \frac{3\,H_0\,k^2}{4\,\Mp^3\,k^{9/2}}\, 
\biggl[{\alpha_{k}^{\ast}}^{3}\,I_{2}(k) 
-{\beta_{k}^{\ast}}^{3}\,I_{2}^{\ast}(k)
-{\alpha_{k}^{\ast}}^{2}\, \beta_{k}^{\ast}\,J_{2}(k)
\nonumber \\ & &
+\alpha_{k}^{\ast}\,{\beta_{k}^{\ast}}^{2}\,
J_{2}^{\ast}(k)\biggr],
\end{eqnarray}
where $I_2$ and $J_2$ are described by the integrals
\begin{eqnarray}
I_2(k) &=& \frac{A_-}{\sqrt{18}\,H_0^2\,\Mp}
\int _{-k_0^{-1}}^{\eta_{\rm e}}\frac{{\rm d}\tau}{\tau^2}\,
\l(1-\rho^3\,\tau^3\r)(1-i\,k\,\tau)^3
{\rm e}^{3\,i\,k\,\tau},\\
J_2(k) &=& \frac{3\, A_-}{\sqrt{18}\,H_0^2\,\Mp}
\int _{-k_0^{-1}}^{\eta_{\rm e}}
\frac{{\rm d}\tau}{\tau^2}
\l(1-\rho^3\,\tau^3\r)(1-i\,k\,\tau)^2(1+i\,k\,\tau)\,
{\rm e}^{i k \tau},
\end{eqnarray}
with $\rho^3$ given by Eq.~(\ref{eq:rho3}). It is clear that the
structure of these integrals is very similar to that of $I_4$ and
$J_4$ [cf. Eqs.~(\ref{eq:I4}) and (\ref{eq:J4})]. However, in contrast,
the poles in $I_2$ and $J_2$ are located at the origin and, hence, they 
do not fall on the integration path. Due to this reason, the two
integrals $I_2$ and $J_2$ can be performed easily, and the results
can be written in the following form:
\begin{eqnarray}
I_2(k) 
= \frac{A_-\,k_0}{\sqrt{18}\,H_0^2\,\Mp}\,
\l[{\cal I}_2^a(k)
+{\cal I}_2^b(k)\, {\rm e}^{-3\,i\,k/k_0}\right], \\
J_2(k) 
= \frac{3\,A_-\,k_0}{\sqrt{18}\,H_0^2\,\Mp}\,
\l[{\cal J}_2^a(k)+{\cal J}_2^b(k)\, {\rm e}^{-i\,k/k_0}\right],
\end{eqnarray}
where the scale dependent quantities ${\cal I}_2^a$, 
${\cal I}_2^b$, ${\cal J}_2^a$ and ${\cal J}_2^b$ are given by
\begin{eqnarray}
{\cal I}_2^a(k) 
&=& -\frac{1}{k_0\,\eta_{\rm e}}+
\frac{k}{81\,k_0}\left(90\,i+\frac{53\,\Delta A\, k_0^3}{A_-\, k^3}\r),\\
{\cal I}_2^b(k) 
&=& -1-\frac{k}{81\,k_0}
\left(90\,i+\frac{53\,\Delta A\,k_0^3}{A_-\,k^3}\r)
-\frac{53\,i\,\Delta A\, k_0}{27\,A_-\, k}
\nonumber \\ & &
+\frac{k^2}{3\,k_0^2}\,\l(1-\frac{\Delta A}{A_-}\r)
+\,
\frac{13\, i\,\Delta A\,k}{9\,A_-\,k_0}+\frac{22\,\Delta A}{9\,A_-},\\
{\cal J}_2^a(k) 
&=& -\frac{1}{k_0\,\eta_{\rm e}}-
\frac{k}{k_0}\, \l(2\,i+\frac{27\,\Delta A\,k_0^3}{A_-\,k^3}\r),\\
{\cal J}_2^b(k) 
&=& -1+\frac{k}{k_0}
\l(2\,i+\frac{27\,\Delta A\,k_0^3}{A_-\,k^3}\r)
-\frac{k^2}{k^2_0}
+\frac{27\,i\,\Delta A\,k_0}{A_-\,k}
\nonumber \\ & &
-\frac{14\,\Delta A}{A_-}
-\,\frac{5\,i\,\Delta A\,k}{A_-\,k_0}
+\frac{\Delta A\,k^2}{A_-\, k_0^{2}}.
\end{eqnarray}

\par

Note that the coefficients ${\cal I}_2^a$ and ${\cal J}_2^a$
actually diverge in the limit $\eta_{\rm e}\rightarrow 0$.  This
implies that $I_2$ and $J_2$ will diverge too, but, as we shall soon
demonstrate, the corresponding contribution to $\cG_2^{-}$ remains
perfectly finite.  It is also worth remarking that the coefficients 
${\cal I}_2^a$, ${\cal I}_2^b$, ${\cal J}_2^a$ and ${\cal J}_2^b$
depend on the wavenumber only through the ratio $k/k_0$, as expected.  
Using the above expressions, one obtains that
\begin{eqnarray}
\cG_2^-(k) 
&=& \frac{A_-\,k^2\,k_0}{4\,\sqrt{2\, k^9}\,\Mp^4\,\,H_0}\;
\Biggl[{\cal I}_2^a(k)\,\alpha_k^{\ast}{}^3
-3\,{\cal J}_2^a(k)\,\alpha_k^{\ast}{}^2\,
\tilde{\beta}_k^{\ast}\,{\rm e}^{-2\,i\,k/k_0}
\nonumber \\ &+ &
{\cal K}_2(k)\,{\rm e}^{-3\,i\,k/k_0}
+3{\cal J}_2^a{}^{\ast}(k)\alpha_k^{\ast}\,
\tilde{\beta}_k^{\ast}{}^2\,{\rm e}^{-4\,i\,k/k_0}
-{\cal I}_2^a{}^{\ast}(k)
\tilde{\beta}_k^{\ast}{}^3{\rm e}^{-6\,i\,k/k_0}\Biggr],
\nonumber \\
\end{eqnarray}
where the new coefficient ${\cal K}_2$ is given by
\begin{equation}
{\cal K}_2(k)
\equiv 
{\cal I}_2^b(k)\,\alpha_k^{\ast}{}^3
-{\cal I}_2^b{}^{\ast}(k)\,\tilde{\beta}_k^{\ast}{}^3
+3\,\alpha_k^{\ast}\,\tilde{\beta}_k^{\ast}\,
\l[{\cal J}_2^b{}^{\ast}(k)\,\tilde{\beta}_k^{\ast}
- {\cal J}_2^b(k)\,\alpha_k^{\ast}\r].
\end{equation}
We can now evaluate the resulting $G_{2}{}^-$.
The remarkable fact is that, all the terms proportional to
$1/\eta_{\rm e}$ cancel out and, therefore, the final result for
$G_{2}{}^-$ turns out to be finite in the limit $\eta_{\rm e}
\rightarrow 0$.  We find that $G_{2}{}^-$ can be expressed as
\begin{eqnarray}
k^6\, G_{2}^-(k) 
&=& \frac{-i}{32\, \sqrt{2\,\epsilon_{1-}^{3}\l(\eta_{\rm e}\r)}}\,
\frac{k_0}{k}\frac{A_-\, H_0^2}{\Mp^5}
\Biggl(\l[{\cal I}_2^a(k)-{\cal I}_2^a{}^{\ast}(k)\r]\,
\left(\alpha_k^3\,\alpha_k^{\ast}{}^3
-\tilde{\beta}_k^3\,\tilde{\beta}_k^{\ast}{}^3\right)
\nonumber \\ & &
+\, 9\,\l[{\cal J}_2^a(k)
-{\cal J}_2^a{}^{\ast}(k)\right]\alpha_k\,\alpha_k^{\ast}\,
\tilde{\beta}_k\,\tilde{\beta}_k^{\ast}\,
\l(\alpha_k\,\alpha_k^{\ast}
-\tilde{\beta}_k\,\tilde{\beta}_k^{\ast}\r)
\nonumber \\ & &
+\, 6\,i\,\Re 
\l[{\cal K}_2(k)\,\alpha_k\,\tilde{\beta}_k^2
+{\cal K}_2^{\ast}(k)\,\alpha_k^{\ast}{}^2\,\tilde{\beta}_k^{\ast}\r]\,
\sin \l(\frac{k}{k_0}\r)
\nonumber \\ & &
+\, 6\, i\, \Im
\l[{\cal K}_2(k)\,\alpha_k\,\tilde{\beta}_k^2
+{\cal K}_2^{\ast}(k)\,\alpha_k^{\ast}{}^2\,\tilde{\beta}_k^{\ast}\r]\,
\cos \l(\frac{k}{k_0}\r)
\nonumber \\ & &
-6i\Re 
\l\{\l[{\cal I}_2^a{}^{\ast}(k)-{\cal J}_2^a(k)\r]
\alpha_k\tilde{\beta}_k^{\ast}
\l(\alpha_k^2\alpha_k^{\ast}{}^2
-\tilde{\beta}_k^2\tilde{\beta}_k^{\ast}{}^2\r)\r\}
\sin\l(\frac{2k}{k_0}\r)
\nonumber \\ & &
+6i \Im 
\l\{\l[{\cal I}_2^a{}^{\ast}(k)-{\cal J}_2^a(k)\r]
\alpha_k\tilde{\beta}_k^{\ast}
\l(\alpha_k^2\alpha_k^{\ast}{}^2
-\tilde{\beta}_k^2\tilde{\beta}_k^{\ast}{}^2\r)\r\}
\cos\l(\frac{2k}{k_0}\r)
\nonumber \\ & &
-\, 2\,i\,\Re \l[{\cal K}_2(k)\,\alpha_k^3
+{\cal K}_2^*(k)\,\tilde{\beta}_k^{\ast}{}^3\r]\,
\sin\left(\frac{3\,k}{k_0}\right)
\nonumber \\ & &
+\,2\,i\,\Im \l[{\cal K}_2(k)\,\alpha_k^3
+{\cal K}_2^*(k)\,\tilde{\beta}_k^{\ast}{}^3\r]\,
\cos\l(\frac{3\,k}{k_0}\r)
\nonumber \\ & &
+6i\Re \l\{\l[{\cal I}_2^a(k)-{\cal J}_2^a(k)\r]
\alpha_k^{\ast}{}^2\tilde{\beta}_k^2
\left(\alpha_k\alpha_k^{\ast}
-\tilde{\beta}_k\tilde{\beta}_k^{\ast}\r)\r\}
\sin\left(\frac{4k}{k_0}\right)
\nonumber \\ & &
+6i\Im
\l\{\l[{\cal I}_2^a(k)-{\cal J}_2^a(k)\right]
\alpha_k^{\ast}{}^2\tilde{\beta}_k^2
\l(\alpha_k\alpha_k^{\ast}
-\tilde{\beta}_k\tilde{\beta}_k^{\ast}\right)\r\}
\cos\l(\frac{4k}{k_0}\r)\Biggr).
\nonumber \\
\end{eqnarray}

\par

\begin{figure}
\begin{center}
\includegraphics[width=15.0cm]{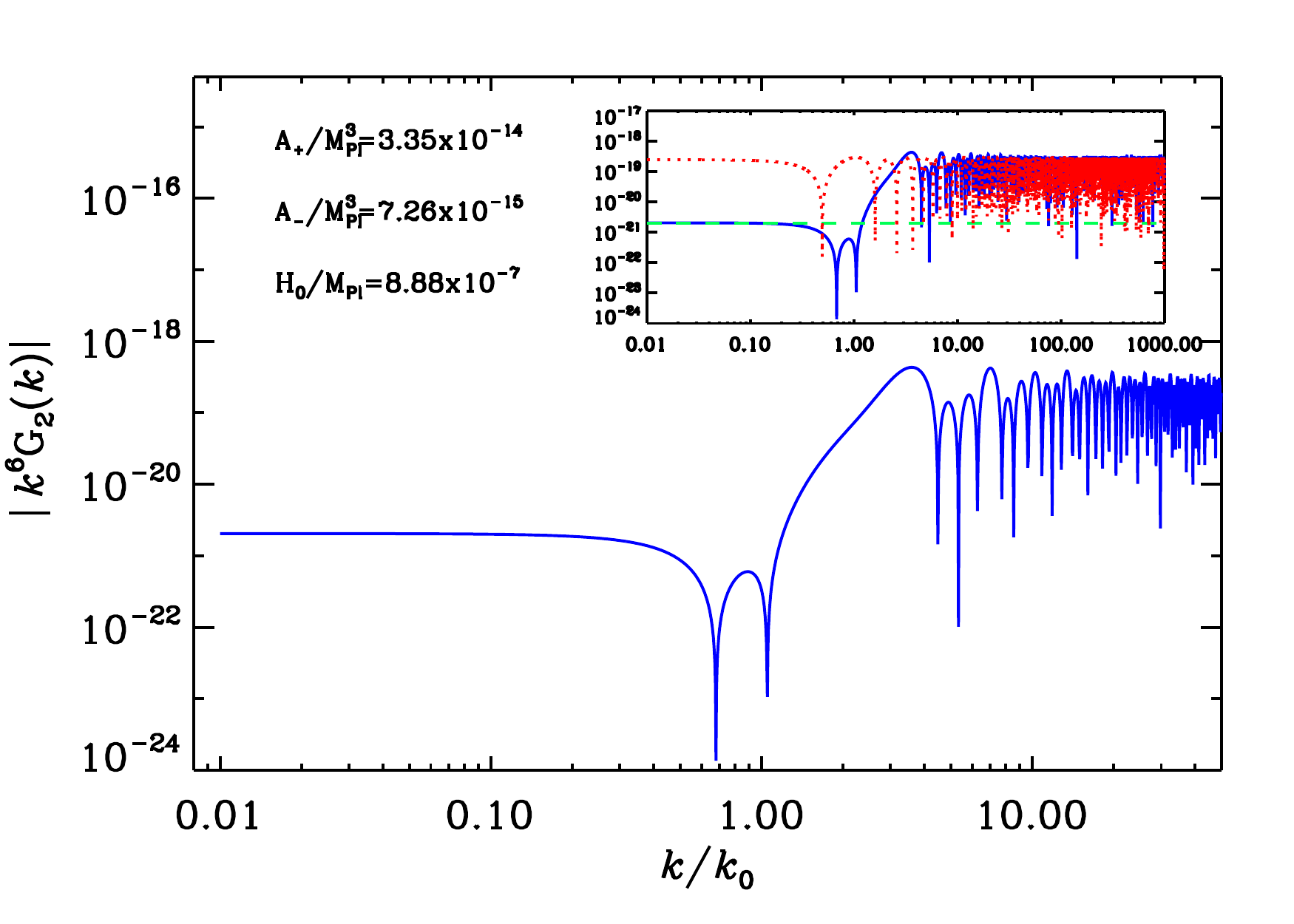}
\end{center}
\caption{The absolute value of the quantity $k^6\, G_2$ has been
  plotted as a function of $k/k_0$ (the blue curve).  In plotting
  this figure, we have worked with the same values of the parameters
  that we had considered in the previous figures.  As in the last
  plot, the green and the red curves in the inset represent the
  limiting forms for $k\ll k_0$ and $k\gg k_0$, respectively [see
  Eqs.~(\ref{eq:g2zero}) and~(\ref{eq:g2infinity})].}\label{fig:abs-G2}
\end{figure}

Let us now study the asymptotic behavior of $G_2^-$.  
In the limit $k/k_0\rightarrow 0$, we find that $k^6\, G_2^{-}$ 
goes to a constant given by
\begin{equation}
\lim _{k/k_0\to 0} k^6\, G_{2}^{-}(k)
= \frac{3}{16\,\sqrt{2\,\epsilon_{1-}^{3}(\eta _{\rm e})}}\,  
\frac{A_-^5\, H_0^{2}}{\Mp^5\,A_+^4}.
\end{equation}
Whereas, as $k/k_0\to \infty$, it has the following behavior
\begin{eqnarray}
\lim _{k/k_0\to \infty}k^6\, G_{2}^{-}(k)
&=& \frac{1}{16\sqrt{2\epsilon_{1-}^{3}(\eta_{\rm e})}}\, 
\frac{A_+H_0^2}{\Mp^5}
\Biggl[\frac{10A_-}{9A_+}
-\frac{3}{2}\l(1-\frac{A_-}{A_+}\r)\cos\left(\frac{k}{k_0}\right)
\nonumber \\ & &
-\frac{k}{3k_0}\sin\l(\frac{3k}{k_0}\r)
-\l(\frac{107}{18}-\frac{29A_-}{6A_+}\r)
\cos\left(\frac{3k}{k_0}\right)\Biggr].
\end{eqnarray}
As we had mentioned earlier, $k^6\, G_2^{-}$ too diverges at large
$k$. Upon comparing with Eq.~(\ref{eq:G2p-lk}), it is clear that the
divergence involving the $\sin\, (3\,k/k_0)$ term in the above
expression is of the same form, but with an opposite sign.  Therefore,
the total contribution due to the second term,
viz. $G_{2}(k)=G_{2}^{+}(k)+G_{2}^{-}(k)$ remains finite even at large
$k$. We find that, as $k/k_0\to 0$, $k^6\, G_{2}$ turns strictly
scale invariant with the amplitude
\begin{equation}
\label{eq:g2zero}
\lim_{k/k_0\to 0}k^6\, G_{2}(k)
=\frac{1}{16\, \sqrt{2\,\epsilon_{1-}^{3}(\eta_{\rm e})}}\;
\frac{A_-^{3}\, H_0^2}{\Mp^5\, A_{+}^{2}}\left(-\frac{17}{9}
+\frac{3\,A_-^2}{A_+^2}\right).
\end{equation}
In the limit of large $k/k_0$, the diverging terms cancel exactly 
and we are left with the following finite expression:
\begin{eqnarray}
\label{eq:g2infinity}
\lim_{k/k_0\to \infty}\, k^6\, G_{2}(k)
&=&\frac{1}{16\, \sqrt{2\,\epsilon_{1-}^{3}(\eta_{\rm e})}}\;
\frac{A_+\,H_0^2}{\Mp^5}\;
\Biggl[\frac{10\,A_-}{9\,A_+}
\nonumber  \\ & &
-\f{10}{3}\,
\l(1-\frac{A_-}{A_+}\right)\cos\l(\frac{3\,k}{k_0}\r)\Biggr],
\end{eqnarray}
which is scale invariant with superimposed oscillations.  
Notice that the superimposed oscillations again involve a cosine 
function with the argument $3\,k/k_0$. 
These asymptotic behavior are also evident from 
Fig.~\ref{fig:abs-G2}, wherein we have plotted the absolute value 
of the quantity $k^6\, G_{2}$.


\subsection{The contributions due to the first and the third terms}

Barring the ${\bf k}$ dependent factors, the contributions $\cG_1$ and
$\cG_3$ due to the first and the third terms in the interaction
Hamiltonian~(\ref{eq:Hint}) involve similar integrals
[cf.~(\ref{eq:cG1}) and~(\ref{eq:cG3})]. 
One finds that, their contributions can be combined and evaluated 
together in the equilateral limit.
As in the case of the second term, we shall first evaluate the contribution 
before the transition and then turn our attention to the calculation of the 
contribution after the transition.
 

\subsubsection{Before the transition}

To begin with, note that, before the transition, the mode $f_{k}$ is
given by Eq.~(\ref{eq:fk-bt}), while its derivative can be computed to
be
\begin{equation}
f_k^{+}{}'(\eta)
=\frac{i\, H_0}{2\, \Mp\, \sqrt{{k^3}\,\epsilon_{1+}}}
\l[-{\cal H}\,\l(\epsilon_{1+}+\f{\epsilon_{2+}}{2}\r)\,
\l(1+i\,k\,\eta\r)
+k^2\,\eta\r] {\rm e}^{-i\,k\,\eta}.
\end{equation}
As is usually done while working in the slow roll approximation,
we shall ignore the first term within the square brackets 
involving~$\epsilon_{1+}$ and~$\epsilon_{2+}$.
Upon using the mode~(\ref{eq:fk-bt}) and its derivative above in 
the integrals~(\ref{eq:cG1}) and~(\ref{eq:cG3}), and upon modifying 
the lower limit of the integrals as in Eq.~(\ref{eq:cG2p}), we find 
that the integrals can be evaluated easily without actually having to 
take the equilateral limit.
We obtain that 
\begin{eqnarray}
\cG_1^{+}(\vka,\vkb,\vkc)
&=&-\frac{H_0}{4\,\Mp^3}\,
\sqrt{\frac{\epsilon_{1+}}{(k_1\,k_2\,k_3)^{3}}}\;
{\rm e}^{-i\,k_{_{\rm T}}/k_0}
\Biggl[\frac{k_2^{2}k_3^{2}}{ik_{_{\rm T}}}\,
\left(1+\frac{k_1}{k_{_{\rm T}}}+\frac{i\,k_1}{k_0}\r)
\nonumber \\ & &
+ {\rm two~permutations}\Biggr]
\end{eqnarray}
and
\begin{eqnarray}
\cG_3^{+}(\vka,\vkb,\vkc)
&=&\frac{H_0}{4\,\Mp^3}
\sqrt{\frac{\epsilon_{1+}}{(k_1 k_2 k_3)^{3}}}
{\rm e}^{-i\,k_{_{\rm T}}/k_0}
\Biggl[\frac{\vka\cdot \vkb}{k_2^2}
\frac{k_2^{2}\,k_3^{2}}{i\,k_{_{\rm T}}}
\l(1+\frac{k_1}{k_{_{\rm T}}}
+\frac{i k_1}{k_0}\r)
\nonumber \\ & &
+ {\rm five~permutations}\Biggr],
\end{eqnarray}
which reduce to the standard slow roll results as 
$k_0\to \infty$~\cite{maldacena-2003,ng-ncsf}.
In the equilateral limit, the above expressions simplify to
\begin{eqnarray}
\cG_1^{+}(k)
&=\frac{i\,H_0\,k^3}{4\,\Mp^3}\,
\sqrt{\frac{\epsilon_{1+}}{k^{9}}}\,
\left(\frac{4}{3}+\frac{i\,k}{k_0}\r)\,{\rm e}^{-3\,i\,k/k_0},\\
\cG_3^{+}(k)
&=-\frac{2\,i\,H_0\,k^3}{4\,\Mp^3}\,
\sqrt{\frac{\epsilon_{1+}}{k^{9}}}\,
\left(\frac{4}{3}+\frac{i\,k}{k_0}\right){\rm e}^{-3\,i\,k/k_0},
\end{eqnarray}
so that we have
\begin{equation}
\cG_1^{+}(k)+\cG_3^{+}(k)
=-\frac{i\,H_0\, k^3}{4\,\Mp^3}\,
\sqrt{\frac{\epsilon_{1+}}{k^{9}}}\,
\l(\frac{4}{3}+\frac{i\,k}{k_0}\r)\,
{\rm e}^{-3\,i\,k/k_0}.
\end{equation}
The corresponding $G_{1}^{+}+G_{3}^{+}$ is given by
Eq.~(\ref{eq:defG}) which involves the mode function 
at late times [cf. Eq.~(\ref{eq:fk-lt})],
and thereby the Bogoliubov coefficients. 
Upon using the expressions~(\ref{eq:alphak-sm})~and~(\ref{eq:betak-sm})
for $\alpha_k$ and $\beta_k$, we find that we can write
\begin{eqnarray}
k^6\,\l[G_{1}^{+}(k)+G_{3}^{+}(k)\r]
&=&-\frac{1}{48\sqrt{2\,\epsilon_{1-}^{3}(\eta_{\rm e})}}\,
\frac{A_+\, H_0^2}{\Mp^5}\nonumber\\ 
& &\times
\Biggl\{-3 \Im \left[\left(\frac{4}{3}+\frac{i k}{k_0}\right) 
\l(\alpha_{k}^2\, {\tilde \beta}_{k}
+\alpha_{k}\, {\tilde \beta}_{k}^2\r)\r] 
\sin \l(\frac{k}{k_0}\r)\nn\\
& &-\,3\,\Re \l[\l(\frac{4}{3}+\frac{i\,k}{k_0}\r)\, 
\l(\alpha_{k}^2\, {\tilde \beta}_{k}
-\alpha_{k}\, {\tilde \beta}_{k}^2\r)\r]\, 
\cos \l(\frac{k}{k_0}\r)\nn\\
& &
+\,\Im \left[\left(\frac{4}{3}+\frac{i\,k}{k_0}\right)\, 
\left(\alpha_{k}^3+{\tilde \beta}_{k}^3\right)\right]\, 
\sin \l(\frac{3\,k}{k_0}\r)\nn\\
& &
+\,\Re \l[\l(\frac{4}{3}+\frac{i\, k}{k_0}\r) 
\l(\alpha_{k}^3-{\tilde \beta}_{k}^3\right)\r]\,
\cos \l(\frac{3\,k}{k_0}\r)\Biggr\}.
\end{eqnarray}

\par 

We again encounter the by-now usual structure containing sine and 
cosine functions with arguments $k/k_0$ and $3\,k/k_0$. 
As we had carried out earlier, it is interesting to study the 
asymptotic behavior of the above expression at small and large 
scales.
In the limit of small $k/k_0$, one has
\begin{equation}
\lim _{k/k_0\to 0}\,
k^6\,\l[G_{1}^{+}(k)+G_{3}^{+}(k)\r]
=-\frac{1}{36\sqrt{2\epsilon_{1-}^{3}(\eta _{\rm e})}}\, 
\frac{A_-^3H_0^2}{\Mp^5A_{+}^{2}},
\end{equation}
while in the limit of large $k/k_0$, one obtains
\begin{eqnarray}
\lim _{k/k_0\to \infty}
k^6\l[G_{1}^{+}(k)+G_{3}^{+}(k)\r]
&=&-\frac{1}{48\, \sqrt{2\,\epsilon_{1-}^{3}(\eta _{\rm e})}}\,
\frac{A_+\, H_0^2}{\Mp^5}
\Biggl[\f{9}{2}\,\left(1-\frac{A_-}{A_+}\right)\,
\nonumber \\ & & \times
\cos \left(\frac{k}{k_0}\right)
+\frac{k}{k_0}\,
\sin \left(\frac{3\,k}{k_0}\right)
\nonumber \\ & &
+\left(\frac{35}{6}-\frac{9\,A_-}{2\,A_+}\r)\,
\cos \left(\frac{3\,k}{k_0}\right)\Biggr].\label{eq:G13p-lk}
\end{eqnarray}
It should be mentioned here that, as in the case of $k^6\, G_2^{+}$,
the above asymptotic form too diverges linearly at large $k$.  Again,
as in the earlier case, the contribution post-transition will cancel
this divergent term leading to an overall finite $k^6\,
\left(G_1+G_3\right)$.


\subsubsection{After the transition}

We shall now evaluate the contribution due to first and the third
terms arising from the post-transition phase. 
As the calculation proves to be somewhat heavy and lengthy (but 
manageable), we shall relegate some of the details of the 
calculation to an appendix. 
The starting point is similar to what we have already encountered 
in the case of the first sub-dominant term:~upon substituting 
the mode~(\ref{eq:fk-at}) and its derivative~(\ref{eq:fkp-at}) in 
the integrals~(\ref{eq:cG1}) and~(\ref{eq:cG2}), we find that, we 
can write
\begin{eqnarray}
\cG_1^{-}(k)+\cG_3^{-}(k)
&=& \frac{3\,H_0}{4\,\Mp^3\,k^{9/2}} 
\biggl[{\alpha_{k}^{\ast}}^{3}\,I_{13}(k) 
-{\beta_{k}^{\ast}}^{3}\,I_{13}^{\ast}(k)
-{\alpha_{k}^{\ast}}^{2}\, \beta_{k}^{\ast}\,J_{13}(k)
\nonumber \\ & &
+\alpha_{k}^{\ast}\,{\beta_{k}^{\ast}}^{2}\,
J_{13}^{\ast}(k)\biggr],\;\label{eq:cG13m-ov}
\end{eqnarray}
where $I_{13}$ and $J_{13}$ are described by the integrals
\begin{eqnarray}
I_{13}(k) 
&=& \frac{A_-}{\sqrt{18}\,H_0^2\,\Mp}
\int _{-k_0^{-1}}^{\eta_{\rm e}}\,\frac{{\rm d}\tau}{1-\rho^3\,\tau^3}\,
\l(1-ik\tau\r)
\biggl[3\,\rho^3\,\tau\,\l(1-i\,k\,\tau\r)
\nonumber \\ & & 
+k^2\,\l(1-\rho^3\,\tau^3\r)\biggr]^2\,{\rm e}^{3\,i\,k\,\tau},
\label{eq:I13}\\
J_{13}(k) &=& \frac{A_-}{\sqrt{18}\,H_0^2\,\Mp}
\int _{-k_0^{-1}}^{\eta_{\rm e}}\,\frac{{\rm d}\tau}{1-\rho^3\,\tau^3}\,
\l[3\,\rho^3\,\tau\, \l(1-i\,k\,\tau\r)+k^2\,
\l(1-\rho^3\,\tau^3\r)\r]\nonumber\\ 
& & \times
\l[9\,\l(1-i\,k\,\tau\r)\,\l(1+i\,k\,\tau\r)\,\rho^3\,\tau
+k^2 \l(1-\rho^3\,\tau^3\r)\,\l(3-i\,k\,\tau\r)\r]
{\rm e}^{i\,k\,\tau}.\nonumber \\
\label{eq:J13}
\end{eqnarray}
In arriving at the above expressions, we have again ignored the term
involving $\epsilon_{1-}$ within the square brackets in
Eq.~(\ref{eq:fkp-at}), just as we had done while evaluating the
dominant contribution~$G_4$. We had earlier mentioned that, in the
case of~$G_4$, the neglected terms can be expected to be of the same
order as the contributions due to $G_{1}$, $G_{2}$, $G_{3}$ and $G_7$.
The terms that we shall ignore here are possibly of the same order as
the contributions due to $G_{5}$ and $G_{6}$.

\par 

\begin{figure}
\begin{center}
\includegraphics[width=14cm]{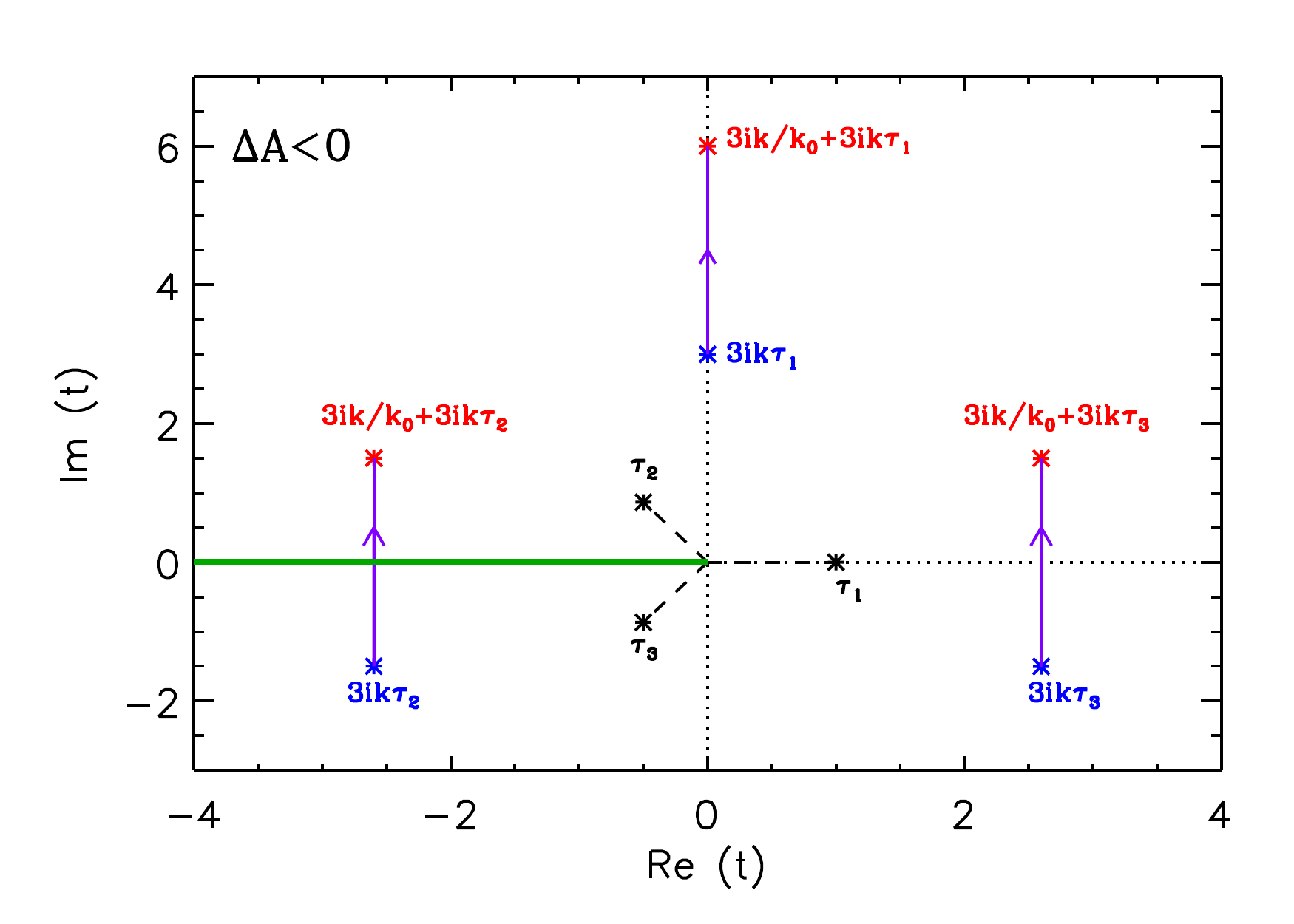}
\caption{The poles corresponding to the 
$\l(1-\rho^3\,\tau^3\r)^{-1}$ term that appears in the integrals
$I_{13}$ and $J_{13}$ [cf. Eqs.~(\ref{eq:I13}) and (\ref{eq:J13})].
We should mention here that, we had encountered the same term in the 
integrals $I_{4}$ and $J_{4}$ as well and, in Fig.~\ref{fig:poleplus}, 
we had illustrated the positions of the poles when $\Delta A>0$.
The location of the three poles $\tau_{1}$, $\tau_{2}$ and $\tau_{3}$ 
in the complex~$\tau$-plane, for a case wherein $\Delta A<0$, have been 
marked in the above figure with black asterisks [in this context, see
Eq.~(\ref{eq:taum})].
The solid blue line segments denote the three new integration paths 
corresponding to $\tau_1$, $\tau_2$ and $\tau_3$ in the complex $t_n$ 
plane, with $t_n$ being related to $\tau_n$ through Eq.~(\ref{eq:tn1}).
The solid green line running along the negative real axis represents 
the branch cut associated with the function~$E_1(z)$.}
\label{fig:poleminus}
\end{center}
\end{figure}

Let us now analyze these integrals in some detail. Firstly, a concern
could arise that these integrals may contain a pole---due to the
$\l(1-\rho^3\,\tau^3\r)^{-1}$ term---along the path of integration,
which runs along the real axis from $-k_0^{-1}$ to $\eta_{\rm e}$,
with the latter approaching zero from the negative direction. 
Actually, the integrals $I_4$ and $J_4$ [cf. Eqs.~(\ref{eq:I4})
and~(\ref{eq:J4})] that we had to carry out earlier in the 
calculation of $G_4$ had also contained the same term, and we had 
discussed the issue of the poles in some detail then.
In the case $\Delta A>0$, the three poles corresponding to 
$\l(1-\rho^3\,\tau^3\r)^{-1}$ are actually located at
\begin{equation}
\tau_1=\frac{{\rm e}^{i\,\pi/3}}{\vert \rho\vert}, \quad 
\tau_2=\frac{{\rm e}^{i\,\pi}}{\vert \rho\vert}, \quad
\tau_3=\frac{{\rm e}^{5\,i\,\pi/3}}{\vert \rho\vert},
\label{eq:taup}
\end{equation}
in the complex $\tau$ plane.
In fact, we had illustrated the positions of these poles for a 
situation wherein $\Delta A>0$ in Fig.~\ref{fig:poleplus}.
Whereas, when $\Delta A<0$, the poles are found to be at
\begin{equation}
\tau_1=\frac{1}{\vert \rho\vert}, \quad 
\tau_2=\frac{{\rm e}^{2\,i\,\pi/3}}{\vert \rho\vert}, \quad
\tau_3=\frac{{\rm e}^{4\,i\,\pi/3}}{\vert \rho\vert},\label{eq:taum}
\end{equation}
and their location is illustrated in Fig.~\ref{fig:poleminus}. 
It should be clear from these two figures that no poles arise on the 
path of the integration.

\par

Secondly, in order to compute the integrals, it is convenient to write 
the term containing the poles as follows:
\begin{eqnarray}
\frac{1}{1-\rho^3\,\tau^3}
&=-\frac{1}{\rho^3}\;\frac{1}{\tau^3-1/\rho^3}
=-\frac{1}{\rho^3}\;
\frac{1}{\l(\tau-\tau_1\r)\,\l(\tau-\tau_2\r)\,\l(\tau-\tau_3\r)}\nn\\
&=-\frac{\vert \rho\vert^2}{\rho^3}\;
\l(\frac{b_1}{\tau-\tau_1}+\frac{b_2}{\tau-\tau_2}
+\frac{b_3}{\tau-\tau_3}\right),
\end{eqnarray}
where $b_1$, $b_2$ and $b_3$ are given by
\begin{eqnarray}
b_1 
&= \frac{1}{\vert \rho\vert^2\,\l(\tau_1-\tau_2\r)\,\l(\tau_1-\tau_3\r)}
=\frac{1}{3}, \\
b_2 
&= -\frac{1}{\vert \rho\vert^2\,\l(\tau_1-\tau_2\r)\,\l(\tau_2-\tau_3\r)}
=-\frac{1}{6}\,\left(1-i\,\sqrt{3}\right), \\
b_3 
&= \frac{1}{\vert \rho \vert^2\,\l(\tau_1-\tau_3\r)\,\l(\tau_2-\tau_3\r)}
=-\frac{1}{6}\,\left(1+i\,\sqrt{3}\right)=b_2^{\ast}.
\end{eqnarray}
These expressions are valid when $\Delta A<0$.  If $\Delta A>0$, then
one has $b_1=-[\l(1+i\,\sqrt{3}\r)/6]$, $b_2=(1/3)$ and
$b_3=-[\l(1-i\,\sqrt{3}\r)/6]$, i.e. simply a permutation of the
previous values.

\par

Lastly, the numerators of the integrals $I_{13}$ and $J_{13}$ are
polynomials in $\tau$ of order seven. As a consequence, one can write
\begin{eqnarray}
I_{13}(k)
&=-\frac{A_-\,\vert \rho\vert^2}{\sqrt{18}\,H_0^2\,\Mp\, \rho^3}\;
\sum _{n=1}^3\, b_n\,\int_{-k_0^{-1}}^{\eta_{\rm e}}
{\rm d}\tau\, \frac{P_1(\tau)}{\tau-\tau_n}\;
{\rm e}^{3\,i\,k\,\tau},
\end{eqnarray}
where $P_1(\tau)$ is the seventh order polynomial given by
\begin{equation}
P_1(\tau)\equiv \l(1-i\,k\,\tau\r)\,
\l[3\,\rho^3\,\tau\,\l(1-i\,k\,\tau\r)
+k^2\,\l(1-\rho^3\,\tau^3\r)\r]^2.
\end{equation}
Let us now perform the following change of variable 
from~$\tau$ to $t_n$:
\begin{equation}
t_n\equiv 
-3\,i\,k\,\l(\tau-\tau_n\r).\label{eq:tn1}
\end{equation}
It should stressed here that, because the change of the variable 
depends on $\tau_n$, it is different for the three integrals 
in the above expression for~$I_{13}$.
The change of the variable leads to
\begin{eqnarray}
I_{13}(k)
&=&-\frac{A_-\,\vert \rho\vert^2}{\sqrt{18}\,H_0^2\,\Mp\, \rho^3}\,
\sum _{n=1}^3\,
b_n\,{\rm e}^{3\,i\,k\tau_n}\,
\nonumber \\ & & \times
\int_{3\,i\,k/k_0+3\,i\,k\,\tau_n}^{-3\,i\,k\,\eta_{\rm e}+3\,i\,k\,\tau_n}
\frac{{\rm d}t_n}{t_n}\, {\rm e}^{-t_n}\,
P_1\l(\frac{-t_n}{3\,i\,k}+\tau_n\r).
\end{eqnarray}
We find that this integral can be evaluated by initially expressing
$P_1$ as a polynomial in $t_n$, and then carrying out the integral
term by term. Since the computation proves to be somewhat longwinded,
as we had mentioned before, we shall relegate the details of the
calculation to an appendix (see~\ref{app:13}), and simply quote the
final result here.  We obtain that
\begin{eqnarray}
I_{13}(k)
=-\frac{A_-\,\vert \rho\vert^2\,k^4}{\sqrt{18}\,H_0^2\,\Mp\, \rho^3}\,
\l[{\cal I}_{13}^a(k)+{\cal I}_{13}^b(k)\, {\rm e}^{-3\,i\,k/k_0}\r],
\label{eq:I13-fv}
\end{eqnarray}
where the scale dependent quantities ${\cal I}_{13}^a$ and
${\cal I}_{13}^b$ can be expressed as
\begin{eqnarray}
{\cal I}_{13}^a(k) 
&=& -\sum _{n=1}^3b_n
\biggl[c_{n0}\, {\cal E}_1\l(3ik\tau_n\right)
+\sum _{m=1}^7\,c_{nm}\;(m-1)!\;
e_{m-1}\l(3ik\tau_n\r)\biggr],\\
{\cal I}_{13}^b(k) 
&=& \sum _{n=1}^3\,b_n\,
\biggl[c_{n0}\,
{\cal E}_1\l(\frac{3\,i\,k}{k_0}+3\,i\,k\,\tau_n\right)
\nonumber \\ & &
+\,\sum _{m=1}^7\,c_{nm}\;(m-1)!\;
e_{m-1}\l(\frac{3\,i\,k}{k_0}+3\,i\,k\,\tau_n\r)\biggr],
\end{eqnarray}
with the coefficients $c_{nm}$ being given by
Eqs.~(\ref{eq:cn0})--(\ref{eq:cn7}).  In these expressions, 
$e_n(z)$ denotes the exponential sum function~(\ref{eq:esf}), 
while ${\cal E}_1(z)\equiv {\rm e}^z\,E_1(z)$, with $E_1(z)$ 
being the exponential integral function~(\ref{eq:eif}).  
Despite the fact that the calculation is more involved, it is 
clear that the structure of $I_{13}$ resembles that of $I_2$.

\par

However, there is one particular point that we have overlooked in the
above considerations.  When one performs the change of variable from
$\tau$ to $t_n=-3\,i\,k\, \l(\tau-\tau_n\r)$, one basically modifies
the contour along which the integrations are performed.  Since the
change in the variable from $\tau$ to $t_n$ involves $\tau_n$, the
modified path evidently depends on $\tau_n$.  The effect of
multiplying by $3\,i\,k$ is to rotate the path by an angle of $\pi/2$
in the anti-clockwise direction.  Therefore, the contours start from
$3\,i\,k\,\tau_n$ (corresponding to $\tau=0$) and run vertically until
$3\,i\,k/k_0$ has been added to $3\,i\,k\,\tau_n$.  The three new
integration paths corresponding to $\tau_1$, $\tau_2$ and $\tau_3$ are
displayed (as the solid blue line segments) in the complex $t_n$ plane
in Fig.~\ref{fig:poleminus}.  Note that the function $E_1(z)$ has a
branch cut running from $-\infty$ to $0$ (represented as the solid
green line in the figure).  While the paths associated with $t_1$ and
$t_3$ are further away, there seems to be the possibility that the
path corresponding to $t_2$ may cross the branch cut. It is clear from
the figure that this will occur if the imaginary part of $t_2$
corresponding to $\tau =-k_0^{-1}$ is positive, i.e. when
\begin{equation}
\Im \left[t_2\l(\tau=-k_0^{-1}\r)\right]
=\frac{3\,k}{k_0}\,
\left[1-\frac12\left(\frac{A_-}{\vert \Delta A\vert}\right)^{1/3}\right]
>0.
\end{equation}
For $\Delta A<0$, this condition implies that $A_+/A_- > 9/8=1.125$.
If this condition is satisfied, then one crosses the branch cut of the
function $E_1(z)$ and, in such a case, one must add a suitable
contribution to the integral~$I_{13}$.  We find that the additional 
contribution amounts to redefining the above ${\cal I}_{13}^a$ by
\begin{eqnarray}
{\cal I}_{13}^a(k) 
&=& -\sum _{n=1}^3\,b_n\,
\Biggl[c_{n0}\, {\cal E}_1\l(3\,i\,k\,\tau_n\r)
+\sum _{m=1}^7\,c_{nm}\,(m-1)!\,
e_{m-1}\left(3\,i\,k\,\tau_n\right)\Biggr]
\nonumber \\ & &
+\,(2\,i\,\pi)\,b_2\,c_{20}\,{\rm e}^{3\,i\,k\tau_2}.
\end{eqnarray}
It is important to point out here that whether the additional term
arises or not depends only on the ratio $A_+/A_-$ and not on the
wavenumber~$k$.

\par

We now need to calculate the integral~$J_{13}$, which, as in the case 
of $I_{13}$, can be expressed as
\begin{eqnarray}
J_{13}(k)
=-\frac{A_-\,\vert \rho\,\vert^2}{\sqrt{18}\,H_0^2\,\Mp\, \rho^3}
\sum _{n=1}^3\,b_n\,\int_{-k_0^{-1}}^{\eta_{\rm e}}\,
{\rm d}\tau\, \frac{P_2(\tau)}{\tau-\tau_n}\;
{\rm e}^{i\,k\,\tau},\label{eq:J13-ov}
\end{eqnarray}
where $P_2(\tau)$ is a seventh order polynomial given by
\begin{eqnarray}
P_2(\tau)
&\equiv \l[3\,\rho^3\,\tau\l(1-i\,k\,\tau\r)
+k^2\,\l(1-\rho^3\,\tau^3\r)\r]\nn\\
&\times\;\l[9\,\l(1-i\,k\,\tau\r)\,\l(1+i\,k\,\tau\r)\,\rho^3\,\tau
+k^2\,\l(1-\rho^3\,\tau^3\r)\,\l(3-i\,k\,\tau\r)\r].
\end{eqnarray}
To evaluate the above integral, we can now apply the same strategy
that we have followed above for $I_{13}$ (for details,
see~\ref{app:13}).  We obtain that
\begin{equation}
J_{13}(k)
=-\frac{A_-\,\vert \rho\vert^2\,k^4}{\sqrt{18}\,H_0^2\,\Mp\, \rho^3}
\l[{\cal J}_{13}^a(k)+{\cal J}_{13}^b(k)\,{\rm e}^{-i\,k/k_0}\r],
\label{eq:J13-fv}
\end{equation}
where ${\cal J}_{13}^{a}$ and ${\cal J}_{13}^{b}$ can be written as
\begin{eqnarray}
{\cal J}_{13}^a(k)
&=& -\sum _{n=1}^3\,b_n\,
\Biggl[d_{n0}\,{\cal E}_1\l(i k\tau_n\r)
+\sum_{m=1}^7\,d_{nm}\;(m-1)!\;
e_{m-1}\l(i k\tau_n\r)\Biggr],\\
{\cal J}_{13}^b(k) 
&=& \sum _{n=1}^3\,b_n\,
\Biggl[d_{n0}\,{\cal E}_1\l(\frac{i\,k}{k_0}+i\,k\, \tau_n\right)
\nonumber \\ & &
+\sum _{m=1}^7\,d_{nm}\;(m-1)!\;e_{m-1}\l(\frac{i\,k}{k_0}
+i\,k\,\tau_n\right)\Biggr],
\end{eqnarray}
while the coefficients $d_{nm}$ are given by
Eqs.~(\ref{eq:dn0})--(\ref{eq:dn7}). 
Just as the structure of $I_{13}$ had resembled~$I_2$, evidently, the 
form of $J_{13}$ resembles that of $J_2$. 
We again encounter here the issue
regarding the path associated with $t_2$ crossing the branch cut of
the exponential integral function $E_{1}(z)$.  It is straightforward
to show that the previous arguments apply for the present case as
well.  As a result, when $A_+/A_->9/8$, instead of ${\cal J}_{13}^a$
above, one has
\begin{eqnarray}
{\cal J}_{13}^a(k) 
&=& -\sum _{n=1}^3b_n\,\Biggl[d_{n0}\,{\cal E}_1\left(ik\tau_n\right)
+\sum _{m=1}^7\,d_{nm}\,(m-1)!\,e_{m-1} \left(ik\tau_n\right)\Biggr]\nn\\
& &
+(2\,i\,\pi)\,b_2\,d_{20}\,{\rm e}^{i\,k\,\tau_2}.
\end{eqnarray}

\par

The above expressions for $I_{13}$ and $J_{13}$ then determine
$\cG_{1}^{-}+\cG_{3}^{-}$ [cf. Eq.~(\ref{eq:cG13m-ov})], which in turn
can be used to arrive at the corresponding $G_{1}^{-}+G_{3}^{-}$.  We
find that the calculation proceeds in exactly the same fashion as in
the evaluation $G_{2}^{-}$, because, as we have already pointed out, 
the final expressions for the integrals $I_{13}$ and $J_{13}$ are very 
similar to that of $I_2$ and $J_2$.  As a consequence, one can write
\begin{eqnarray}
k^6\, \l[G_{1}^{-}(k)+G_{3}^{-}(k)\r]
&=& \frac{i}{32\sqrt{2\epsilon_{1-}^{3}\l(\eta_{\rm e}\r)}}\,
\frac{A_-\vert\rho\vert^{2}H_0^2k}{\Mp^5\rho^{3}}
\nonumber \\ & &
\times\,\Biggl(\l[{\cal I}_{13}^a(k)-{\cal I}_{13}^a{}^{\ast}(k)\r]
\left(\alpha_k^3\,\alpha_k^{\ast}{}^3
-\tilde{\beta}_k^3\,\tilde{\beta}_k^{\ast}{}^3\r)
\nonumber \\ & +&
3\,\l[{\cal J}_{13}^a(k)-{\cal J}_{13}^a{}^{\ast}(k)\r]\,
\alpha_k\,\alpha_k^{\ast}\,
\tilde{\beta}_k\,\tilde{\beta}_k^{\ast}\,
\l(\alpha_k\,\alpha_k^{\ast}-\tilde{\beta}_k\,\tilde{\beta}_k^{\ast}\r)
\nonumber \\ & +&
6\,i\,\Re \l[{\cal K}_{13}(k)\,\alpha_k\,\tilde{\beta}_k^2
+{\cal K}_{13}^{\ast}(k)\,\alpha_k^{\ast}{}^2\,
\tilde{\beta}_k^{\ast}\r]\,
\sin \l(\frac{k}{k_0}\r)
\nonumber \\ & +&
6i\Im \l[{\cal K}_{13}(k)\alpha_k\tilde{\beta}_k^2
+{\cal K}_{13}^{\ast}(k)\alpha_k^{\ast}{}^2\,
\tilde{\beta}_k^{\ast}\r]
\cos \left(\frac{k}{k_0}\right)
\nonumber \\ & -&
2\Re \biggl\{\l[3\,{\cal I}_{13}^a{}^{\ast}(k)-{\cal J}_{13}^a(k)\r]\,
\alpha_k\,\tilde{\beta}_k^{\ast}\,
\nonumber \\ & \times&
\left(\alpha_k^2\,\alpha_k^{\ast}{}^2
-\tilde{\beta}_k^2\,\tilde{\beta}_k^{\ast}{}^2\right)\biggr\}\,
\sin\left(\frac{2\,k}{k_0}\right)
\nonumber \\ & +&
2i\Im \biggl\{\l[3\,{\cal I}_{13}^a{}^{\ast}(k)
-{\cal J}_{13}^a(k)\r]\, \alpha_k\,\tilde{\beta}_k^{\ast}\,
\nonumber \\ & \times&
\left(\alpha_k^2\,\alpha_k^{\ast}{}^2
-\tilde{\beta}_k^2\,\tilde{\beta}_k^{\ast}{}^2\right)\biggr\}
\cos\l(\frac{2\,k}{k_0}\r)
\nonumber \\ & -&
2i\Re \l[{\cal K}_{13}(k)\,\alpha_k^3
+{\cal K}_{13}^*(k)\,\tilde{\beta}_k^{\ast}{}^3\r]\,
\sin\l(\frac{3\,k}{k_0}\right)
\nonumber \\ & +&
2i\Im \l[{\cal K}_{13}(k)\,\alpha_k^3
+{\cal K}_{13}^*(k)\,\tilde{\beta}_k^{\ast}{}^3\r]
\cos\left(\frac{3\,k}{k_0}\right)
\nonumber \\ &+ &
2i\Re \biggl\{\l[3\,{\cal I}_{13}^a(k)-{\cal J}_{13}^a(k)\r]\,
\alpha_k^{\ast}{}^2\,\tilde{\beta}_k^2\,
\nonumber \\ & \times &
\left(\alpha_k\,\alpha_k^{\ast}
-\tilde{\beta}_k\,\tilde{\beta}_k^{\ast}\r)\biggr\}
\sin\l(\frac{4\,k}{k_0}\r)
\nonumber \\ & +&
2i\Im \biggl\{\l[3\,{\cal I}_{13}^a(k)-{\cal J}_{13}^a(k)\r]\,
\alpha_k^{\ast}{}^2\tilde{\beta}_k^2\,
\nonumber \\ & \times &
\left(\alpha_k\,\alpha_k^{\ast}
-\tilde{\beta}_k\,\tilde{\beta}_k^{\ast}\right)\biggr\}\,
\cos\l(\frac{4\,k}{k_0}\r)\Biggr),
\end{eqnarray}
where the coefficient ${\cal K}_{13}$ has been defined to be
\begin{equation}
{\cal K}_{13}
\equiv 
{\cal I}_{13}^b(k)\,\alpha_k^{\ast}{}^3
-{\cal I}_{13}^b{}^{\ast}(k)\,\tilde{\beta}_k^{\ast}{}^3
+\alpha_k^{\ast}\,\tilde{\beta}_k^{\ast}\,
\l[{\cal J}_{13}^b{}^{\ast}(k)\,\tilde{\beta}_k^{\ast}
-{\cal J}_{13}^b(k)\,\alpha_k^{\ast}\r].
\end{equation}
Obviously, in the calculation of the quantity $G_1^{-}+G_3^{-}$,
the coefficient ${\cal K}_{13}$ plays the same role that ${\cal K}_2$ 
had played in the evaluation of $G_2^{-}$. 
Therefore, it does not come as a surprise that the structure of
${\cal K}_{13}$ is identical to that of ${\cal K}_2$. 
Moreover, the expression of $G_1^-+G_3^-$ itself bears a lot of 
resemblance to $G_2^-$ with, in particular, the `harmonics' 
$k/k_0$, $2\,k/k_0$, $3\,k/k_0$ and $4\,k/k_0$, all being present. 

\par

Let us now calculate the limiting forms of the above expression for
small and large~$k/k_0$. As $k/k_0\to 0$, we find that
$G_{1}^{-}+G_{3}^{-}$ behaves as follows:
\begin{eqnarray}
\lim _{k/k_0\to 0}
k^6\, \l[G_{1}^{-}(k)+G_{3}^{-}(k)\r]
&=& -\f{1}{320\, \sqrt{2\, \epsilon_{1-}^3(\eta_{\rm e})}}\;
\f{A_-^3\,H_0^2}{\Mp^5\, A_+^2}\;\l(\f{k}{k_0}\r)^{2}
\nonumber \\ & &
\times\,\l(15+ \f{2\, A_-}{A_+} + \f{3\, A_-^2}{A_+^2}\r),
\end{eqnarray}
i.e. it goes to zero quadratically for small wavenumbers.  Therefore,
in this limit, it is the term before the transition that dominates
and, hence, the complete contribution due to the first and the third
terms reduces to
\begin{eqnarray}
\lim _{k/k_0\to 0}
k^6\,\l[G_{1}(k)+G_{3}(k)\r]
&=&\lim _{k/k_0\to 0}
k^6\,\l[G_{1}^{+}(k)+G_{3}^{+}(k)\r]
\nonumber \\ 
&=&-\frac{1}{36\, \sqrt{2\,\epsilon_{1-}^{3}(\eta _{\rm e})}}\, 
\frac{A_-^3\, H_0^2}{\Mp^5\, A_{+}^{2}}.
\end{eqnarray}
\begin{figure}
\begin{center}
\includegraphics[width=15.0cm]{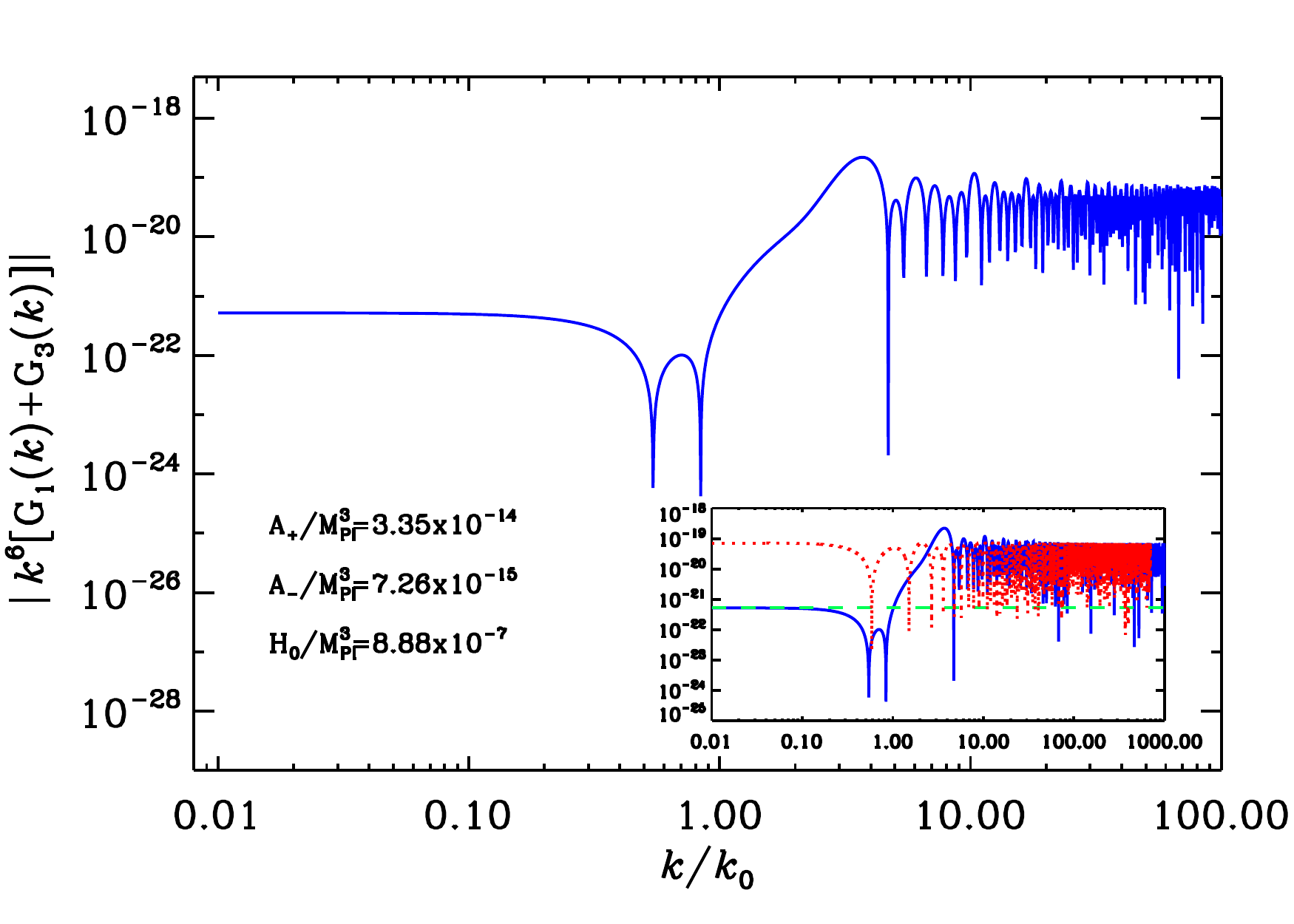}
\end{center}
\caption{The absolute value of the quantity $k^6\, (G_1+G_3)$ has been
  plotted as a function of $k/k_0$ (the blue curve). We have again
  worked with the same values of the parameters as in the earlier
  figures.  As in the plots before, the green and the red curves in
  the inset denote the asymptotic forms.}
\label{fig:abs-G13}
\end{figure}
In the limit $k/k_0\to \infty$, one finds that
\begin{eqnarray}
\lim _{k/k_0\to \infty}
k^6\,\l[G_{1}^{-}(k)+G_{3}^{-}(k)\r]
&=&-\frac{1}{36 \sqrt{2\,\epsilon_{1-}^{3}(\eta _{\rm e})}}\,
\frac{A_-\, H_0^2}{\Mp^5}
\Biggl\{1
\nonumber \\ & &
+\frac{27\Delta A}{8A_-}
\cos \left(\frac{k}{k_0}\right)
-\frac{36}{48}\frac{A_+}{A_-}\frac{k}{k_0}\sin \l(\frac{3\,k}{k_0}\r)
\nonumber \\ & &
+\frac{1}{8}\Biggl[\frac{\Delta A}{A_-}
\left(50+27\frac{\Delta A}{A_+}\right)
-8-27\frac{\Delta A}{A_+}\Biggr]
\nonumber \\ & & \times
\cos \l(\frac{3\,k}{k_0}\r)
\Biggr\}.
\end{eqnarray}
As usual, only trigonometric functions with the argument $3\,k/k_0$ 
remain in the final expression of the limit. Also, note that the
coefficient of the $\sin\, (3\, k/k_0)$ term diverges linearly at
large $k$, exactly as the contribution before the transition had
[cf.~Eq.~(\ref{eq:G13p-lk})], albeit with an opposite sign. Hence,
the complete contribution remains finite and, we find that, in the
limit $k/k_0\to \infty$, $G_{1}+G_{3}$ goes to
\newpage
\begin{eqnarray}
\lim _{k/k_0\to \infty}
k^6\,\l[G_{1}(k)+G_{3}(k)\r]
&=&-\frac{1}{36\, \sqrt{2\,\epsilon_{1-}^{3}(\eta _{\rm e})}}\,   
\frac{A_-\, H_0^2}{\Mp^5}\,
\Biggl(1+\Biggl(-1
\nonumber \\ & &
-\frac{27}{8}\left(\frac{A_-}{A_+}-1\right)
+\left(1-\frac{A_+}{A_-}\right)
\Biggl[\frac{50}{8}
\nonumber \\ & &
+\frac{27}{8}\left(\frac{A_-}{A_+}-1\right)
\Biggr]
+\frac{A_+}{A_-}\left(\frac{35}{8}-\frac{27}{8}
\frac{A_-}{A_+}\right)\Biggr)
\nonumber \\ & & \times
\cos\l(\f{3\,k}{k_0}\r)\Biggr),
\end{eqnarray}
As before, at large wavenumbers, we obtain a scale invariant 
amplitude which is modulated by super-imposed oscillations of 
the form $\cos\, (3\,k/k_0)$.
In Fig.~\ref{fig:abs-G13}, we have plotted the absolute value of 
the quantity $[k^6\, (G_1+G_3)]$. The figure clearly reflects the
limiting forms that we have obtained above.


\subsection{The contribution due to the fifth and the sixth terms}

Let us now turn to the evaluation of the contribution due to the
remaining two terms in the interaction Hamiltonian~(\ref{eq:Hint}),
viz. the fifth and the sixth.  As in the case of $\cG_{1}$ and~$\cG_{3}$, 
we find that, the terms $\cG_{5}$ and $\cG_{6}$ too contain the
same types of integrals.  Hence, in the equilateral limit, they can be 
evaluated together.


\subsubsection{Before the transition}
 
Since $\epsilon_{1}$ is a constant before the transition, the
integrals encountered in $\cG_{5}$ and $\cG_{6}$ prove to be exactly
of the same form as in the cases of $\cG_{1}$ and $\cG_{3}$.
Therefore, they can be evaluated in the same fashion and, in the most
generic situation of $\vka\ne\vkb\ne\vkc$, one obtains that
\begin{eqnarray}
\cG_5^{+}(\vka,\vkb,\vkc)
&=&\frac{iH_0}{16\Mp^3k_{_{\rm T}}}
\sqrt{\frac{\epsilon_{1+}^3}{(k_1k_2k_3)^{3}}}
{\rm e}^{-ik_{_{\rm T}}/k_0}
\Biggl[(\vka\cdot\vkb)k_3^{2}
\nonumber \\ & & \times
\left(1+\frac{k_1}{k_{_{\rm T}}}+\frac{i\,k_1}{k_0}\r)
+ {\rm five~permutations}\Biggr]
\end{eqnarray}
and
\begin{eqnarray}
\cG_6^{+}(\vka,\vkb,\vkc)
&=&\frac{iH_0}{16\Mp^3k_{_{\rm T}}}
\sqrt{\frac{\epsilon_{1+}^3}{(k_1k_2k_3)^{3}}}
{\rm e}^{-ik_{_{\rm T}}/k_0}
\Biggl[k_1^2 (\vkb\cdot \vkc)
\nonumber \\ & & \times
\l(1+\frac{k_1}{k_{_{\rm T}}}
+\frac{i\,k_1}{k_0}\r)
+ {\rm two~permutations}\Biggr].
\end{eqnarray}
In the equilateral limit, we find that 
\begin{eqnarray}
\cG_5^{+}(k)+\cG_6^{+}(k)
&=&\frac{3\,i\,H_0\, k^3}{16\,\Mp^3}\,
\sqrt{\frac{\epsilon_{1+}^{3}}{k^{9}}}\,
\l(\frac{4}{3}+\frac{i\,k}{k_0}\r)\,
{\rm e}^{-3\,i\,k/k_0}
\nonumber \\ 
&=&-\frac{3\,\epsilon_{1+}}{4}\,
\l[\cG_{1}^{+}(k)+\cG_{3}^{+}(k)\r].\;\;
\end{eqnarray}
From the last equality, we can then immediately conclude that 
$G_5^{+}(k)+G_6^{+}(k)=-3\,\epsilon_{1+}\l[G_{1}^{+}(k)+G_{3}^{+}(k)\r]/4$. 
As a consequence, the asymptotic behavior of $G_5^{+}+G_6^{+}$ can
be arrived at from the forms of $G_1^{+}+G_3^{+}$ we had obtained
earlier. As $k/k_0\to 0$, one has
\begin{eqnarray}
\lim _{k/k_0\to 0}\,
k^6\,\l[G_{5}^{+}(k)+G_{6}^{+}(k)\r]
&=& \frac{\epsilon_{1+}}{48\, 
\sqrt{2\,\epsilon_{1-}^{3}(\eta _{\rm e})}}\, 
\frac{A_-^3\, H_0^2}{\Mp^5\, A_{+}^{2}}
\nonumber \\
&=&\frac{1}{864\, 
\sqrt{2\,\epsilon_{1-}^{3}(\eta _{\rm e})}}\, 
\frac{A_-^3}{\Mp^7\, H_0^{2}},
\end{eqnarray}
while, as $k/k_0\to\infty$, one obtains
\begin{eqnarray}
\lim _{k/k_0\to \infty}\,
k^6\,\l[G_{5}^{+}(k)+G_{6}^{+}(k)\r]
&=&\frac{1}{1152\, 
\sqrt{2\,\epsilon_{1-}^{3}(\eta_{\rm e})}}\,
\frac{A_+^{3}}{\Mp^7\,H_0^2}
\nonumber \\ & &
\times\,\Biggl[\f{9}{2}\,\l(1-\frac{A_-}{A_+}\r)\,
\cos \left(\frac{k}{k_0}\right)
\nonumber \\ & &
+\frac{k}{k_0}
\sin \left(\frac{3\,k}{k_0}\right)
\nonumber \\ & &
+\l(\f{35}{6}-\f{9\,A_-}{2\,A_+}\r)\,
\cos \left(\frac{3\,k}{k_0}\right)\Biggr].\;\;
\end{eqnarray}
Note that, while $k^6\,(G_{5}^{+}+G_{6}^{+})$ goes to a constant at
small $k/k_0$, it diverges linearly (modulated by oscillations) at
large $k/k_0$, as in the case of $k^6\,(G_{1}^{+}+G_{3}^{+})$. Also,
as we shall illustrate, just as in the earlier cases, the diverging
term will be exactly canceled by a corresponding term that arises
post-transition.


\subsubsection{After the transition}

The next step is to perform the calculation after the transition. 
The calculation proceeds exactly as in the case of $\cG_1^-+\cG_3^-$. 
We find that, we can write
\begin{eqnarray}
\cG_5^{-}(k)+\cG_6^{-}(k) 
&=& -\frac{9\,H_0}{16\,\Mp^3\,k^{9/2}}
\biggl[{\alpha_{k}^{\ast}}^{3}\,I_{56}(k) 
-{\beta_{k}^{\ast}}^{3}\,I_{56}^{\ast}(k)
-{\alpha_{k}^{\ast}}^{2}\, \beta_{k}^{\ast}\,J_{56}(k)
\nonumber \\ & &
+\alpha_{k}^{\ast}\,{\beta_{k}^{\ast}}^{2}\, 
J_{56}^{\ast}(k)\biggr],
\end{eqnarray}
where $I_{56}$ and $J_{56}$ are described by the integrals
\begin{eqnarray}
I_{56}(k) 
&=& \frac{A_-^3}{54\,\sqrt{2}\,H_0^6\,\Mp^3}
\int _{-k_0^{-1}}^{\eta_{\rm e}}{\rm d}\tau\, \l(1-\rho^3\,\tau^3\r)\,
\l(1-i\,k\,\tau\r)
\nonumber \\ & &
\times\,\l[3\,\rho^3\,\tau\,\l(1-i\,k\,\tau\r)
+k^2\,\l(1-\rho^3\,\tau^3\r)\r]^2\,{\rm e}^{3\,i\,k\,\tau},\\
J_{56}(k) 
&=& \frac{A_-^3}{54\,\sqrt{2}\,H_0^6\,\Mp^3}
\int _{-k_0^{-1}}^{\eta_{\rm e}}{\rm d}\tau\, \l(1-\rho^3\,\tau^3\r)\,
\biggl[3\,\rho^3\,\tau\, \l(1-i\,k\,\tau\r)
\nonumber \\ & &
+k^2\,\l(1-\rho^3\,\tau^3\r)\biggr]
\biggl[9\,\l(1-i\,k\,\tau\r)\l(1+i\,k\,\tau\r)\,\rho^3\,\tau
\nonumber \\ & &
+k^2\,\l(1-\rho^3\,\tau^3\r)\,\l(3-i\,k\,\tau\r)\biggr]\,{\rm e}^{i\,k\,\tau}.
\end{eqnarray}
These integrals are very similar to the ones considered before but,
crucially, they do not contain any poles.  This makes their explicit
calculation considerably easier. Upon following the same strategy that
we had adopted in the cases of $I_{13}$ and $J_{13}$, one obtains that
\begin{equation}
I_{56}(k)
=\frac{A_-^3\,k^3}{54\,\sqrt{2}\,H_0^6\,\Mp^3}\,
\l[{\cal I}_{56}^a(k)+{\cal I}_{56}^b(k)\,{\rm e}^{-3\,i\,k/k_0}\r].
\end{equation}
with
\begin{eqnarray}
{\cal I}_{56}^a(k) 
&= -\sum _{n=0}^{10}\,\l(\frac{i}{3}\r)^{n+1}\,n!\;f_n,\\
{\cal I}_{56}^b(k)
&= \sum _{n=0}^{10}\,\l(\frac{i}{3}\r)^{n+1}\, n!\;f_n\,
e_{n}\l(\frac{3\,i\,k}{k_0}\r),
\end{eqnarray}
where the coefficients $f_n$ are given by
Eqs.~(\ref{eq:f0})--(\ref{eq:f10}), while $e_{n}(z)$ is the
exponential sum function~(\ref{eq:esf}). Similarly, we find that
$J_{56}$ can be obtained to be
\begin{equation}
J_{56}(k)
=\frac{A_-^3\,k^3}{54\,\sqrt{2}\,H_0^6\,\Mp^3}\,
\l[{\cal J}_{56}^a(k)+{\cal J}_{56}^b(k)\,{\rm e}^{-i\,k/k_0}\r]
\end{equation}
with
\begin{eqnarray}
{\cal J}_{56}^a(k)
= -\sum _{n=0}^{10}\,i^{n+1}\, n!\;g_n,\\
{\cal J}_{56}^b(k) 
= \sum _{n=0}^{10}\, i ^{n+1}\, n!\;g_n\,
e_{n}\l(\frac{i\,k}{k_0}\right),
\end{eqnarray}
and the coefficients $g_n$ being given by
Eqs.~(\ref{eq:g0})--(\ref{eq:g10}).

\par

At this stage, the calculation of $G_{5}^{-}+G_{6}^{-}$ 
progresses as before and the final expression reads
\begin{eqnarray}
k^{6}\,[G_{5}^{-}(k)+G_{6}^{-}(k)] 
&=& \frac{i}{768\sqrt{2\epsilon_{1-}^{3}(\eta_{\rm e})}}
\frac{A_-^3}{\Mp^7H_0^2}
\nn\\ & &
\times\Biggl(\l[{\cal I}_{56}^a(k)-{\cal I}_{56}^a{}^{\ast}(k)\r]\,
\l(\alpha_k^3\,\alpha_k^{\ast}{}^3
-\tilde{\beta}_k^3\,\tilde{\beta}_k^{\ast}{}^3\right)\nn\\
&+ &
3\,\l[{\cal J}_{56}^a(k)-{\cal J}_{56}^a{}^{\ast}(k)\r]\,
\alpha_k\,\alpha_k^{\ast}\, \tilde{\beta}_k\,\tilde{\beta}_k^{\ast}\,
\l(\alpha_k\,\alpha_k^{\ast}
-\tilde{\beta}_k\,\tilde{\beta}_k^{\ast}\r)\nn\\ 
&+ &
6\,i\,\Re \l[{\cal K}_{56}(k)\,\alpha_k\,\tilde{\beta}_k^2
+{\cal K}_{56}^{\ast}(k)\,\alpha_k^{\ast}{}^2\,
\tilde{\beta}_k^{\ast}\r]\,
\sin \l(\frac{k}{k_0}\r)\nn\\
&+ &
6\,i\,\Im \l[{\cal K}_{56}(k)\,\alpha_k\,\tilde{\beta}_k^2
+{\cal K}_{56}^{\ast}(k)\,\alpha_k^{\ast}{}^2\,
\tilde{\beta}_k^{\ast}\r]\,
\cos \left(\frac{k}{k_0}\r)\nn\\
& -&
2\,i\,\Re \biggl\{\l[3\,{\cal I}_{56}^a{}^{\ast}(k)
-{\cal J}_{56}^a(k)\r]\nn\\
& &\times\alpha_k\,\tilde{\beta}_k^{\ast}\,
\left(\alpha_k^2\,\alpha_k^{\ast}{}^2
-\tilde{\beta}_k^2\,\tilde{\beta}_k^{\ast}{}^2\right)\biggr\}\,
\sin\left(\frac{2\,k}{k_0}\right)\nn\\ 
& +&
2\,i\,\Im \biggl\{\l[3\,{\cal I}_{56}^a{}^{\ast}(k)
-{\cal J}_{56}^a(k)\r]\nn\\
& &\times \alpha_k\,\tilde{\beta}_k^{\ast}\,
\left(\alpha_k^2\,\alpha_k^{\ast}{}^2
-\tilde{\beta}_k^2\,\tilde{\beta}_k^{\ast}{}^2\right)\biggr\}\,
\cos\left(\frac{2\,k}{k_0}\right)\nn\\
&- &
2\,i\,\Re \l[{\cal K}_{56}(k)\,\alpha_k^3
+{\cal K}_{56}^*(k)\,\tilde{\beta}_k^{\ast}{}^3\r]\,
\sin\left(\frac{3\,k}{k_0}\right)\nn\\
&+ &
2\,i\,\Im \l[{\cal K}_{56}(k)\,\alpha_k^3
+{\cal K}_{56}^*(k)\,\tilde{\beta}_k^{\ast}{}^3\r]\,
\cos\left(\frac{3\,k}{k_0}\right)\nn\\
&+ &
2\,i\,\Re \biggl\{\l[3\,{\cal I}_{56}^a(k)-{\cal J}_{56}^a(k)\r]
\alpha_k^{\ast}{}^2\,\tilde{\beta}_k^2\,
\left(\alpha_k\,\alpha_k^{\ast}
-\tilde{\beta}_k\,\tilde{\beta}_k^{\ast}\right)\biggr\}\,
\nonumber \\ & & \times
\sin\left(\frac{4\,k}{k_0}\right)\nn\\
&+ &
2\,i\,\Im \biggl\{\l[3\,{\cal I}_{56}^a(k)-{\cal J}_{56}^a(k)\r]\,
\alpha_k^{\ast}{}^2\,\tilde{\beta}_k^2\,
\l(\alpha_k\,\alpha_k^{\ast}
-\tilde{\beta}_k\,\tilde{\beta}_k^{\ast}\r)\biggr\}\,
\nonumber \\ & & \times
\cos\left(\frac{4\,k}{k_0}\right)\Biggr),
\end{eqnarray}
where the coefficient ${\cal K}_{56}$ has been defined to be
\begin{equation}
{\cal K}_{56}(k)
= {\cal I}_{56}^b(k)\,\alpha_k^{\ast}{}^3
-{\cal I}_{56}^b{}^{\ast}(k)\,\tilde{\beta}_k^{\ast}{}^3\,
+\alpha_k^{\ast}\,\tilde{\beta}_k^{\ast}
\l[{\cal J}_{56}^b{}^{\ast}(k) \tilde{\beta}_k^{\ast}
-{\cal J}_{56}^b(k) \alpha_k^{\ast}\r].
\end{equation}
Needless to add, the coefficient ${\cal K}_{56}$ is similar to 
${\cal K}_2$ and ${\cal K}_{13}$. 
Further, as far as the structure of $G_{5}^{-}+G_{6}^{-}$ is 
concerned, it again involves the various trigonometric functions 
that we had observed in the earlier cases.

\par

\begin{figure}
\begin{center}
\includegraphics[width=15.0cm]{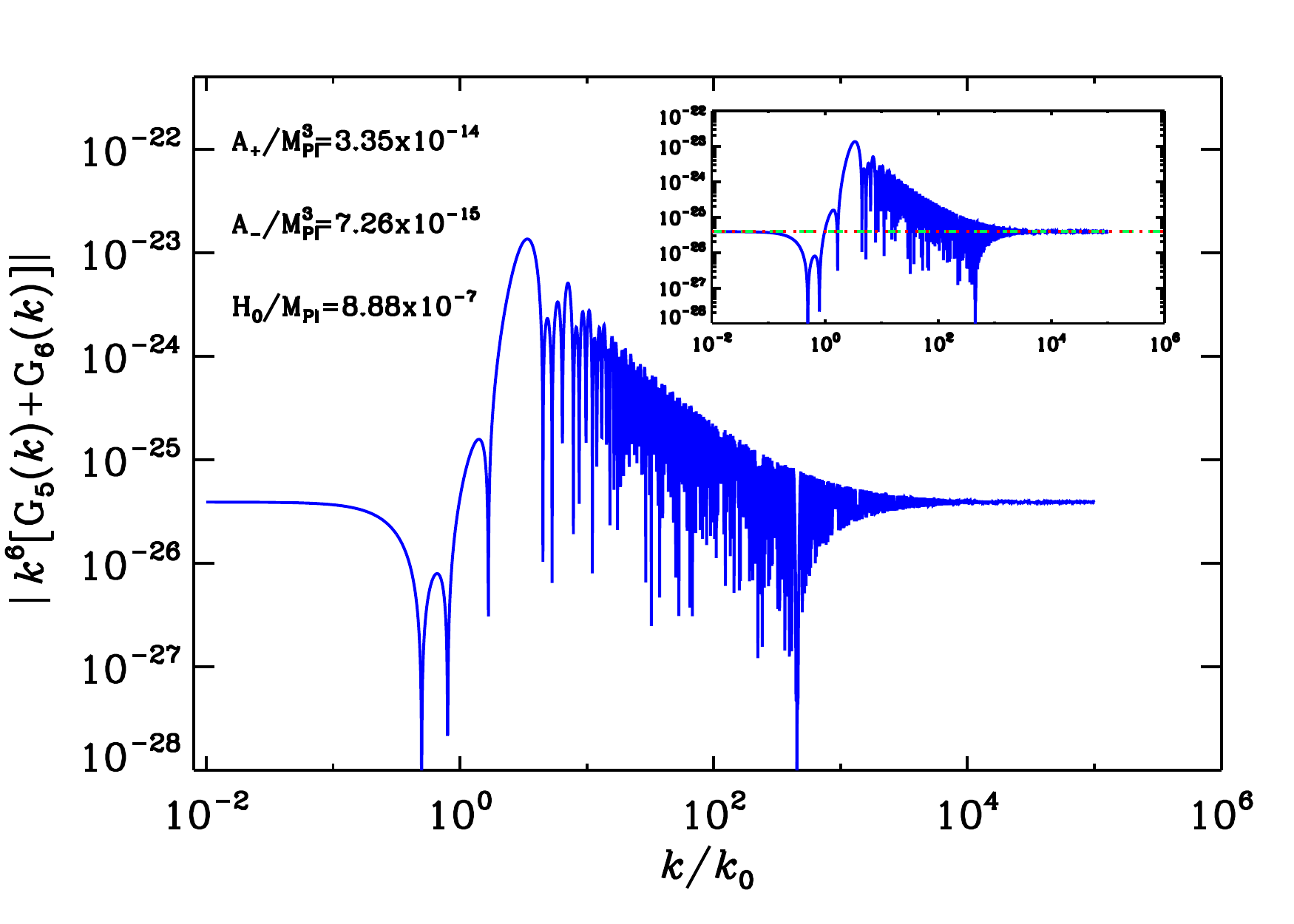}
\end{center}
\caption{The absolute value of the quantity $k^6\, (G_{5}+G_{6})$ (in
  blue) and the asymptotic behavior (in green and in red in the inset)
  has been plotted for the same set of parameters as in the previous
  figures.  The inset highlights the fact that the quantity goes to
  the same scale invariant value at small and large wavenumbers.}
\label{fig:abs-G56}
\end{figure}

Let us now evaluate the asymptotic forms of the above expression.  
As $k/k_0\to 0$, we find that
\begin{eqnarray}
\lim _{k/k_0\to 0}
k^6\,\l[G_{5}^{-}(k)+G_{6}^{-}(k)\r]
&=& \f{1}{887040\, \sqrt{2\, \epsilon_{1-}^3(\eta_{\rm e})}}\,
\f{A_-^3}{H_0^2\,\Mp^7}\;\l(\f{k}{k_0}\r)^{2}
\nonumber \\ & &
\times\,\Biggl(550+ \f{600\, A_-}{A_+}
+ \f{675\, A_-^2}{A_+^2}
+ \f{194\, A_-^3}{A_+^3} 
\nonumber \\ & &
+\f{291\,A_-^4}{A_+^4} \Biggr).
\end{eqnarray}
In other words, this term vanishes quadratically for small
wavenumbers, just as the contributions due to the first and the third
terms, post-transition, had.  So, in this limit, the total contribution 
due to the fifth and the sixth terms reduces to the strictly scale 
invariant value arising before the transition, i.e. we have
\begin{eqnarray}
\lim _{k/k_0\to 0}
k^6\,\l[G_{5}(k)+G_{6}(k)\r]
&=&\lim _{k/k_0\to 0}
k^6\,\l[G_{5}^{+}(k)+G_{6}^{+}(k)\r]\nn\\
&=&\frac{1}{864\, 
\sqrt{2\,\epsilon_{1-}^{3}(\eta _{\rm e})}}\, 
\frac{A_-^3}{\Mp^7\, H_0^{2}}.
\end{eqnarray}
In the limit $k/k_0\to \infty$, we find that 
\begin{eqnarray}
\lim _{k/k_0\to \infty}
k^6\,\l[G_{5}^{-}(k)+G_{6}^{-}(k)\r]
&=&\frac{1}{864\, 
\sqrt{2\,\epsilon_{1-}^{3}(\eta _{\rm e})}}\, 
\frac{A_-^3}{\Mp^7\, H_0^{2}}
\Biggl[1
\nonumber \\ & &
-\frac{27\, A_+^3}{8\,A_-^3}\,\l(1-\f{A_-}{A_+}\r)\,
\cos \left(\frac{k}{k_0}\right)
\nonumber \\ & &
-\,\frac{3\,A_+^3}{4\,A_-^3}\,
\frac{k}{k_0}\,\sin \l(\frac{3\,k}{k_0}\r)
\nonumber \\ & &
-\frac{A_+^3}{8\,A_-^3}\left(35-\frac{27\,A_-}{A_+}\right)
\cos \l(\frac{3\,k}{k_0}\r)\Biggr].
\end{eqnarray}
If we now combine the asymptotic forms before and after the
transition, for large $k/k_0$, we find that the terms proportional to
$k/k_0$ cancel, and one is led to the expression
\begin{equation}
\lim _{k/k_0\to \infty}
k^6\,\left[G_{5}(k)+G_{6}(k)\r]
=\frac{1}{864\, 
\sqrt{2\,\epsilon_{1-}^{3}(\eta _{\rm e})}}\, 
\frac{A_-^3}{\Mp^7\, H_0^{2}},
\end{equation}
which is exactly the same as the scale invariant quantity that we had
encountered in the limit of small $k/k_0$. In particular, it is
interesting to note that, in this case, no superimposed oscillations
are present. These asymptotic behavior are evident in
Fig.~\ref{fig:abs-G56}, wherein we have plotted the absolute value of
$k^6\, (G_{5}+G_{6})$.

\subsection{The contribution due to the field redefinition}

Let us now turn to the contribution $G_7$ due to the field
redefinition~(\ref{eq:G7}) which in the equilateral limit simplifies
to
\begin{equation}
G_7(k)=\f{3\,\epsilon_{2-}(\eta_{\rm e})}{2}\; 
\vert f_k^{-}(\eta_{\rm e})\vert^4
=6\,\,\epsilon_{1-}(\eta_{\rm e})\;
\vert f_k^{-}(\eta_{\rm e})\vert^4,
\end{equation}
where the last equality follows from the fact that, at late times, for
the linear potential of our interest,
$\epsilon_{2-}=4\,\epsilon_{1-}$.  Upon using the
expression~(\ref{eq:fk-lt}) for $f_k^{-}$ at late times, we obtain
that
\begin{equation}
k^6\,G_7(k)
=\f{3\, H_0^4}{8\,\Mp^4\,\epsilon_{1-}(\eta_{\rm e})}\;
\vert\alpha_k-\beta_k\vert^4.
\end{equation}
Using the expressions~(\ref{eq:alphak-sm}) and~(\ref{eq:betak-sm}) for
$\alpha_k$ and $\beta_k$, we find that, we can write
\begin{figure}
\begin{center}
\includegraphics[width=15.0cm]{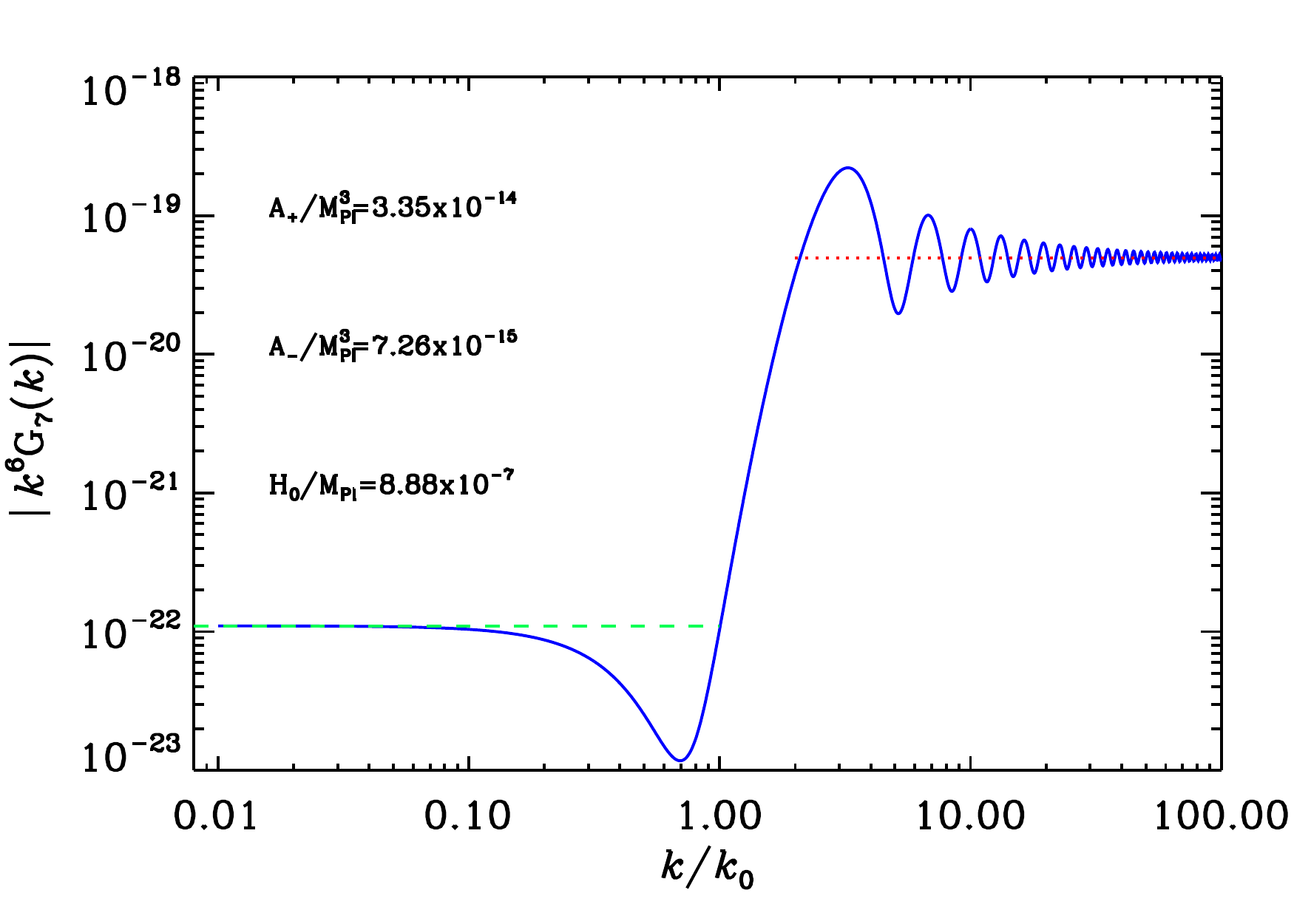}
\end{center}
\caption{The absolute value of the quantity $k^6\, G_7$ (in blue) and
  its asymptotic forms (in green and in red) have been plotted for the
  same set of parameters as in the previous figures.}
\label{fig:abs-G7}
\end{figure}
\begin{eqnarray}
k^6\, G_7(k)
&=&\f{3\, H_0^4}{8\,\Mp^4\,\epsilon_{1-}(\eta_{\rm e})}\;
\Biggl\{\l(\alpha_k\, \alpha_k^*+\tilde{\beta}_k\, \tilde{\beta}_k^*\r)^2
+2\,\alpha_k\, \alpha_k^*\, \tilde{\beta}_k\, \tilde{\beta}_k^*\nn\\
& &
-\,4\,\Im\l[\l(\alpha_k\, \alpha_k^*
+\tilde{\beta}_k\, \tilde{\beta}_k^*\r)\,
\l(\alpha_k\, \tilde{\beta}_k^*\r)\r]\, 
\sin \l(\frac{2\,k}{k_0}\r)\nn\\
& &
-\,4\,\Re\l[\l(\alpha_k\, \alpha_k^*
+\tilde{\beta}_k\, \tilde{\beta}_k^*\r)\,
\l(\alpha_k\, \tilde{\beta}_k^*\r)\r]\, 
\cos \l(\frac{2\,k}{k_0}\r)\nn\\
& &
+\,2\, \Im\l(\alpha_k^2\,\tilde{\beta}_k^{\ast}{}^2\r)
\sin \left(\frac{4\,k}{k_0}\right)
+ 2\, \Re\l(\alpha_k^2\,\tilde{\beta}_k^{\ast}{}^2\r)
\cos \left(\frac{4\,k}{k_0}\right)\Biggr\}.
\end{eqnarray}
Far away from the characteristic scale $k_0$, this quantity turns
strictly scale invariant, just as the power spectrum does. Its
asymptotic forms are found to be
\begin{equation}
\lim_{k/k_0\,\to\, 0}\;
k^6\, G_7(k)
= \f{3\, A_-^4\, H_0^4}{8\,A_+^4\, \Mp^4\,
\epsilon_{1-}(\eta_{\rm e})}\;
=\f{27\, A_-^2\, H_0^8}{4\,A_+^4\, \Mp^2}
\end{equation}
and
\begin{equation}
\lim_{k/k_0\,\to\, \infty}\;
k^6\, G_7(k)
= \f{3\, H_0^4}{8\, \Mp^4\,\epsilon_{1-}(\eta_{\rm e})}
=\f{27\, H_0^8}{4\,A_-^2\, \Mp^2}.
\end{equation}
In Fig.~\ref{fig:abs-G7}, we have plotted the absolute value of the
above expression for~$k^6\, G_7$. The figure clearly illustrates
that the quantity turns scale invariant asymptotically.

\par

This concludes our calculation of the complete bi-spectrum.
With all the expressions at our disposal, we can now relate these 
results to the observable parameter $\fnl$, and also discuss the 
various possible conclusions and implications.


\section{$\fnl$ in the Starobinsky model}\label{sec:l-fnl}

In this section, we shall discuss two issues. 
We shall firstly focus on whether the Starobinsky model can lead 
to as large a value for the non-Gaussianity parameter~$\fnl$ as 
the currently quoted mean values.
Recall that, as we had mentioned in the introduction, the recent
CMB data constrains the parameter to be $\fnl=32\pm 21$ in the 
local limit.
However, at this stage, it is important to stress that the local 
non-Gaussianity parameter is a priori different from a scale 
dependent~$\fnl$ in the equilateral case. 
In fact, the relevance of the constraints on the local parameter 
to the $\fnl^{\rm eq}$ that we obtain below is not entirely clear. 
This point needs to be borne in mind when we compare the~$\fnl$ 
in the equilateral limit with the observational constraints quoted 
above. 
Then, in the second part of this section, we shall focus on an issue 
related to the hierarchy of the various contributions to the bi-spectrum.


\subsection{Can $\fnl^{\rm eq}$ be large in the Starobinsky model?}

\begin{figure}
\begin{center}
\includegraphics[width=15.0cm]{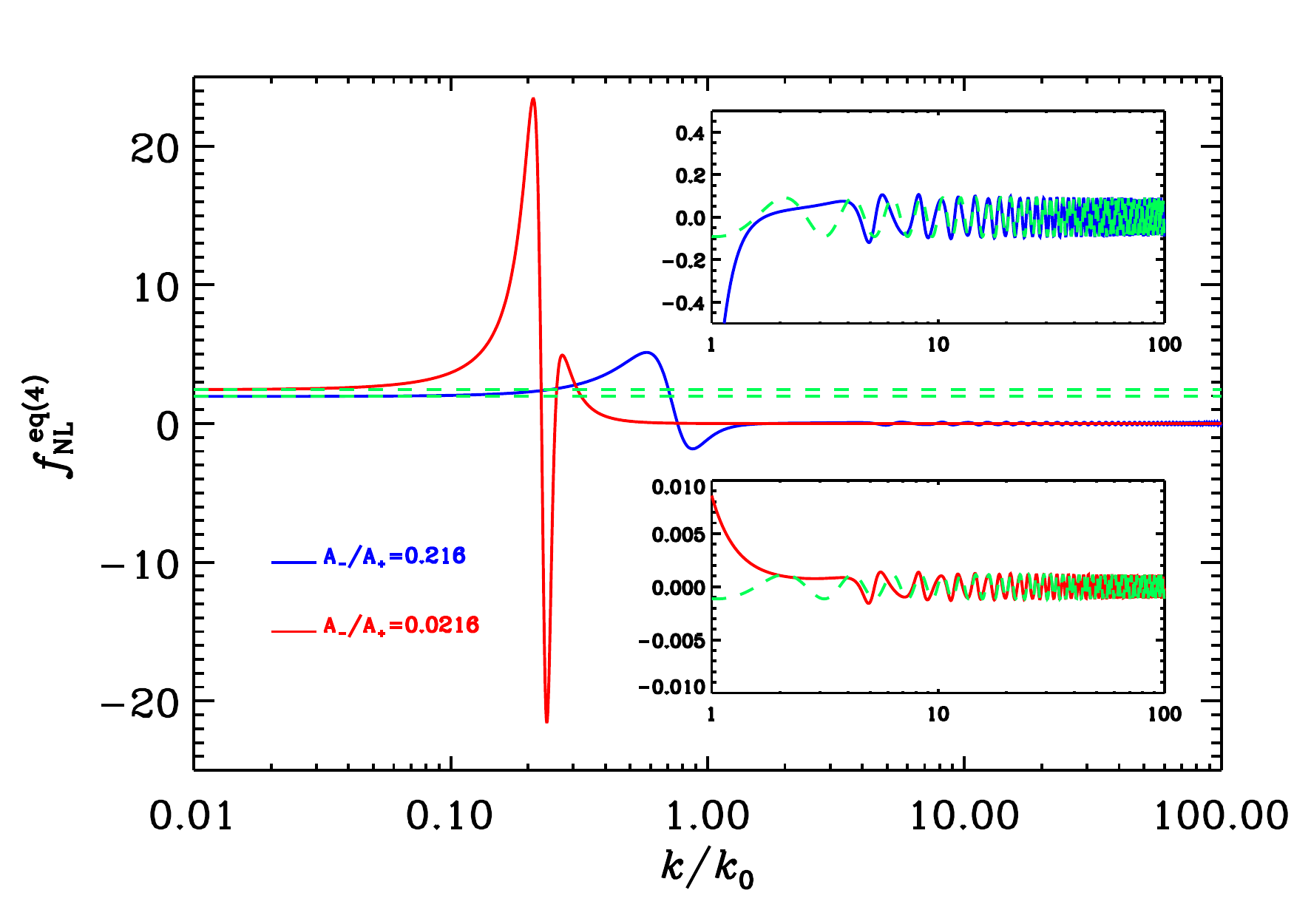}
\end{center}
\caption{\label{fig:fnl} The non-Gaussianity parameter in the
  equilateral limit, i.e.  $\fnl^{\rm eq}$, due to the dominant term
  in the Starobinsky model.  The blue curve corresponds to the values
  of the parameters that we had considered in the previous figures.
  The red curve corresponds a larger $A_{+}$ (we have set $A_{+}/\Mp^3
  =3.35\times10^{-13}$), but with the remaining parameters being the
  same as for the blue curve.  The dashed green lines represent the
  corresponding asymptotic values for small $k/k_0$ [given by
  Eq.~(\ref{eq:fnl-sk})]. The insets exhibit the oscillations about
  zero, with the dashed green curves highlighting the behavior at
  large $k/k_0$.}
\end{figure}

\begin{figure}
\begin{center}
\includegraphics[width=15.0cm]{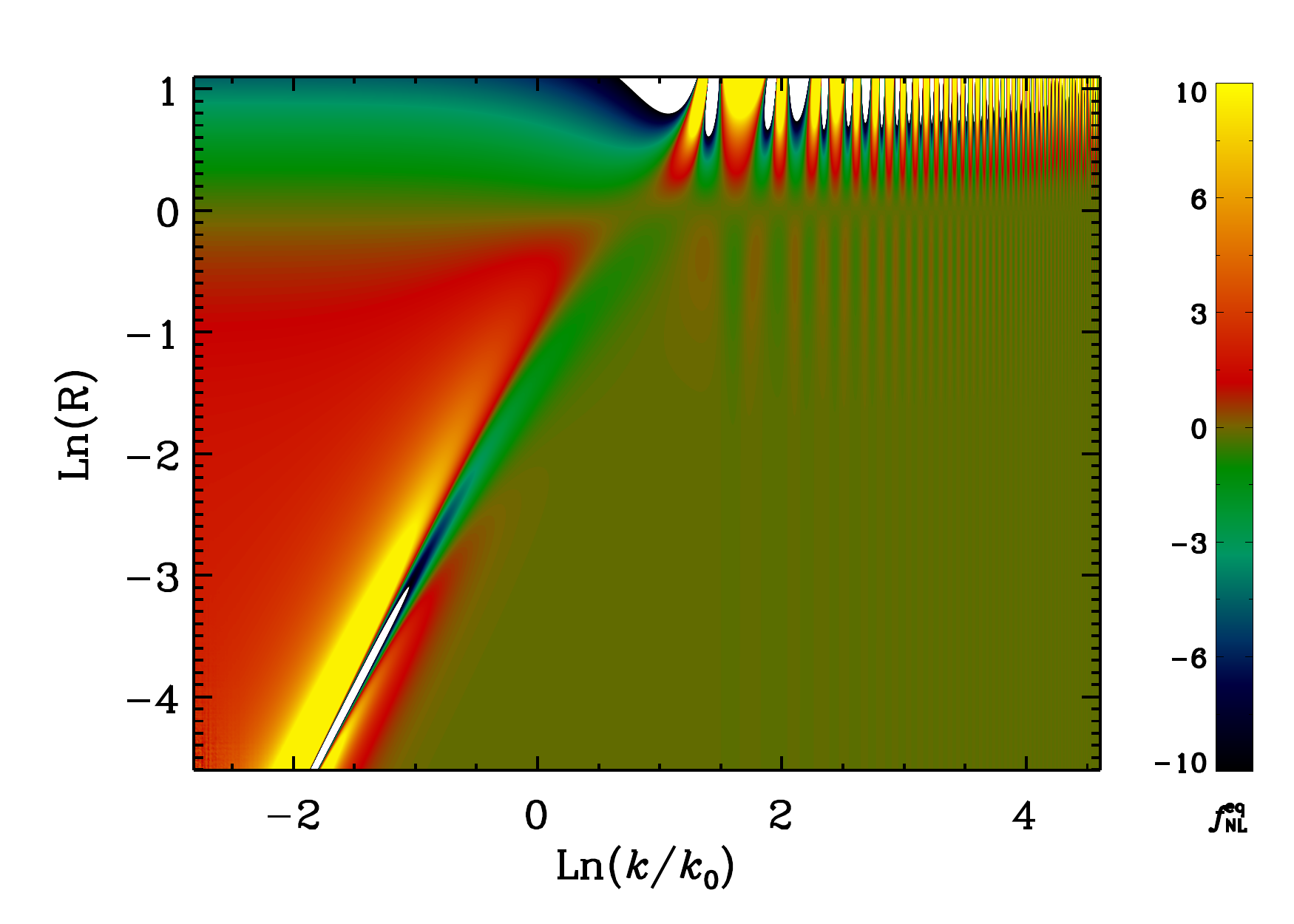}
\end{center}
\caption{\label{fig:fnl-2d} A two-dimensional contour plot of the
  non-Gaussianity parameter $\fnl^{\rm eq}$ due to the dominant term
  in the Starobinsky model, plotted in the plane of $k/k_0$ and
  $R=A_-/A_+$.  The white contours indicate regions wherein 
   $\fnl^{\rm eq}$ can
  be as large as $50$.  Note that, provided $R$ is reasonably small,
  $\fnl^{\rm eq}$ can be of the order of $25$ or so, as is indicated
  by the currently observed mean value.}
\end{figure}

Let us now discuss as to how large can the non-Gaussianity 
parameter $\fnl^{\rm eq}$ be in the Starobinsky model. Since it is 
the fourth term that seems to often provide the dominant contribution 
to the bi-spectrum (in this context, however, see, the following 
sub-section), the quantity $\fnl^{\rm eq}$ can be evaluated based 
on this contribution.  
On substituting the expression~(\ref{eq:G4}) for $k^6\,G_4$ in the
definition~(\ref{eq:fnl-el}) of $\fnl$ in the equilateral limit, we
obtain that
\begin{eqnarray}
\fnl^{{\rm eq}(4)}&=&
-\frac{10}{9} 
\frac{81}{8\,(2\pi)^4\,\sqrt{2\,\epsilon_{1-}^{3}(\eta_{\rm e})}}\, 
\l(\f{k_0}{k}\r)^3\;
\f{\Delta A\, H_0^6}{A_-^2\,\Mp^3}\;
\l[{\cal P}_{_{\rm S}}(k)\r]^{-2}
\nonumber \\ & & \times
\Biggl[{\cal A}_1(k)\,\sin\l(\f{k}{k_0}\right)
+{\cal A}_2(k)\,\cos\l(\f{k}{k_0}\r)
+{\cal A}_3(k)\,\sin\l(\f{3\, k}{k_0}\r)
\nonumber \\ & & 
+{\cal A}_4(k)\,\cos\l(\f{3\,k}{k_0}\r)\Biggr],\label{eq:fnl4eq}
\end{eqnarray}
with the power spectrum ${\cal P}_{_{\rm S}}$ being given by
Eq.~(\ref{eq:sps}), whereas the coefficients ${\cal A}_1$--${\cal A}_4$ 
are given by Eqs.~(\ref{eq:cA1})--(\ref{eq:cA4}). 
Let us now consider the various qualitative aspects of this 
result\footnote{As we were completing this manuscript, a preprint 
appeared, which deals with a similar but different model, and also 
evaluates the non-Gaussianities analytically~\cite{ng-ae}. We should 
emphasize that our effort is more complete, as we evaluate the complete
bi-spectrum rather than just focus on the dominant contribution.}. 
It is important to notice that, as already remarked in the section 
devoted to the calculation of~$G_4$, $\fnl$ is not simply given by an 
expression proportional to $\sin\,(3\,k/k_0+\varphi_0)$, $\varphi_0$ 
being a phase (in this context, see, Ref.~\cite{ng-reviews}), but 
contains four different terms that oscillate with 
different frequencies. It is only in the limit $k/k_0\rightarrow \infty$ 
that a simple behavior as the one previously mentioned is recovered. 
Moreover, interestingly, we find that the expression for 
$\fnl^{{\rm eq}(4)}$ above depends on the slopes $A_+$ and $A_-$ of the 
potential only through the ratio $R\equiv A_-/A_+$, and it does not 
depend on the parameter $H_0$ at all. As should be clear by now, both the 
power spectrum and the bi-spectrum [or, equivalently, $G_n(k)$] exhibit a 
step in the Starobinsky model.  Note that, in all the earlier figures, 
we had worked with the parameters $A_+$ and $A_-$ such that $R=0.216<1$.  
As is obvious from the figures, the step always rises towards the larger
wavenumbers in such a case.  In contrast, one finds that the step in
the power spectrum and the bi-spectrum reverse direction when $R>1$,
with the lower level of the step being located at large
wavenumbers. The height of the step depends on the extent of the
difference in the slopes of the potential on either side of the
discontinuity.  The further is $R$ from unity, the larger proves to be
height of the step. This property becomes explicit in the limit $k/k_0
\to 0$ wherein the dominant contribution to $\fnl$ due to the term $G_4$
has the following simple form:
\begin{equation}
\lim_{k/k_0\,\to\, 0}\;
\fnl^{{\rm eq}(4)}=\f{5}{2}\, (1-R).\label{eq:fnl-sk}
\end{equation}
In Fig.~\ref{fig:fnl}, we have plotted the expression~(\ref{eq:fnl4eq}) 
for $\fnl^{{\rm eq}(4)}$ for two different values of the ratio of the
slopes of the potential.
And, Fig.~\ref{fig:fnl-2d} contains a two-dimensional contour plot
$\fnl^{{\rm eq}(4)}$ in the plane of $R$ and $k/k_0$.
It is clear from these figures that, in the Starobinsky model, the
non-Gaussianity parameter $\fnl^{\rm eq}$ can be as large as the
currently indicated mean value.

\par

But, the question that immediately springs mind is whether 
$\fnl^{\rm eq}$ can be large for values of the parameters of the 
Starobinsky model for which the power spectrum proves to be consistent 
with the data. In plotting all the figures, we have ensured that 
the parameters that we have been working with lead to the COBE
normalization of the power spectrum at suitably small scales.  Given a
$H_0$, COBE normalization restricts the parameter~$A_-$ to a fixed
value [cf.~Eq.~(\ref{eq:ps-ss})], while leaving the parameter $A_+$
or, equivalently, $R$, unconstrained. Though the actual power spectrum
that arises in the Starobinsky model in itself has not been compared
with the CMB data, comparisons of some variations thereof have been
been carried out.  These variations have essentially involved
introducing, by hand, an overall multiplicative factor to incorporate
a suitable spectral tilt or a sharp cut-off at the lower wavenumbers
(see the fourth reference in Refs.~\cite{quadrupole}). While the tilt
ensures a good fit to the data at the higher multipoles, a sharp
cut-off can improve the fit to the outliers at the lower multipoles
that we had discussed in some detail in the introductory section.  One
finds that, in the presence of a suitable spectral tilt, the best fit
value of $R$ proves to be about $0.73^{+0.25}_{-0.14}$ (see the fourth
reference in Refs.~\cite{quadrupole}; also, according to the article, 
the best fit value for $k_0$ is $k_0\simeq 3.1^{+5.8}_{-2.8} \times
10^{-4}\;\mbox{Mpc}^{-1}$) and, therefore, the domain $R<1$ is favored
by the data. This implies that $R=0.216$ is about $3.6\,\sigma$ from 
the mean value, whereas $R=0.00216$ is further away, being at 
$5.21\,\sigma$, from the mean value. Clearly, had these results been
obtained for the exact case of the Starobinsky model (i.e. in the 
absence of the overall tilt that has been introduced by hand), 
low values of $R$, such as $R=0.00216$, would be ruled out. 
In summary, as is evident from Fig.~\ref{fig:fnl}, the Starobinsky 
model can result in a power spectrum that remains reasonably
consistent with the data while at the same time lead to an $\fnl$ 
that is as large as a few, at most $\simeq {\cal O}(10)$. 
Indeed, it would be interesting to strengthen such a conclusion by
carrying out an explicit comparison of the power spectrum in the 
Starobinsky model with the data and also studying the exact 
implications of an oscillatory, scale dependent, $\fnl$ for 
the non-Gaussianities in the CMB.

\par

Though the Starobinsky model, with parameters that are consistent with 
the data, in itself, may not lead to a large $\fnl$, we believe that 
the Starobinsky model with a much smaller $R$, and hence a rather large 
$\fnl$ can be considered to effectively capture certain aspects of other 
models that are known to lead to a better fit to the data than the 
conventional, power law, primordial spectrum.
An example of such a model would be the punctuated inflationary 
scenario~\cite{pi}.
In punctuated inflation, a step in the power spectrum arises exactly as 
in the Starobinsky model with $R<1$.
However, the step turns out to be sharper (than in the case of the best 
fit value of $R\simeq 0.73$), and the oscillations at the higher level 
of the step, before the spectrum turns scale invariant, are fewer.
The Starobinsky model with a much smaller~$R$ can broadly be considered 
to mimic the sharper step that arises in punctuated inflation.
Therefore, if one naively extends the results of the Starobinsky model 
to punctuated inflation, it suggests that there can exist scenarios 
which lead to a sharp step in the power spectrum (along with certain
characteristic oscillations), an improved fit to the data as well as a 
reasonably large $\fnl$.


\subsection{The hierarchy of contributions to the bi-spectrum}

As we have argued before, prior experience suggests that, when
departures from the slow roll arise, it is the fourth term in the
interaction Hamiltonian~(\ref{eq:Hint}) that leads to the most
significant contribution to the bi-spectrum.  However, such a
conclusion has largely been based on numerical analysis, which
typically involves working with specific values (or, at the most, a
limited range of values) for the parameters of the model concerned.
Since we have been able to arrive at analytic expressions for all the
different contributions to the bi-spectrum for the Starobinsky model,
it is interesting to investigate whether such a conclusion indeed
applies for a wide range of the parameters of the model or if there
exist certain values of the parameters for which other terms can
possibly contribute as much as or even more than the fourth term.

\par

Let us quickly recall some essential properties of the various
contributions to the bi-spectrum, arising out of the interaction
Hamiltonian and the field redefinition.  While the contributions due
to the first, the second and the third terms in the interaction
Hamiltonian involve $\epsilon_1^2$, the fourth term contains
$\epsilon_{1}\,\epsilon_{2}'$, whereas the fifth and the sixth terms
depend on~$\epsilon_1^3$ [cf.~Eqs.~(\ref{eq:cG1})--(\ref{eq:cG6})].
Also, note that, apart from the difference in the dependence on the
slow roll parameter~$\epsilon_1$, the first, the third, the fifth and
the sixth terms depend on the curvature perturbation and its
derivative in the same fashion, viz. they depend linearly on the
curvature perturbation and quadratically on its derivative.  In
contrast, the second term does not depend on the derivative of the
curvature perturbation at all, while the fourth term depends linearly
on the derivative and quadratically on the curvature perturbation
itself.  Further, the additional seventh term arising due to the field
redefinition depends on the late time behavior of $\epsilon_2$ and the
fourth power of the curvature perturbation.  In a slow roll
inflationary scenario, one can anticipate that the first, the second,
the third and the seventh terms lead to contributions of similar
magnitudes to the bi-spectrum, while the contributions due to the
fourth, the fifth and the sixth terms can be expected to be suitably
suppressed in amplitude due to the presence of the additional slow
roll parameter.  In fact, these expectations are broadly corroborated
by the results obtained~\cite{maldacena-2003,ng-ncsf}.

\par

\begin{figure}
\begin{center}
\includegraphics[width=15.0cm]{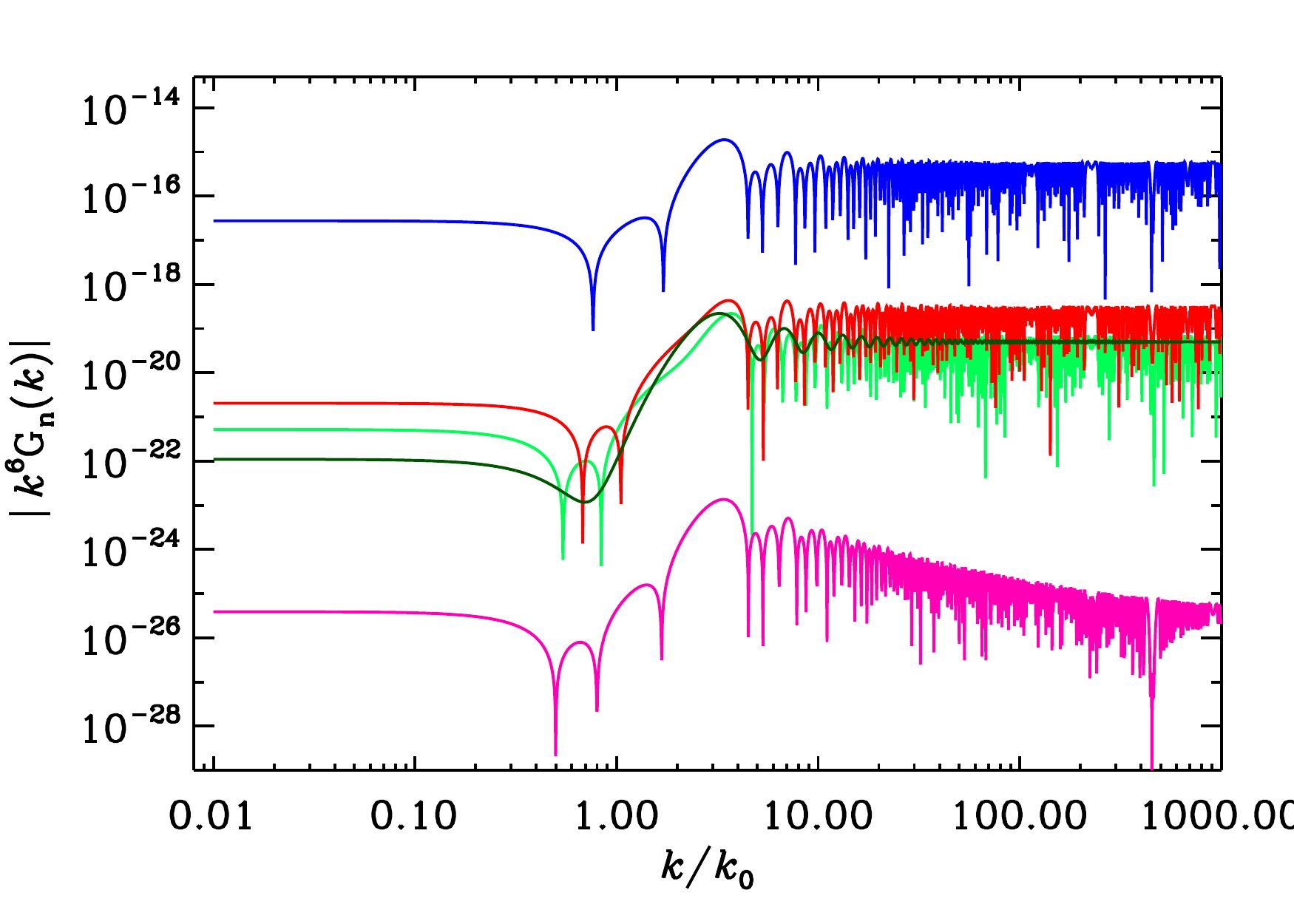}
\caption{The contributions due to the different terms to the
  bi-spectrum---viz. $k^6$ times the absolute values of $G_1+G_3$ (in
  light green), $G_2$ (in red), $G_4$ (in blue), $G_5+G_6$ (in purple)
  and $G_7$ (in dark green)---that we had plotted in the earlier 
  figures (i.e. in Figs.~\ref{fig:abs-G13}, \ref{fig:abs-G2},
  \ref{fig:abs-G4}, \ref{fig:abs-G56} and~\ref{fig:abs-G7}, respectively) 
  for the Starobinsky model, have been assembled here to illustrate the 
  hierarchy in a fast roll regime.
  The hierarchy of the various contributions to the bi-spectrum is evident 
  from the figure.}\label{fig:Gall}
\end{center}
\end{figure}

When there arise deviations from slow roll, as we have repeatedly
mentioned, it is the fourth term that is expected to dominate, with
the first, the second, the third and the seventh terms proving to be
of roughly similar order, but smaller in amplitude than the fourth
term.  The contributions of the fifth and the sixth terms can be
expected to be further smaller in amplitude.  These expectations are
confirmed by Fig.~\ref{fig:Gall}, wherein we have plotted the absolute
values of all the contributions to the bi-spectrum in the equilateral
limit that we have focused on.
It is worth pointing out here that the slightly higher amplitude of
the second term in contrast to the contributions due to the first and
the third terms can be attributed to the fact that the second term
does not involve any derivative of the curvature perturbation.
We shall return to this point a little later in our discussion.

\par

Since the hierarchy of the different contributions changes as one
moves from the slow roll to a fast roll regime, it is interesting to
identify the domain where the hierarchy shifts in the Starobinsky
model.  Recall that, a crucial assumption of the Starobinsky model is
that it is the constant~$V_0$ in the potential which is dominant near 
the discontinuity. Once this condition is satisfied, the departures 
from slow roll can essentially be described in terms of the ratio $R$. 
Upon plotting the various contributions for different $R$, we find that, 
deviations from the slow roll hierarchy begin to occur even for an 
$\vert R-1\vert$ as small as~$10^{-5}$.

\par

\begin{figure}
\begin{center}
\includegraphics[width=15.0cm]{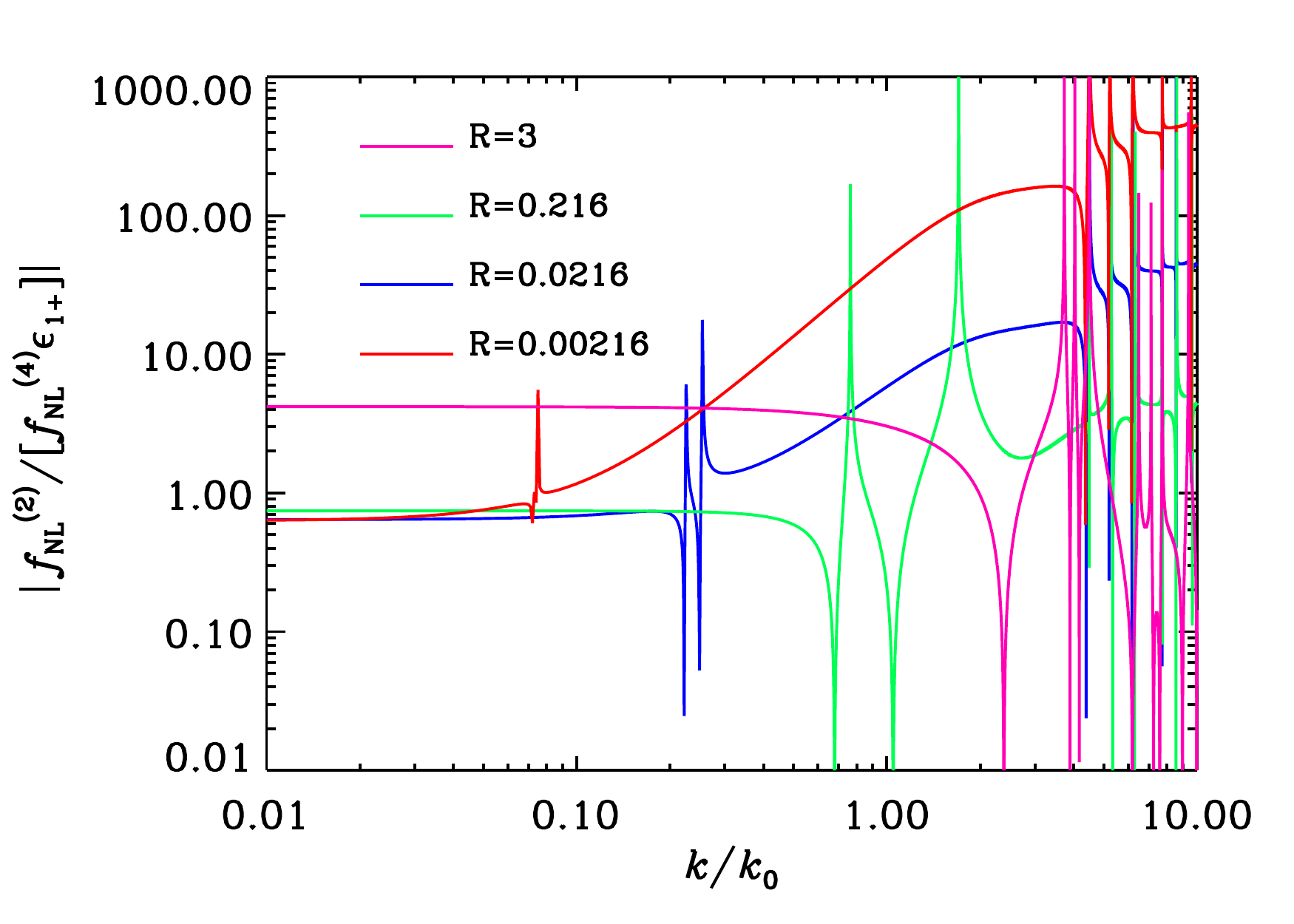}
\caption{The absolute value of the quantity $Q$, which essentially
  reflects the amplitude of the contribution due to the second term
  $G_2$ in contrast to the fourth term $G_4$, has been plotted as a
  function of $k/k_0$ for a few different values of the ratio
  $R=A_-/A_+$. We should stress again that $Q$ depends only on $R$ and
  $k/k_0$. One finds that, deviations from slow roll begin to occur 
  for $\vert R-1\vert$ as small as $10^{-5}$. The upward spikes 
  correspond to the wavenumbers at which $G_4$ vanishes, so that the 
  contribution due to $G_2$ turns dominant. Even apart from these specific
  locations, it is evident that $Q$ can be as large as $10^2$ in a
  domain which admits fast roll. Since $\epsilon_{1+}$ can be possibly
  as large as $10^{-2}$ with the assumptions and
  approximations of the Starobinsky model continuing to remain valid,
  this indicates that $G_2$ can be of the order of $G_4$ in a fast
  roll regime.}\label{fig:Q}
\end{center}
\end{figure}

With the analytic results that we have at hand, it is also worthwhile
to investigate whether the fourth term remains the dominant term for
all values of the parameters of the Starobinsky model when deviations
from slow roll occur or if there exist regimes where the hierarchy is
mixed or even, possibly, absent.  Motivated by such an aim, let us
enquire if, say, the second term $G_2$ can probably be as large as the
fourth term $G_4$ for any set or a range of parameters of the
Starobinsky model.  Let us consider, for instance, the ratio
\begin{equation}
Q(k)=\f{G_2(k)}{G_4(k)\,\epsilon_{1+}\;}
=\f{\fnl^{{\rm eq}(2)}(k)}{\fnl^{{\rm eq}(4)}(k)\,
\epsilon_{1+}},
\end{equation}
where $\fnl^{{\rm eq}(2)}$ and $\fnl^{{\rm eq}(4)}$ are the
contributions to the non-Gaussianity parameter $\fnl$ due to the terms
$G_2$ and $G_4$ in the equilateral limit.  As we have discussed, the
first slow roll parameter before the transition, viz. $\epsilon_{1+}$,
is expected to be small.  In plotting the earlier figures, we have
worked with parameters such that $\epsilon_{1+}\simeq10^{-4}$
(cf.~Fig.~\ref{fig:potstaro}).  But, the various assumptions and the
approximations of the Starobinsky model will remain valid even if
$\epsilon_{1+}$ is larger, say, about $10^{-2}$ or so.  If we choose
$\epsilon_{1+} \simeq10^{-2}$, we can have $G_2\simeq G_4$, provided
$Q\simeq 10^2$.  Interestingly, we find that the quantity $Q$ can be
expressed completely in terms of $R$ and, as in the case of all the
other perturbed quantities, its dependence on the wavenumber arises
only through the ratio $k/k_0$. In Fig.~\ref{fig:Q}, we have plotted
the absolute value of the quantity $Q$ as a function of $k/k_0$
for a few different values of the ratio~$R$.  It is clear from the
figure that, there does exist ranges of parameters of the Starobinsky
model for which $Q\simeq 10^2$ and, therefore, $G_2 \simeq G_4$, in
domains that could admit departures from slow roll (i.e. when $R$
deviates sufficiently from unity). To our knowledge, this conclusion
is new, since, in the literature, $G_4$ has always been considered to 
be the dominant term when deviations from slow roll occur.
Moreover, we believe that this is the first time that this phenomenon 
has been explicitly illustrated. This suggests that, it is plausible 
that the hierarchy is different in different regions of the parameter 
space of a model. However, a cautionary remark needs to be added to 
arriving at such a conclusion. Recall that, in our evaluation of 
$\cG_4$, we had neglected certain terms in the quantity $f_k'{}^{-}$ 
[see our remarks that immediately follow Eq.~(\ref{eq:fkp-at})]. As a 
result, a concern could arise that, for the parameters discussed
above, the neglected terms contribute sufficiently to restore the 
hierarchy wherein $G_4$ is the dominant term during fast roll.
But, we have also analyzed the issue using a numerical 
code~\cite{hazra-2011} and our preliminary results seem to confirm 
the above conclusion, viz. that the neglected terms are not 
significant enough to restore the commonly expected fast roll 
hierarchy. 

\par

\begin{figure}
\begin{center}
\includegraphics[width=13.0cm]{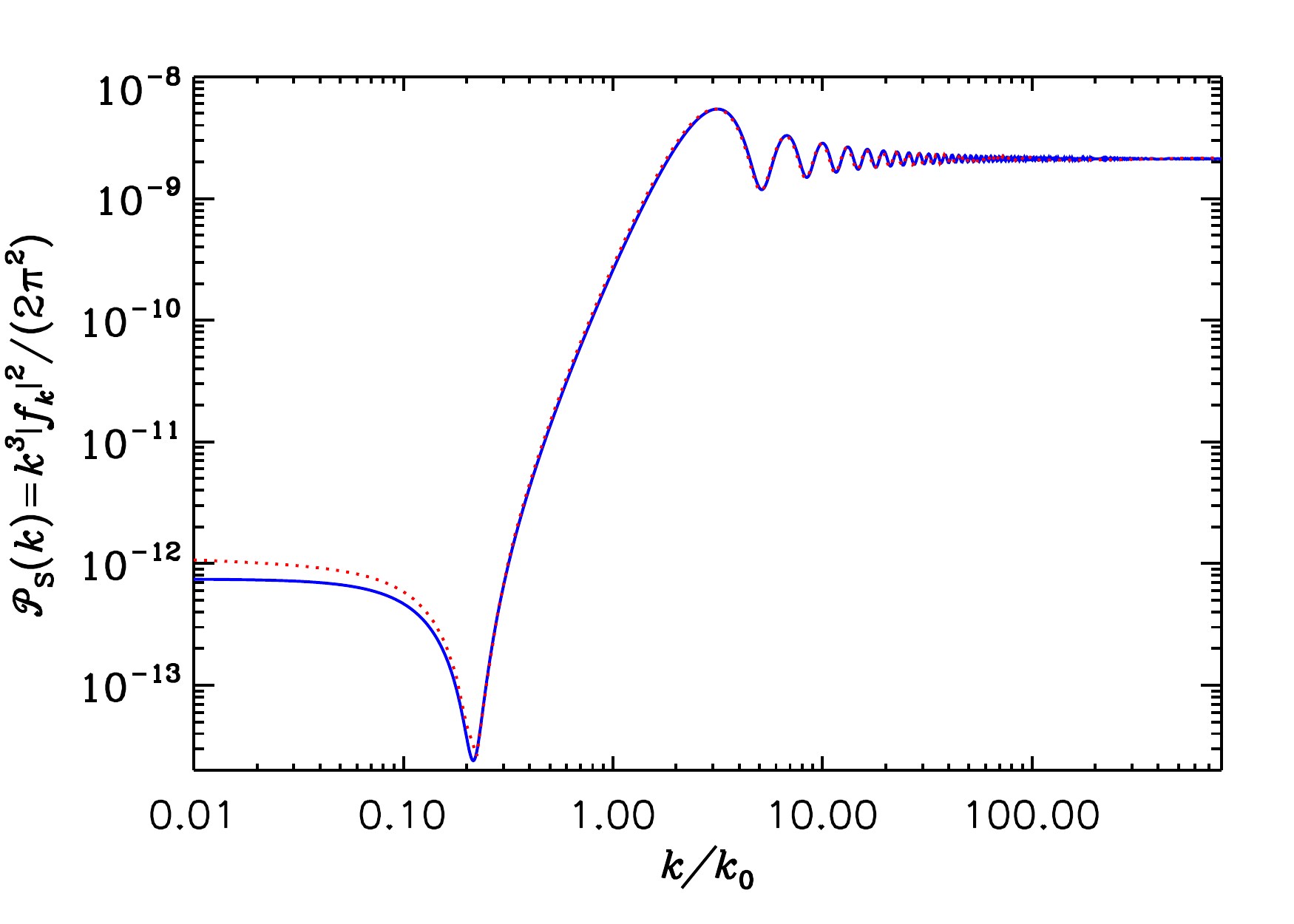}
\vskip -15pt
\includegraphics[width=13.0cm]{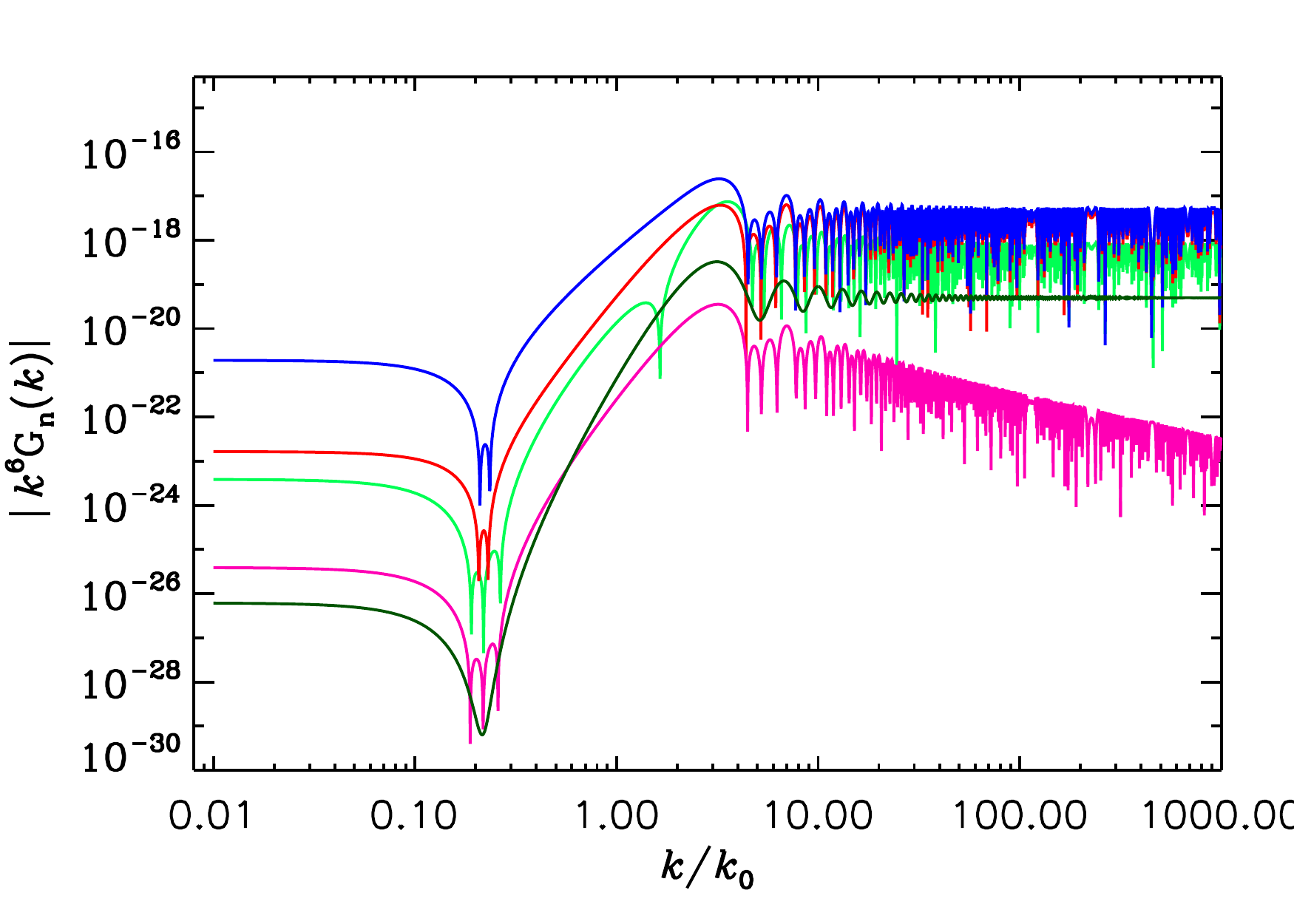}
\vskip -15pt
\caption{The power spectrum (on top) and the contributions due to the
  different terms to the bi-spectrum (below) for a case wherein
  $Q\simeq 10^2$.  The power spectrum and the different contributions
  to the bi-spectrum have been plotted in the same fashion as in
  Figs.~\ref{fig:spec} and~\ref{fig:Gall}. In arriving at these
  plots, we have worked with the same values of $H_0$ and $A_-$ as 
  in Figs.~\ref{fig:spec} and \ref{fig:Gall}, but we have set 
  $A_+/\Mp^3=3.87\times10^{-13}$, which corresponds to $R=0.0188$.
  The match between the analytical and the numerical results in the
  case of the power spectrum is rather good, which indicates that the
  assumptions and approximations of the Starobinsky model are quite
  valid for the parameters that we are working with here.
  Interestingly, in contrast to Fig.~\ref{fig:Gall}, where $G_4$ was
  the dominant term, here we find that the contribution due to $G_2$
  is of the same order as $G_4$ at large wavelengths.}
\label{fig:Gall-H} 
\end{center}
\end{figure}

In order to reinforce and also to explicitly illustrate the above 
conclusion, it is important to arrive at a specific set of parameters 
of the Starobinsky model which allow fast roll as well as lead to 
$G_2$ and $G_4$ that are of the same order. But, in order to converge
on a viable set of parameters, a few points concerning certain aspects
of the Starobinsky model and the results that we have obtained requires 
some emphasis. In particular, as $k/k_0\to 0$, we find that $Q$ has the 
following simple form:
\begin{equation}
\lim_{k/k_0\,\to\, 0}\;
Q(k)= \f{17 - 27\,R^2}{27 - 27\,R}.
\end{equation}
In this limit, $Q\simeq 17/27\simeq 0.63$ for small $R$, while
$Q\simeq R$, when $R$ is large. These behavior are clearly reflected
in Fig.~\ref{fig:Q}.  So, if we require $Q$ to be large, we can either
work with a high enough $R$ and focus on small wavenumbers, or choose
an $R$ that is suitably smaller than unity which permits sufficient
deviation from slow roll and also possibly leads to $G_2 \simeq G_4$
at large wavenumbers. However, recall that the regime $R<1$ seems to be
favored by the CMB data. With these points in mind, let us now focus on 
arriving at a set of parameters in the $R < 1$ domain which leads to 
$Q\simeq 10^2$.  Since, given a $H_0$, $A_-$ is already determined by 
COBE normalization, choosing an $\epsilon_{1+}$ fixes $A_{+}$.  We 
find that, if we work with the earlier values of $H_0$ and $A_-$ (listed 
in the caption of Fig.~\ref{fig:potstaro}), while set $\epsilon_{1+}\simeq 
1/75\simeq 0.013$, corresponding to $A_+/\Mp^3=3.87\times 10^{-13}$ and
$R=0.0188$, the hierarchy of the contributions is indeed different,
with $G_2$ being of the same order as $G_4$ at large wavenumbers. This
point is evident from Fig.~\ref{fig:Gall-H} wherein we have plotted
the various contributions $G_n$ for these values of the parameters. Of
course, such a low value of $R$ lies about $5\, \sigma$ away from the 
best fit value of $R\simeq 0.73$ that we had mentioned earlier. Given 
that the Starobinsky model in itself has not been directly constrained 
(only certain variations around it have been tested precisely), this
could indicate that the above choice of $R$ is possibly not completely 
ruled out by the data. However, in the case of the Starobinsky model, 
it seems to us that, an altered hierarchy of the contributions to the
bi-spectrum is systematically associated with regions of the parameter 
space that is not favored by the CMB data. If this is indeed true, 
this obviously tones down the importance of this modification to the
hierarchy in the Starobinsky model. But, as we had discussed in the previous 
sub-section, the Starobinsky model with a smaller~$R$ may be considered 
to mimic other models that provide a good fit to the data. It is 
possible that one may encounter such altered hierarchies in these 
models. 

\par

Before we conclude, let us briefly touch upon the possible reasons 
for the variations to the hierarchy.
We find that the understanding gained on the hierarchy of the 
different contributions to the bi-spectrum has been largely 
based on the behavior of the slow roll parameters.
Needless to mention, in addition to the slow roll parameters, the 
various contributions involve the curvature perturbation and its 
derivative as well.
In a slow roll inflationary scenario, the amplitude of the curvature 
perturbation evolves monotonically once it leaves the Hubble radius. 
In particular, while the amplitude of the mode $f_k$ quickly goes 
to a constant value, its time derivative $f_k'$ dies down as
${\rm e}^{-2\,N}$ at super Hubble scales. 
Because of such a monotonic behavior of the curvature perturbation 
during slow roll, in such situations, the hierarchy is largely 
dependent on the slow parameters themselves.   
However, when departures from slow arise, it is known that the modes 
which leave the Hubble radius just before or during the periods of 
deviations from slow roll can evolve strongly around the time of
Hubble exit.
In fact, at super Hubble scales, the amplitude of these modes can 
be enhanced or suppressed compared to their amplitudes at Hubble 
exit (in this context, see, for instance, Refs.~\cite{e-dfsr}).
Therefore, it is seems that, when departures from slow roll occur,
the hierarchy can be different based on a mixture of the non-trivial
evolution of the slow roll parameters and the curvature perturbation.
We believe that this is an interesting aspect that demands further 
investigation. 


\section{Summary and outlook}\label{sec:so}

As we had discussed in some detail in the introductory section, models
that lead to features in the primordial spectrum gain importance due
to the fact that certain features can lead to a better fit to the CMB
data than the conventional, nearly scale invariant scalar power
spectrum, as is generated by slow roll inflation.  The generation of
features in the primordial spectrum requires one or more periods of of
fast roll.  While the scalar power spectrum and the bi-spectrum can be
evaluated in a model independent fashion in the slow roll
approximation, such a model independent approach seems difficult when
departures from slow roll arise.  This is essentially due to the fact
that the deviations from slow roll can occur in a multitude of forms
and it is impossible to describe all possible deviations in terms of a
limited number of variables or parameters (for a broader effort, that
attempts to capture a certain class of departures from slow roll, see
Ref.~\cite{recent-fips}). In fact, in the literature, numerical
computations are often resorted to in order to investigate scenarios
involving fast roll.

\par

In such a situation, the Starobinsky model provides a welcome relief
as it allows the background as well as the scalar power spectrum to be
evaluated analytically to a very good approximation, even though it
contains departures from slow roll. In this work, we have shown that
the scalar bi-spectrum can also be computed analytically.
Simultaneously, we have also been working towards numerically
computing the scalar bi-spectrum and the non-Gaussianity parameter
$\fnl$ in the Starobinsky model. Preliminary investigations indicate
that the analytic expressions we have obtained match the numerical
results quite well~\cite{hazra-2011}, which reflect the extent of the
accuracy of the assumptions and approximations of the Starobinsky
model.

\par

Interestingly, we have found that, in the Starobinsky model, for
certain values of the parameters, the non-Gaussianity parameter $\fnl$
in the equilateral limit can be as large as indicated by the currently
observed mean values, with the possible limitation that this
occurs in a region of the parameter space that is not favored by the
current CMB data.  We had focused here on the equilateral
limit and, clearly, it is imperative that we extend the analysis to
other configurations such as, say, the squeezed
limit~\cite{martin-wip}.  Further, there exist a few other models in
the literature which allow the background and the perturbations to be
evaluated analytically, and it is worthwhile to evaluate the
bi-spectrum in these models too.  We are currently investigating these
issues.


\ack{JM and LS wish to thank the Harish-Chandra Research Institute,
  Allahabad, India, and the Institut d'Astrophysique de Paris, France,
  for hospitality, respectively, where part of this work was carried
  out. JM would like to thank Guillaume Faye for help on numerical
  issues.  LS wishes to thank David Seery for discussions.}


\appendix


\section{Evaluation of the integrals}

This appendix contains some additional information pertaining to the 
evaluation of certain non-trivial integrals we had encountered in the 
text.


\subsection{Evaluation of $I_{13}$ and $J_{13}$}\label{app:13}

In this sub-section, we shall provide a few details regarding the 
evaluation of the integrals $I_{13}$ and $J_{13}$.

\par

Let us start with the expression~(\ref{eq:I13}) for $I_{13}$.  As we
had mentioned, the integral can be evaluated by writing~$P_1$ as a
polynomial in the variable $t_n$, and computing the integral term by
term.  If we write
\begin{equation}
P_1\left(\frac{-t_n}{3\,i\,k}+\tau_n\right)
=k^4\,\sum_{m=0}^7c_{nm}\,t_n^m,\label{eq:P1-e}
\end{equation}
we find that the coefficients $c_{nm}$ are given by
\begin{eqnarray}
c_{n0} 
&= -\,27\,i\,\varepsilon\,\delta^{3}
-27\, \varepsilon\,\delta^{2}\,{\rm e}^{i\,\theta _n}
+9\,\l(i\,\varepsilon\,\delta +\delta^{4}\r)\,{\rm e}^{2\,i\,\theta_n},
\label{eq:cn0}\\
c_{n1} 
&= -\,42\,i\,\varepsilon\,\delta^{3}
+\l(-27\,\varepsilon\,\delta^{2}+6\,i\,\delta^{5}\r)\,{\rm e}^{i\,\theta_n}
+\l(6\,i\,\varepsilon\,\delta +27\,\delta ^{4}\right){\rm e}^{2\,i\,\theta_n},\\
c_{n2} 
&= -\,22\,i\,\varepsilon\,\delta^3-\delta^6
-9\,\l(\varepsilon\,\delta^2-i\,\delta^5\r)\,{\rm e}^{i\,\theta_n}
+\l(i\,\varepsilon\,\delta+22\,\delta^4\r)\,{\rm e}^{2\,i\,\theta_n},\\
c_{n3} 
&= -\,\frac{132}{27}\,i\,\varepsilon\,\delta^3-\delta^6
+\l(-\varepsilon\,\delta^2+\frac{44}{9}\,i\,\delta^5\right)\,
{\rm e}^{i\theta_n}
+\frac{70}{9}\,\delta^4\,{\rm e}^{2\,i\,\theta_n},\\
c_{n4} 
&= -\,\frac{33}{81}\,i\,\varepsilon\,\delta ^3-\frac{11}{27}\,\delta^6
+\frac{35}{27}\,i\,\delta^5\,{\rm e}^{i\,\theta_n}
+\frac{35}{27}\,\delta^4\,{\rm e}^{2\,i\,\theta_n},\\
c_{n5} 
&= -\,\frac{7}{81}\,\delta^6
+\frac{14}{81}\,i\,\delta^5\,{\rm e}^{i\,\theta_n}
+\frac{7}{81}\,\delta^4\,{\rm e}^{2\,i\,\theta_n},\\
c_{n6} 
&= -\,\frac{7}{729}\,\delta^6
+\frac{7}{729}\,i\,\delta^5\,{\rm e}^{i\,\theta_n}\\
c_{n7} &= -\,\frac{1}{2187}\,\delta^6,\label{eq:cn7}
\end{eqnarray}
where $\delta\equiv (\vert\rho\vert/k)=(\vert \Delta
A\vert/A_-)^{1/3}\, (k_0/k)$ and $\varepsilon=1$ if $\Delta A<0$ and
$\varepsilon=-1$ if $\Delta A>0$.  (The quantity $\varepsilon$ that we
have introduced here should not be confused with the slow roll
parameters.)  Moreover, when $\Delta A>0$, then we have
$\theta_1=\pi/3$, $\theta_2 =\pi$, and $\theta_3 =5\,\pi/3$. On the
other hand, if $\Delta A<0$, then we find that $\theta_1=0$,
$\theta_2=2\,\pi/3$ and $\theta_3=4\,\pi/3$. Upon using the
expression~(\ref{eq:P1-e}), $I_{13}$ can be written as
\begin{eqnarray}
I_{13}(k)
&=&-\frac{A_-\,\vert \rho\vert^2\,k^4}{\sqrt{18}\,H_0^2\,\Mp\, \rho^3}\,
\sum_{n=1}^3\, \sum_{m=0}^7\,b_n\,c_{nm}\,{\rm e}^{3\,i\,k\,\tau_n}\,
\nonumber \\ & & \times
\int_{3\,i\,k/k_0+3\,i\,k\,\tau_n}^{-3\,i\,k\,\eta_{\rm e}+3\,i\,k\,\tau_n}\,
{\rm d}t_n\,t_n^{m-1}\,{\rm e}^{-t_n}
\end{eqnarray}
which can be integrated to yield
\begin{eqnarray}
I_{13}(k)
&=&-\frac{A_-\,\vert \rho\vert^2\,k^4}{\sqrt{18}\,H_0^2\,\Mp\, 
\rho^3}
\sum _{n=1}^3\,\sum _{m=0}^7\,b_n\,c_{nm}\,{\rm e}^{3\,i\,k\,\tau_n}
\Biggl[\Gamma \left(m,\frac{3\,i\,k}{k_0}+3\,i\,k\,\tau_n\right)
\nonumber \\ & &
-\Gamma \left(m,-3\,i\,k\,\eta_{\rm e}+3\,i\,k\,\tau_n\right)\Biggr],
\end{eqnarray}
where $\Gamma(m,z)$ is the incomplete Gamma function (see, for
example, Ref.~\cite{gradshteyn-1980})
\begin{equation}
\Gamma(m,z)
\equiv \int_z^{\infty}{\rm d}t\;{\rm e}^{-t}\,t^{m-1}.
\end{equation}
If $m$ in a non-vanishing integer, the incomplete Gamma function
reduces to an elementary function~\cite{gradshteyn-1980}, viz.
\begin{equation}
\Gamma (m,z)=(m-1)!\; {\rm e}^{-z}\; e_{m-1}(z),\label{eq:Gf}
\end{equation}
where $e_{n}(z)$ is the exponential sum function defined as
\begin{equation}
e_n(z)\equiv \sum _{\ell=0}^{n}\,\frac{z^\ell}{\ell!}.\label{eq:esf}
\end{equation}
However, if $m=0$, then
\begin{equation}
\Gamma(0,z)\equiv E_1(z)
=\int_z^{\infty}\f{{\rm d}t}{t}\,{\rm e}^{-t},\label{eq:eif}
\end{equation}
where $E_1(z)$ is the exponential integral function~\cite{gradshteyn-1980}. 
For this reason, it is convenient to split the sum over $m$ into two 
parts as follows:
\begin{eqnarray}
I_{13}(k)
&=&-\frac{A_-\,\vert \rho\vert^2\,k^4}{\sqrt{18}\,H_0^2\,\Mp\,\rho^3}\,
\sum _{n=1}^3\,b_n\,{\rm e}^{3\,i\,k\,\tau_n}\,
\Biggl\{c_{n0}\,E_1\l(\frac{3\,i\,k}{k_0}+3\,i\,k\,\tau_n\r)
\nonumber \\ & &
-c_{n0}\,E_1\l(3\,i\,k\,\tau_n\right)
+\,\sum _{m=1}^7\,c_{nm}\,
\Biggl[\Gamma \left(m,\frac{3\,i\,k}{k_0}+3\,i\,k\,\tau_n\right)
\nonumber \\ & &
-\Gamma \left(m,3\,i\,k\,\tau_n\right)\Biggr]\Biggr\},
\end{eqnarray}
where we have taken the limit $\eta_{\rm e}\rightarrow 0$. 
In terms of the exponential sum function $e_n(z)$ and the exponential 
integral function $E_1(z)$, we can rewrite the above expression for
$I_{13}$ as
\begin{eqnarray}
I_{13}(k)
&=&-\frac{A_-\,\vert \rho\vert^2\,k^4}{\sqrt{18}\,H_0^2\,\Mp\, \rho^3}
\Biggl\{-\sum _{n=1}^3\,b_n\,\Biggl[c_{n0}\,{\rm e}^{3\,i\,k\tau_n}\,
E_1\l(3\,i\,k\,\tau_n\r)
\nonumber \\ & &
+\sum _{m=1}^7\,c_{nm}\,(m-1)!\;e_{m-1}\l(3\,i\,k\,\tau_n\r)\Biggr]
\nonumber \\ & &
+\,{\rm e}^{-3\,i\,k/k_0}
\sum _{n=1}^3\,b_n\,
\Biggl[c_{n0}\,{\rm e}^{3\,i\,k/k_0+3\,i\,k\,\tau_n}\,
E_1\l(\frac{3\,i\,k}{k_0}+3\,i\,k\,\tau_n\right)
\nonumber \\ & &
+\sum _{m=1}^7\,c_{nm}\;(m-1)!\,
e_{m-1}\l(\frac{3\,i\,k}{k_0}+3\,i\,k\,\tau_n\r)\Biggr]\Biggr\}
\end{eqnarray}
which is the expression~(\ref{eq:I13-fv}) we have quoted in the text.


\par

Let us now turn to the evaluation of $J_{13}$ described by the
integral~(\ref{eq:J13-ov}).  In this case, we can perform the change
of variable $t_n\equiv -i\,k\,\left(\tau-\tau_n\right)$, which leads
to
\begin{eqnarray}
J_{13}(k)
&=&-\frac{A_-\,\vert \rho\vert^2}{\sqrt{18}\,H_0^2\,\Mp\,\rho^3}\,
\sum _{n=1}^3\,b_n\,{\rm e}^{i\,k\,\tau_n}\,
\nonumber \\ & & \times
\int_{i\,k/k_0+i\,k\,\tau_n}^{-i\,k\,\eta_{\rm e}+i\,k\,\tau_n}\,
\frac{{\rm d}t_n}{t_n}\,{\rm e}^{-t_n}\,P_2\l(\frac{-t_n}{i\,k}+\tau_n\r).
\end{eqnarray}
We can write, as in the case of the function $P_1$,
\begin{equation}
P_2\left(\frac{-t_n}{i\,k}+\tau_n\right)
=k^4\,\sum_{m=0}^7\,d_{nm}\,t_n^m,
\end{equation}
where the coefficients $d_{nm}$'s are given by
\begin{eqnarray}
d_{n0} 
&=& -27\,i\,\varepsilon\,\delta^{3}+27\,\varepsilon\,\delta ^{2}\,
{\rm e}^{i\,\theta _n}
+27\,\l(-i\,\varepsilon\,\delta +\delta^{4}\r)\,{\rm e}^{2\,i\,\theta_n},
\label{eq:dn0}\\
d_{n1} 
&=& 54\,i\,\varepsilon\, \delta^{3}
+\l(99\,\varepsilon\,\delta ^{2}+54\,i\,\delta^{5}\r)\,
{\rm e}^{i\,\theta_n}
+\l(-18\,i\,\varepsilon\,\delta +81\,\delta^{4}\r)\,
{\rm e}^{2\,i\,\theta_n},\\
d_{n2} 
&=& 162i\varepsilon\delta^3-27\delta^6
+\l(45\varepsilon\delta^2+81i\delta^5\r)
{\rm e}^{i\,\theta_n}
+\l(9i\varepsilon \delta-54\delta^4\r)
{\rm e}^{2i\theta_n},\qquad\\
d_{n3} 
&=& 60\,i\,\varepsilon\,\delta^3-27\,\delta^6
+\l(-27\,\varepsilon\,\delta^2-36\,i\,\delta^5\r)\,
{\rm e}^{i\,\theta_n}
-150\,\delta^4\,{\rm e}^{2\,i\,\theta_n},\\
d_{n4} 
&=& -33\,i\,\varepsilon\,\delta^3+9\,\delta^6
-75\,i\,\delta^5\,{\rm e}^{i\,\theta_n}
-45\,\delta^4\,{\rm e}^{2\,i\,\theta_n},\\
d_{n5} 
&=& 15\,\delta^6
-18\,i\,\delta^5\,{\rm e}^{i\,\theta_n}
+21\,\delta^4\,{\rm e}^{2\,i\,\theta_n},\\
d_{n6} 
&=& 3\,\delta^6
+7\,i\,\delta^5\,{\rm e}^{i\,\theta_n},\\
d_{n7} &=& -\delta^6.\label{eq:dn7}
\end{eqnarray}
It is clear that the coefficients $d_{mn}$ and $c_{mn}$ have a very
similar structure. The integral $J_{13}(k)$ can then be rewritten as
\begin{eqnarray}
J_{13}(k)
&=&-\frac{A_-\,\vert \rho\vert^2\,k^4}{\sqrt{18}\,H_0^2\,\Mp\,\rho^3}\,
\sum _{n=1}^3 \sum _{m=0}^7\,b_n\,d_{nm}\,{\rm e}^{i\,k\,\tau_n}\,
\nonumber \\ & & \times
\int_{i\,k/k_0+i\,k\,\tau_n}^{-i\,k\,\eta_{\rm e}+i\,k\,\tau_n}
{\rm d}t_n\,t_n^{m-1}\,{\rm e}^{-t_n}, 
\end{eqnarray}
which can be integrated, as earlier, to yield 
\begin{eqnarray}
J_{13}(k)
&=&-\frac{A_-\,\vert \rho\vert^2\,k^4}{\sqrt{18}\,H_0^2\,\Mp\,\rho^3}\,
\sum _{n=1}^3\sum _{m=0}^7\,b_n\,d_{nm}\,{\rm e}^{i\,k\,\tau_n}
\nonumber \\ & &
\times\, \l[\Gamma\l(m,\frac{i\,k}{k_0}+i\,k\,\tau_n\r)
-\Gamma \l(m,-i\,k\,\eta _{\rm e}+i\,k\,\tau_n\r)\r].
\end{eqnarray}
The rest of the calculation in arriving at the final 
expression~(\ref{eq:J13-fv}) proceeds just as in the 
case of $I_{13}$. 


\subsection{The coefficients $f_n$ and $g_n$}\label{app:56}

The coefficients $f_n$ are given by
\begin{eqnarray}
f_0 &= 1,\label{eq:f0}\\
f_1 &= -i+\frac{6\,\rho^3}{k^3},\\
f_2 &= -\frac{12\,i\,\rho^3}{k^3}+\frac{9\,\rho^6}{k^6} ,\\
f_3 &= -\frac{9\,\rho^3}{k^3}-\frac{27\,i\,\rho^6}{k^6},\\
f_4 &= \frac{3\,i\,\rho^3}{k^3}-\frac{39\,\rho^6}{k^6},\\
f_5 &= \frac{33\,i\,\rho^6}{k^6}-\frac{9\,\rho^9}{k^9},\\
f_6 &= \frac{15\,\rho^6}{k^6}+\frac{27\,i\,\rho^9}{k^9},\\
f_7 &= -\frac{3\,i\,\rho^6}{k^6}+\frac{33\,\rho^9}{k^9},\\
f_8 &= -\frac{21\,i\,\rho^9}{k^9},\\
f_9 &= -\frac{7\,\rho^9}{k^9},\\
f_{10} &= \frac{i\,\rho^9}{k^9},\label{eq:f10}
\end{eqnarray}
while $g_n$ are given by
\begin{eqnarray}
g_0 &= 3,\label{eq:g0}\\
g_1 &= -i+\frac{18\,\rho^3}{k^3},\\
g_2 &= -\frac{12\,i\,\rho^3}{k^3}+\frac{27\,\rho^6}{k^6},\\
g_3 &= -\frac{3\,\rho^3}{k^3}-\frac{27\,i\,\rho^6}{k^6},\\
g_4 &= \frac{3\,i\,\rho^3}{k^3}-\frac{9\,\rho^6}{k^6},\\
g_5 &= -\frac{3\,\,i\,\rho^6}{k^6}-\frac{27\,\rho^9}{k^9},\\
g_6 &= -\frac{3\,\rho^6}{k^6}+\frac{27\,i\,\rho^9}{k^9},\\
g_7 &= -\frac{3\,i\,\rho^6}{k^6}-\frac{9\,\rho^9}{k^9},\\
g_8 &= \frac{15\,i\,\rho^9}{k^9},\\
g_9 &= \frac{3\,\rho^9}{k^9},\\
g_{10} &= \frac{i\,\rho^9}{k^9}.\label{eq:g10}
\end{eqnarray}


\end{document}